\let\ifarxiv=\iftrue     
\pdfoutput=1
\documentclass[12pt,a4paper]{article}

\ifarxiv\ifnum\pdfoutput=1\else
\PassOptionsToPackage{hypertex}{hyperref}
\PassOptionsToPackage{draft}{graphicx}
\usepackage{showkeys}
\fi\fi

\usepackage{amsmath,amssymb}
\usepackage[bookmarks=true,hyperfigures=true]{hyperref}
\usepackage{graphicx}
\usepackage[nosort]{cite}
\usepackage[bulletsep]{collref}

\usepackage[a4paper,text={173mm,216mm},centering]{geometry}

\allowdisplaybreaks[3]

\numberwithin{equation}{section}

\usepackage[font=small,labelfont=bf,width=0.85\textwidth]{caption}

\let\oldbfseries=\bfseries
\let\oldmdseries=\mdseries
\let\oldnormalfont=\normalfont
\renewcommand{\bfseries}{\oldbfseries\boldmath}
\renewcommand{\mdseries}{\oldmdseries\unboldmath}
\renewcommand{\normalfont}{\oldnormalfont\unboldmath}

\makeatletter
\newlength{\apb@width}
\newcommand{\autoparbox}[2][c]{\settowidth{\apb@width}{#2}\parbox[#1]{\apb@width}{#2}}
\newcommand{\includegraphicsbox}[2][]{\autoparbox{\includegraphics[#1]{#2}}}
\makeatother

\DeclareMathSymbol{\Gamma}{\mathalpha}{letters}{"00}
\DeclareMathSymbol{\Delta}{\mathalpha}{letters}{"01}
\DeclareMathSymbol{\Theta}{\mathalpha}{letters}{"02}
\DeclareMathSymbol{\Lambda}{\mathalpha}{letters}{"03}
\DeclareMathSymbol{\Xi}{\mathalpha}{letters}{"04}
\DeclareMathSymbol{\Pi}{\mathalpha}{letters}{"05}
\DeclareMathSymbol{\Sigma}{\mathalpha}{letters}{"06}
\DeclareMathSymbol{\Upsilon}{\mathalpha}{letters}{"07}
\DeclareMathSymbol{\Phi}{\mathalpha}{letters}{"08}
\DeclareMathSymbol{\Psi}{\mathalpha}{letters}{"09}
\DeclareMathSymbol{\Omega}{\mathalpha}{letters}{"0A}

\ifx\genfrac\sdflkaj\else\fi
\newcommand{\sfrac}[2]{{\textstyle\frac{#1}{#2}}}
\newcommand{\half}{\sfrac{1}{2}}
\newcommand{\ihalf}{\sfrac{i}{2}}
\newcommand{\quarter}{\sfrac{1}{4}}
\newcommand{\gammafn}{\mathrm{\Gamma}}

\newcommand{\alg}[1]{\mathfrak{#1}}
\newcommand{\grp}[1]{\mathrm{#1}}
\newcommand{\cas}{C\indup{A}}

\newcommand{\indup}[1]{_{\mathrm{#1}}}

\newcommand{\supup}[1]{^{\mathrm{#1}}}

\newcommand{\brk}[1]{(#1)}
\newcommand{\lrbrk}[1]{\left(#1\right)}
\newcommand{\bigbrk}[1]{\bigl(#1\bigr)}

\newcommand{\comm}[2]{[#1,#2]}
\newcommand{\lrcomm}[2]{\left[#1,#2\right]}
\newcommand{\bigcomm}[2]{\bigl[#1,#2\bigr]}

\newcommand{\gcomm}[2]{[#1,#2\}}
\newcommand{\abs}[1]{|#1|}

\newcommand{\bigabs}[1]{\bigl|#1\bigr|}
\newcommand{\bigeval}[1]{#1\big|}

\newcommand{\set}[1]{\{#1\}}

\newcommand{\sbra}[1]{\langle #1|}
\newcommand{\sket}[1]{|#1\rangle}
\newcommand{\cbra}[1]{[#1|}
\newcommand{\cket}[1]{|#1]}
\newcommand{\sprod}[2]{\langle#1,#2\rangle}
\newcommand{\sprods}[2]{\langle#1#2\rangle}
\newcommand{\cprod}[2]{[#1,#2]}
\newcommand{\cprods}[2]{[#1#2]}

\newcommand{\superN}{\mathcal{N}}
\newcommand{\Tr}{\mathop{\mathrm{Tr}}}
\newcommand{\sign}{\mathop{\mathrm{sign}}}
\newcommand{\disc}{\mathop{\mathrm{disc}}}

\newcommand{\Li}{\mathop{\mathrm{Li}}\nolimits}
\newcommand{\order}[1]{\mathcal{O}(#1)}
\newcommand{\gen}[1]{\mathrm{#1}}
\newcommand{\genY}[1]{\widehat{\gen{#1}}}
\newcommand{\genD}[1]{\widetilde{\gen{#1}}}
\newcommand{\rng}[2]{^{#1}_{#2}}
\newcommand{\sinv}[2]{t\rng{#1}{#2}}

\newcommand{\gengend}{G}
\newcommand{\gengen}{\gen{\gengend}}
\newcommand{\gengenY}{\genY{\gengend}}
\newcommand{\prefac}{ \left(\frac{i}{4\pi^2\cas}\right)}

\newcommand{\nn}{\nonumber}
\newcommand{\nln}{\nonumber\\}
\newcommand{\nl}[1][0pt]{\nonumber\\[#1]&\hspace{-4\arraycolsep}&\mathord{}}

\newcommand{\earel}[1]{\mathrel{}&\hspace{-2\arraycolsep}#1\hspace{-2\arraycolsep}&\mathrel{}}
\newcommand{\eq}{\earel{=}}

\def\[{\begin{equation}}
\def\]{\end{equation}}
\def\<{\begin{eqnarray}}
\def\>{\end{eqnarray}}

\makeatletter
\def\mr@ignsp#1 {\ifx\:#1\@empty\else #1\expandafter\mr@ignsp\fi}%
\newcommand{\multiref}[1]{\begingroup
\xdef\mr@no@sparg{\expandafter\mr@ignsp#1 \: }%
\def\mr@comma{}%
\@for\mr@refs:=\mr@no@sparg\do{\mr@comma\def\mr@comma{,}\ref{\mr@refs}}%
\endgroup}
\makeatother

\ifx\href\asklfhas\newcommand{\href}[2]{#2}\fi
\newcommand{\arxivlink}[1]{\href{http://arxiv.org/abs/#1}{arxiv:#1}}

\newcommand{\hypref}[2]{\ifx\href\asklfhas #2\else\href{#1}{#2}\fi}

\newcommand{\secref}[1]{Sec.~\multiref{#1}}

\newcommand{\appref}[1]{App.~\multiref{#1}}

\newcommand{\figref}[1]{Fig.~\multiref{#1}}
\renewcommand{\eqref}[1]{(\multiref{#1})}

\ifx\hypersetup\sadfkjashdfkxja\newcommand\hypersetup[1]{}\newcommand{\texorpdfstring}[2]{#1}\fi

\hypersetup{plainpages=false}
\hypersetup{pdfpagemode=UseNone}
\hypersetup{bookmarksnumbered=true}
\hypersetup{pdfstartview=FitH}
\hypersetup{colorlinks=false}
\hypersetup{citebordercolor={.5 1 .5}}
\hypersetup{urlbordercolor={.5 1 1}}
\hypersetup{linkbordercolor={1 .7 .7}}

\begin{document}
\thispagestyle{empty}

\ifarxiv\else\begingroup\raggedleft\footnotesize\ttfamily
AEI-2010-019\\
HU-EP-10/06\\
\arxivlink{1002.1733}\par\vspace{15mm}
\endgroup\fi

\begingroup\centering
{\Large\bfseries One-Loop Superconformal and Yangian Symmetries\ifarxiv$^\ast$\fi\\of Scattering Amplitudes in $\mathcal{N}=4$ Super Yang-Mills\par}%
\hypersetup{pdftitle={One-Loop Superconformal and Yangian Symmetries of Scattering Amplitudes in N=4 Super Yang-Mills}}%
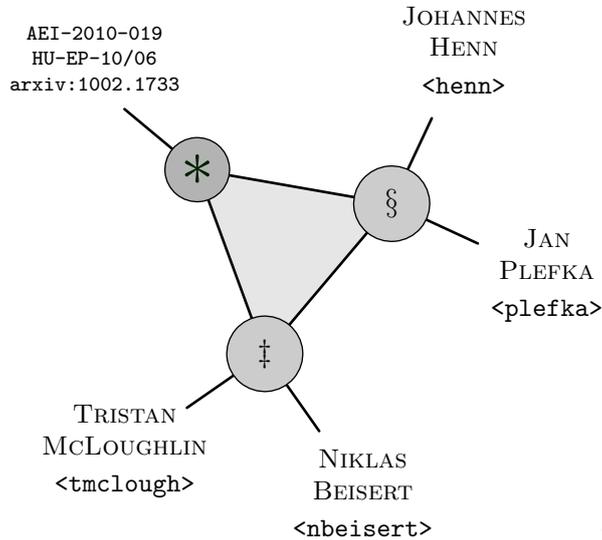
\hypersetup{pdfsubject={}}%
\hypersetup{pdfkeywords={Superconformal Symmetry, Yangian, Scattering Amplitude, N=4 Supersymmetric Gauge Theory, One Loop, Anomaly}}%
\ifarxiv\vspace{8mm}\else\vspace{15mm}\fi

\hypersetup{pdfauthor={Niklas Beisert, Johannes Henn, Tristan McLoughlin, Jan Plefka}}%
\ifarxiv
\includegraphics[scale=1.0]{FigTitle.mps}%
\vspace{5mm}
\else
\begingroup\scshape\large
Niklas Beisert$^\ddag$,
Johannes Henn$^\S$,\\
Tristan McLoughlin$^\ddag$
and
Jan Plefka$^\S$\par
\endgroup
\vspace{5mm}
\fi

\begingroup\ifarxiv\small\fi
$^\ddag$
\textit{Max-Planck-Institut f\"ur Gravitationsphysik, Albert-Einstein-Institut, \phantom{$^\ddag$}\\
Am M\"uhlenberg 1, D-14476 Potsdam, Germany}\par
\ifarxiv\texttt{\ldots@aei.mpg.de\phantom{\ldots}}\fi
\vspace{3mm}

$^\S$
\textit{Institut f\"ur Physik, Humboldt-Universit\"at zu Berlin, \phantom{$^\S$}\\
Newtonstra{\ss}e 15, D-12489 Berlin, Germany}\par
\ifarxiv\texttt{\ldots@physik.hu-berlin.de\phantom{\ldots}}\fi
\endgroup

\ifarxiv\else
\vspace{5mm}

\begingroup\ttfamily
\verb+{nbeisert,tmclough}@aei.mpg.de+,\\
\verb+{henn,plefka}@physik.hu-berlin.de+\par
\endgroup
\fi

\vspace{\fill}

\textbf{Abstract}\vspace{5mm}\par
\begin{minipage}{14.7cm}
Recently it has been argued that tree-level scattering amplitudes in $\mathcal{N}=4$
Yang-Mills theory are uniquely determined by a careful
study of their superconformal and Yangian symmetries.
However, at one-loop order these symmetries are known to become anomalous due to
infrared divergences.
We compute these one-loop anomalies for amplitudes defined through dimensional
regularisation by studying the tree-level symmetry transformations of the unitarity branch cuts,
keeping track of the crucial collinear terms arising from the holomorphic anomaly.
We extract the superconformal anomalies
and show that they may be cancelled through a universal one-loop deformation
of the tree-level symmetry generators which involves only tree-level data.
Specialising to the planar theory we also obtain the analogous
deformation for the level-one Yangian generator of momentum.
Explicit checks of our one-loop deformation are performed for MHV and the 6-point NMHV amplitudes.
\end{minipage}\par
\endgroup
\newpage


\setcounter{tocdepth}{2}
\hrule height 0.75pt
\tableofcontents
\vspace{0.8cm}
\hrule height 0.75pt
\vspace{1cm}

\setcounter{tocdepth}{2}

\section{Introduction}

Maximally supersymmetric $\superN=4$ Yang--Mills theory (SYM)
\cite{Brink:1976bc, Gliozzi:1976qd}
is an important testing ground for the foundations of
four-dimensional quantum field theories.
On the one hand, the model is based on
highly non-trivial interactions
which are reasonably similar to those appearing
in the standard model of particle physics.
On the other hand, a host of surprising features
make the theory much more tractable than many others.
By maximally exploiting these features,
we hope to gain access to previously unexplored regions of the model,
e.g.\ the finite coupling regime.
Such insights would not only be beneficial to
the study of the particular model,
but they could also teach us
about the (qualitative) behaviour of four-dimensional
quantum field theory in general.

Maximal supersymmetry turns out to constrain the model uniquely
up to the choice of a gauge group and two coupling constants.
It improves the quantum behaviour and leads to
various cancellations and simplifications.
For instance, the model is  ``finite''
\cite{Brink:1982wv, Mandelstam:1982cb}
in that its beta function vanishes exactly,
leading to unbroken (super)conformal symmetry
at the quantum level.
Finiteness can be traced back to the large
amount of supersymmetry, but there are also
a number of curious features
not following from maximal supersymmetry,
at least not immediately.
The AdS/CFT correspondence, \cite{Maldacena:1998re}, \cite{Witten:1998qj,Gubser:1998bc},
claiming exact duality to IIB strings on $AdS_5\times S^5$
is doubtlessly the most influential property of $\superN=4$ SYM
which is far from obvious in the standard quantum field theoretical formulation.
Another important insight is that the planar limit is
apparently exactly integrable
\cite{Minahan:2002ve,Beisert:2003yb, Beisert:2003tq, Bena:2003wd}
(for reviews see \cite{Tseytlin:2004cj,Belitsky:2004cz,Beisert:2004ry,Beisert:2004yq,Zarembo:2004hp,Plefka:2005bk,Minahan:2006sk,Arutyunov:2009ga}).
This enables one to compute efficiently the spectrum of
scaling dimensions of local operators using the Bethe ansatz
and related techniques instead of elaborate higher-loop QFT machinery.
Integrability in the guise of dual superconformal symmetry
was also identified as the underlying reason for non-trivial simplifications
observed in the computation of planar scattering amplitudes:
Hints of this hidden, dual conformal symmetry
were first seen in \cite{Drummond:2006rz}, were further studied in
\cite{Drummond:2007aua,Drummond:2007cf,Drummond:2007au}
and extended to dual superconformal symmetry in \cite{Drummond:2008vq}.
It was shown in \cite{Brandhuber:2008pf,Drummond:2008cr} that this is
indeed a symmetry of the tree-level amplitudes.
On the string theory side the dual symmetries were identified as the
symmetries of a T-dual model \cite{Alday:2007hr,Berkovits:2008ic}
and shown to be a part of the integrable hierarchy \cite{Beisert:2008iq,Berkovits:2008ic,Beisert:2009cs}.
For scattering amplitudes, integrability serves thus as an enhancement
of superconformal symmetry to an infinite-dimensional Hopf algebra called the
Yangian \cite{Drummond:2009fd}.
As is commonly the case in integrable systems,
one may hope that the large amount of symmetry
highly constrains physical observables
and that it predicts a unique S-matrix
for planar $\superN=4$ SYM.

Now it is well-known that scattering amplitudes are
not properly defined in a conformal field theory,
so how to make sense of the above statements?
Typically one introduces a regulator,
e.g.\ by going away from $D=4$ to $D=4-2\epsilon$
spacetime dimensions in a dimensional regularisation scheme.
Alternatively one can regulate the
theory by introducing small masses for the particles by going
off-shell \cite{Drummond:2007aua}
or higgsing the theory e.g.\ \cite{Alday:2007hr,Alday:2009zm}.
In every case the regulator breaks conformal symmetry and
scattering amplitudes become well-defined.
After renormalisation one tries to remove the regulator
in order to return to the original model.
For example, for local operators the dimensional regularisation
procedure leads to perfectly
finite answers showing the desired conformal behaviour,
albeit with a non-trivial spectrum of anomalous dimensions.
Scattering amplitudes, however, remain divergent in the limit of
vanishing regulator, hence they are problematic as anticipated.
The main distinction between local operators and scattering amplitudes
is the following: The former introduce short-distance (UV) singularities
due to multiple fields at coincident spacetime points,
while the latter introduce long-distance (IR) singularities
due to collinear massless particles.
We know well how to renormalise UV singularities by redefining
local operators but this is not the case for the IR singularities.
The structure of these divergences is of course well known from the study of QCD
--- they factorise and exponentiate
\cite{Mueller:1979ih, Collins:1980ih, Sen:1982bt, Korchemsky:1988pn, Korchemsky:1988hd, DelDuca:1989jt, Magnea:1990zb, Sterman:2002qn}.
Furthermore in properly defined physical observables, \cite{Kinoshita:1962ur, Lee:1964is},
such as inclusive cross-sections
or hadronic event shapes we expect that
all such divergences cancel (see \cite{Hofman:2008ar} and \cite{Bork:2009ps, Bork:2009nc}
for recent discussions of such observables in the current context).
However,  they cannot be removed from the S-matrix itself.
For the special case of the \emph{dual} conformal symmetry
we do have control over the apparent breaking of the symmetry via the relation
between light-like Wilson loops and amplitudes
\cite{Alday:2007hr,Alday:2007he,Drummond:2007aua,Brandhuber:2007yx,Drummond:2007cf}.
As the IR divergences of the amplitudes are mapped to the UV divergences, due
to cusps, in the Wilson loops an all-order anomalous Ward identity
can be derived \cite{Drummond:2007cf, Drummond:2007au}. This in turn strongly constrains the form of the amplitudes.

Unfortunately we do not know how to make use of the (super)conformal symmetries
or integrability to constrain
the S-matrix at loop level.
Yet it has become clear that
the divergent and finite contributions to
the scattering amplitudes
can be computed unambiguously and have a physical interpretation.
Furthermore conformal symmetry in $\superN=4$ SYM is non-anomalous.
Consequently it is our firm belief that all symmetries
apply to every physical observable such as the S-matrix,
even if some symmetries are obscured by quantum effects.

At first sight superconformal and Yangian symmetry appear to be
good symmetries of the tree level S-matrix whereas
at loop level they are broken beyond repair.
Fortunately, both statements are not true.
Even at tree level the conformal symmetry is superficially broken
at singular configurations where massless particles become collinear.
Interestingly, the existence of collinear particles is closely related to the
subtleties in defining asymptotic states and scattering amplitudes
in a conformal field theory:
Multiple quanta with collinear momenta are physically indistinguishable
from a single particle with the same overall momentum.
Hence the Fock space description for asymptotic states is not quite adequate,
one ought to factor out collinear configurations.
Unfortunately, such a projective space of asymptotic states
is technically hard to realise and instead it is easier to work
in the larger Fock space with some projective structure implied.
That the S-matrix respects the projective structure can be
inferred from the well-studied collinear behaviour
 determined by the splitting functions \cite{Berends:1988zn, Kosower:1999xi}.
Therefore the S-matrix is indeed a proper physical object
even in the presence of conformal symmetry or massless states.
The problem rather lies with conformal symmetry,
because na\"ively it does not respect the
projective structure on Fock space.
Luckily the free representation of superconformal symmetry
on scattering amplitudes can be deformed in such a way as
to make it compatible with the projective structure.
In particular this makes the tree level S-matrix exactly conformal.

One can see that na\"ive conformal symmetry is broken by a
holomorphic anomaly due to collinear particles
\cite{Cachazo:2004by, Cachazo:2004dr,Bargheer:2009qu, Korchemsky:2009hm}.
At tree level this happens exclusively at singular particle configurations
which is why the anomaly can be ignored to large extent.
At loop level the situation is more complicated because particles
running in loops can become collinear with others.
Due to the integration over loop momenta
the anomaly is smeared over all particle configurations
and thus it always requires proper treatment.

In this paper we consider the superconformal and Yangian symmetries
of one-loop  scattering amplitudes in $\superN=4$ SYM.
We will show how to deform the representation of superconformal symmetry,
in a manner generalising the tree-level construction \cite{Bargheer:2009qu},
such that it annihilates one-loop scattering amplitudes
including the IR singularities as well as the finite contributions.
Importantly, we will work in a strictly on-shell framework for the
S-matrix.
As we will see, such a framework is provided by generalised unitarity
which was introduced in \cite{Bern:1994zx,Bern:1994cg}
and further developed in \cite{Bern:1997sc,Britto:2004nc}.
These methods, based on
studying the behaviour of amplitudes across branch cut singularities
\cite{Landau:1959fi,Mandelstam:1958xc,Mandelstam:1959bc, Cutkosky:1960sp}
relate loop-level amplitudes to on-shell tree level amplitudes.
Symmetries of the latter are under full control
and will dictate the structure of the symmetries at one loop
\cite{Brandhuber:2009xz,Korchemsky:2009hm,Sever:2009aa,Brandhuber:2009kh}.
Although any other self-consistent regulator could be used in principle,
we shall choose a dimensional regularisation scheme for convenience.
The majority of perturbative results are formulated in this scheme where
they take a reasonably compact form.

Our proposal shares several features
 with a recent proposal \cite{Sever:2009aa}
which however uses a very different framework
of off-shell amplitudes and a particular
massive particle regularisation scheme.
Although the previous proposal is very elegant and economical,
the application to the on-shell S-matrix including its divergences
appears to be subtle.
In our framework we thus have to choose different deformations
whose action we can however define straight-forwardly on
the IR-singular on-shell S-matrix.

For the reader's convenience we outline the contents of subsequent
sections. We start in \secref{sec:Review} with a review of the
on-shell superspace description of tree-level scattering amplitudes.
We discuss the symmetries of the amplitudes and
how they can be deformed to account for the holomorphic anomaly which
arises in collinear configurations. In \secref{sec:sym1loop} we turn to our
main topic: Symmetries of one-loop amplitudes. After an outline of our general method
we analyse the  portion of the
superconformal  anomaly
for a generic amplitude arising from unitarity cuts. We argue that the
anomaly of the full amplitude
can be trivially lifted from the cut contribution and propose
a set of deformations of the representation
that annihilates all one-loop amplitudes.
Restricting this deformation to the planar limit we check by explicit calculation that
all MHV, \secref{sec:MHVex},
 and the six-point NMHV, \secref{sec:6NMHVex}, amplitudes are indeed invariant with respect
to the deformed generators. In \secref{sec:Yangian} we perform the analogous analysis for
the level one momentum generator of the Yangian and outline the necessary
procedure for the level one supersymmetry generator. In \secref{sec:comparison_prop}
 we discuss the propagator
 $i\epsilon$ prescription  used in our definition of amplitudes.
This allows us to compare our proposal to that
of \cite{Sever:2009aa}. We close with a discussion of our conclusions and
several appendices with further calculational details and conventions.

\section{Tree Amplitudes and Their Symmetries}
\label{sec:Review}
We start by reviewing tree-level scattering amplitudes
and their symmetries in the on-shell superspace formulation.
This provides a context for our later discussion
of one-loop amplitudes and allows us to fix our notation.

\subsection{On-Shell Superspace and Generating Functional}

We will be concerned with the $n$-particle scattering
amplitudes of
$\grp{U}(N\indup{c})$
  $\superN=4$ super-Yang-Mills (see
\cite{Alday:2008yw,Henn:2009bd} for recent, relevant reviews).
These amplitudes can be conveniently expanded
in a basis of colour structures. At tree level only
single-trace structures appear%
\footnote{The trace in this expression can be expanded into
$\grp{U}(N\indup{c})$ structure constants
when making use of the explicit form of the colour-ordered amplitude $A_n$.
As such it becomes valid for generic gauge groups.}
\[\label{eq:AmpPerm}
A_n^{a_1\ldots a_n}(1,\ldots,n)=
\sum_{\sigma\in S_n/\mathbb{Z}_n}
\Tr\lrbrk{ T^{a_{\sigma(1)}} \dots T^{a_{\sigma(n)}} }
A_n (\sigma(1),\dots,\sigma(n) ),
\]
where $T^a$ is an $\grp{U}(N\indup{c})$ generator in the fundamental representation
and $a,b,\ldots$ denote indices for the adjoint representation.

The tree-level amplitudes take a particularly simple form when written as
functions of the on-shell spinor-helicity superspace coordinates
$(\lambda^\alpha_i,{\tilde \lambda}^{\dot\alpha}_i,\eta_i^A)$
\cite{Nair:1988bq}
(see also e.g.\ \cite{Witten:2003nn,Drummond:2008vq,ArkaniHamed:2008gz}).
Here $\alpha,\beta=1,2$ and $\dot\alpha,\dot\beta=1,2$ are fundamental indices
for two distinct $\alg{su}(2)$'s and
$A,B=1,2,3,4$ are indices for $\alg{su}(4)$.
The commuting spinors $\lambda^\alpha$ and $\tilde \lambda^{\dot\alpha}$
parametrise the on-shell momenta
in spinor notation $p^{\beta\dot\alpha}=\sigma^{\beta\dot\alpha}_\mu p^\mu$
\[\label{eq:mom}
p^{\beta{\dot \alpha}}_i=\lambda^{\beta}_i{\tilde \lambda}_i^{\dot \alpha},
\qquad P^{\beta\dot\alpha}
=\sum_{i=1}^n p^{\beta{\dot \alpha}}_i
=\sum_{i=1}^n \lambda^{\beta}_i{\tilde \lambda}_i^{\dot \alpha},
\]
and can be used to form the usual invariants, i.e.\
$\sprod{i}{j}=\varepsilon_{\alpha\gamma}\lambda^\alpha_i\lambda_j^{\gamma}$
and
$\cprod{i}{j}=\varepsilon_{\dot\alpha\dot\gamma}\tilde\lambda^{\dot\alpha}_i\tilde\lambda_j^{\dot\gamma}$.
In Minkowski signature the spinors are related by complex conjugation
and for positive (negative) energy
particles we have $\tilde \lambda=+(-)\bar \lambda$.
The Gra\ss{}mann variable $\eta^A$ allows one to combine the on-shell states of $\superN=4$ SYM
--- the two gluon helicity states  $G^\pm$, the fermions $\Gamma_A$/$\bar\Gamma^A$, and scalars $S_{AB}$  ---
into a single superwavefunction \cite{Mandelstam:1982cb,Brink:1983pd}
\<\label{eq:superwavefunction}
\Phi(\lambda,\tilde\lambda,\eta)
\eq
G^+(\lambda,\tilde\lambda)
+\eta^A \Gamma_A(\lambda,\tilde\lambda)
+\sfrac{1}{2}\eta^A\eta^B S_{AB}(\lambda,\tilde\lambda)
\nl
+\sfrac{1}{6} \varepsilon_{ABCD} \eta^A\eta^B \eta^C {\bar \Gamma}^D(\lambda,\tilde\lambda)
 +\sfrac{1}{24}\varepsilon_{ABCD}\eta^A\eta^B \eta^C\eta^D  G^-(\lambda,\tilde\lambda)~.
\>
Using the $\eta$'s one can form the supermomentum $q_i$ of a particle
which serves as the fermionic partner to the momentum $p_i$
($Q$ is the overall supermomentum)
\[\label{eq:supermom}
q^{\beta A}_i=\lambda^{\beta}_i\eta_i^A,
\qquad
Q^{\beta A}=\sum_{i=1}^n q^{\beta A}_i
=\sum_{i=1}^n \lambda^{\beta}_i \eta_i^{A}.
\]

It will often be convenient to refer to all spinor-helicity superspace
coordinates in a collective fashion through
$\Lambda:=(\lambda^\alpha,{\tilde \lambda}^{\dot\alpha},\eta^A)$.
These can represent particles with both positive and negative energies
depending on the choice $\tilde \lambda=\pm \bar\lambda$.
Related to this we introduce the compact notation $\bar\Lambda$
for flipping the energy as well as all other components of the
momentum $p$ and supermomentum $q$
\[
\bar\Lambda:=(+\lambda,-\tilde\lambda,-\eta),
\qquad
e^{i\varphi}\Lambda:=
(e^{+i\varphi}\lambda,e^{-i\varphi}\tilde\lambda,e^{-i\varphi}\eta).
\]
The second notation $e^{i\varphi}\Lambda$
corresponds to a helicity rotation about the particle axis.
Finally, we introduce the canonical measure on superspace
$d^{4|4}\Lambda:=d^4\lambda\,d^4\eta$.
The bosonic integral $d^4\lambda:=d^2\lambda\,d^2\bar\lambda$
is equivalent to the Lorentz-invariant
on-shell integral and an integral over the particle phase
\[
d^4\lambda=d\varphi\,d^4p\,\delta(p^2)=d\varphi\,\frac{d^3p}{2E}\,.
\]
The fermionic integral $d^4\eta$ implements the sum over all particle types.
We shall assume that integration is over positive \emph{and} negative energies.
To restrict the integral to the forward or backward light-cone we shall
use the notation $d^{4|4}_\pm\Lambda$.

The super-amplitudes $A_n(\Lambda_1,\ldots,\Lambda_n)$
are polynomials in the $\eta_i$'s
whose coefficients are the amplitudes of the various component fields.
For reasons of $\alg{su}(4)$-invariance the $\eta$'s must come in
sets of four leading to the helicity classification
\[
A_n=\sum_{k=2}^{n-2}A_{n,k},\qquad A_{n,k}\sim\eta^{4k}.
\]
The terms $A_{n,m+2}$ are called N$^m$MHV subamplitudes.
Furthermore, conservation of momentum and supermomentum
as well as conformal transformations enforce that
all amplitudes have a common prefactor
\[\label{eq:TreeAmp}
A_{n,k}
=\frac{\delta^4(P)\,\delta^8(Q)}{\sprods{1}{2}\dots\sprods{n}{1}}\,
R_{n,k}.
\]
The remainder functions $R_{n,m+2}$
are homogeneous of degree $4m$ in the $\eta$'s.
The first term $R_{n,2}$ is simply $1$,
so we recover the well-known
formula for MHV amplitudes \cite{Nair:1988bq,Witten:2003nn},
\[\label{eq:MHVtree}
A_{n,\mathrm{MHV}}
:=A_{n,2}
=\frac{\delta^4(P)\,\delta^8(Q)}{\sprods{1}{2}\dots\sprods{n}{1}}~,
\]
and explicit, though somewhat more complicated,  expressions
for all other terms at tree level can be found in \cite{Drummond:2008cr}.

In considering the symmetries
it is useful to combine all amplitudes
into a single generating functional (see also \cite{ArkaniHamed:2009si}).
We introduce a source field $J(\Lambda)$
conjugate to the superspace field $\Phi(\Lambda)$
and we shall use the compressed notation $J_i:=J(\Lambda_i)$
for the source field corresponding to particle $i$.
The generating functional of tree amplitudes reads
\[\label{eq:GenFuncOrdered}
\mathcal{A}[J]=\sum_{n=4}^{\infty}\int \prod_{i=1}^n (d^{4|4}\Lambda_i)\, \frac{1}{n}
\Tr\bigbrk{J_1\dots J_n}A_n(\Lambda_1,\dots,\Lambda_n)~.
\]
The $n$-particle super-amplitude can be extracted
by taking functional derivatives
\[
\check{J}(\Lambda):=\frac{\delta}{\delta J(\Lambda)}
\]
of the generating functional
\[
A^{a_1\ldots a_n}_n(\Lambda_1,\dots,\Lambda_n)=
\bigeval{\check{J}^{a_1}(\Lambda_1)\ldots\check{J}^{a_n}(\Lambda_n) \mathcal{A}[J] }_{J=0},
\]
and we note that the commutative variations naturally account for the sum
over all permutations in \eqref{eq:AmpPerm}.

\subsection{Free Symmetries}

We will be interested in how the superconformal algebra, $\alg{psu}(2,2|4)$,
is realised on the scattering amplitudes. This algebra comprises
the Lorentz rotations $\gen{L}$, $\bar{\gen{L}}$, the internal symmetry rotations
$\gen{R}$, momentum generators $\gen{P}$, special conformal generators
$\gen{K}$, the dilatation generator $\gen{D}$, the Poincar\'{e} supercharges
$\gen{Q}$, $\bar{\gen{Q}}$ and special conformal supercharges $\gen{S}$, $\bar{\gen{S}}$.
Using the on-shell superspace notation the free representation carried
by a single on-shell superparticle
can be written very compactly \cite{Witten:2003nn}
\[\label{eq:StdRep}
\begin{array}[b]{@{}rclcrcl@{}}
\gen{L}^\alpha{}_\beta\eq \lambda^\alpha\partial_{\beta}-\half\delta^\alpha_\beta \lambda^\gamma\partial_\gamma,
&&
\gen{\bar L}^{\dot \alpha}{}_{\dot \beta}\eq
\tilde\lambda^{\dot \alpha}\tilde\partial_{\dot \beta}
-\half\delta^{\dot \alpha}_{\dot \beta} \tilde\lambda^{\dot \gamma}\tilde \partial_{\dot \gamma},
\\[1ex]
\gen{R}^A{}_B\eq \eta^A\partial_B-\quarter \delta^A_B\eta^C\partial_C,
&&
\gen{D}\eq 1+\half\lambda^{\gamma}\partial_{\gamma}+\half\tilde\lambda^{\dot \gamma}\tilde \partial_{\dot \gamma}
,
\\[1ex]
\gen{Q}^{aB}\eq \lambda^\alpha\eta^B ,
&&
\gen{S}_{a B}\eq \partial_\alpha\partial_B  ,
\\[1ex]
\bar{\gen{Q}}^{\dot a}_{B}\eq \tilde\lambda^{\dot \alpha} \partial_B ,
&&
\bar{\gen{S}}_{\dot a}^B\eq \eta^B \tilde\partial_{\dot \alpha} ,
\\[1ex]
\gen{P}^{\beta\dot\alpha}\eq \lambda^\beta\tilde\lambda^{\dot\alpha} ,
&&
\gen{K}_{\beta\dot \alpha}\eq \partial_\beta\tilde\partial_{\dot \alpha},
\end{array}
\]
where we abbreviate $\partial_a=\partial/\partial\lambda^a$,
$\tilde\partial_{\dot a}=\partial/\partial\tilde\lambda^{\dot a}$
and $\partial_A=\partial/\partial\eta^A$.
Furthermore, there is a central charge $\gen{C}$
\[
\gen{C}
=1+\half\lambda^\gamma\partial_\gamma-\half\tilde\lambda^{\dot\gamma}\tilde\partial_{\dot\gamma}
- \half\eta^C\partial_C.
\]
It acts as the constraint that every physical particle
must be uncharged under it,
which follows from \eqref{eq:superwavefunction}
and the helicities of the various fields.

The corresponding representation of a generic generator $\gengen\supup{free}$ on $n$ particles is simply
given by the sum over actions on individual particles $\gengen\supup{free}_i$,
\[
\label{eq:gengen_free}
\gengen\supup{free}=\sum_{i=1}^n\gengen\supup{free}_{i}.
\]
%

\subsection{Exact Tree-Level Symmetries}

\begin{figure}
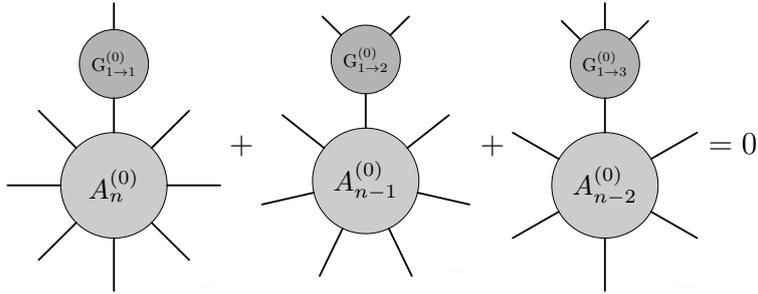
\centering
$\includegraphicsbox[scale=0.7]{FigInvTree1.mps}
+\includegraphicsbox[scale=0.7]{FigInvTree2.mps}
+\includegraphicsbox[scale=0.7]{FigInvTree3.mps}
=0$
\caption{Statement of exact invariance of tree amplitudes
under the deformed superconformal representation.}
\label{fig:DynamicInvarianceTree}
\end{figure}

Invariance of the amplitude $A_n$ under the generator $\gengen$ is the statement
\[
\gengen A_n=0.
\]
As was discussed at length in \cite{Bargheer:2009qu}
the free representation does not exactly
annihilate tree-level amplitudes, but rather must
be deformed by generators which change the number of external legs.
These non-linear contributions to the generators in the interacting
\emph{classical} theory are generically of the form, see \figref{fig:DynamicInvarianceTree},
\[
\gengen=\gengen_{1 \to 1}+\gengen_{1 \to 2}+\gengen_{1 \to 3}~.
\]
The first term $\gengen_{1 \to 1}=\gengen\supup{free}$ is the free generator discussed above,
which simply takes a single leg and returns a single modified leg.
The correction terms compensate for the contributions
occurring at values of the external momenta where particles become collinear.
The deformation $\gengen_{1 \to 2}$ can be found by explicitly calculating
the action of generators involving derivatives
in $\lambda$ or $\tilde\lambda=\sign(E(\lambda))\bar\lambda$ on the $n$-point amplitude, $A_n$
and by carefully accounting for the holomorphic anomaly terms,
which arise in $(3,1)$ spacetime signature,
i.e.\
\[\label{eq:spinoranomaly}
\frac{\partial}{\partial\tilde\lambda^{\dot\alpha}}\,
\frac{1}{\sprod{\lambda}{\mu}}
=
2\pi\sign \bigbrk{E(\lambda)E(\mu)}\,
\varepsilon_{\dot\alpha\dot\gamma}\tilde\mu^{\dot\gamma}\delta^2\bigbrk{\sprod{\lambda}{\mu}}\,.
\]
It was shown in \cite{Bargheer:2009qu} that this anomaly is equivalent
to attaching an anomaly three-vertex $\gengend_3$
to an amplitude with one leg less.
The anomaly is thus cancelled by deforming the na\"{i}ve free generator
(left in \figref{fig:DynamicInvarianceTree})
by a term
$\gengen_{1 \to 2}$ which attaches the same vertex to the amplitude
but with the opposite sign
(middle in \figref{fig:DynamicInvarianceTree}).
For a three-vertex of massless particles the
support must be on configurations with all three momenta collinear
$p^\mu_1\sim p^\mu_2\sim p^\mu_3$, i.e.\
\[\label{eq:gengen_anom}
\gengend_3\sim
\delta^4(\lambda_{\bar1}-e^{-i\varphi}\lambda_3\sin\alpha)\,
\delta^4(\lambda_{\bar2}-e^{-i\vartheta}\lambda_3\cos\alpha).
\]
Furthermore, their colour structure equals
the structure constants $f^{abc}$
\[
\gengend_3^{abc}(\Lambda_1,\Lambda_2,\Lambda_3)
=if^{abc}\gengend_3(\Lambda_1,\Lambda_2,\Lambda_3).
\]
The third term $\gengen_{1 \to 3}$ corresponds to a
four-vertex $\gengend_4$.
Luckily, $\gengen_{1 \to 3}$ arises only in the closure of
the algebra.
For this reason it is not necessary
to specify the vertex $\gengend_4$ explicitly;
it consists of a combination of two $\gengend_3$'s.
To complete the picture it is instructive to note that also
the free representation corresponds to a vertex, but now with two legs
\[\label{eq:twovertex}
\gengend^{ab}_2(\Lambda_1,\Lambda_2)=
\delta^{ab}\gengend_2(\Lambda_1,\Lambda_2),
\qquad
\gengend_2(\Lambda_1,\Lambda_2)=
\gengen\supup{free}_1\delta^{4|4}(\Lambda_1-\bar\Lambda_2).
\]
Here $\gengen\supup{free}_1$ is the free generator acting as a differential operator
on the spinor-helicity superspace with label $1$.

In the language of sources and generating functionals the deformations are very natural
\<
\gengen_{1\to 1} \eq \int (d^{4|4}\Lambda)^2 \, \gengend_2^{ab}(\Lambda_1,\Lambda_2)
J^a(\Lambda_1)\check{J}^b(\bar\Lambda_2),
\nln
\gengen_{1\to 2} \eq \half\int (d^{4|4}\Lambda)^3 \, \sign(E_1E_2)\, \gengend_3^{abc}(\Lambda_1,\Lambda_2,\Lambda_ 3)
J^a(\Lambda_1) J^b(\Lambda_2)\check{J}^c(\bar\Lambda_3).
\>
The sign in $\gengen_{1\to 2}$ for opposite energy states $1$ and $2$
is required to match the sign in \eqref{eq:spinoranomaly}.
Invariance of the amplitude as a whole now becomes the statement
$\gengen\mathcal{A}[J]=0$.

Among the superconformal generators, only
$\gengen=\gen{S}$, $\bar{\gen{S}}$ and $\gen{K}$ receive deformations at
tree level. The latter follows from the algebra and hence we
do not need to consider it further.
The two former generators receive only corrections
of the type $\gengen_{1\to2}$.
In \appref{app:AnomalyVertex} we present a formal derivation of
the corresponding anomaly vertices.
The resulting anomaly vertices read
\<\label{eq:AnomalyMain}
(S_3)_{\alpha B}(\Lambda_1,\Lambda_2,\Lambda_3)\eq
-2\int d^{4|4}\Lambda'\, \delta^{4|4}(\Lambda')\,
\varepsilon_{\alpha\gamma}\lambda_3^\gamma \partial'_{B}
\int d\alpha\,d\varphi\,d\vartheta\,e^{-i\varphi-i\vartheta}
\nl\qquad\cdot
\delta^{4|4}(e^{-i\varphi}\Lambda_3\sin\alpha+e^{i\vartheta}\Lambda'\cos\alpha-\bar\Lambda_{1})
\nl\qquad\cdot
\delta^{4|4}(e^{-i\vartheta}\Lambda_3\cos\alpha-e^{i\varphi}\Lambda'\sin\alpha-\bar\Lambda_{2})+
\begin{array}{c}\mbox{two cyclic}\\\mbox{images}\end{array},
\nln
(\bar S_3)^B_{\dot\alpha}(\Lambda_1,\Lambda_2,\Lambda_3)\eq
-2\int d^{4|4}\Lambda'\,\delta^4(\lambda')\,
\varepsilon_{\dot\alpha\dot\gamma}\tilde\lambda_3^{\dot\gamma}\eta'^B
\int d\alpha\,d\varphi\,d\vartheta\,e^{i\varphi+i\vartheta}
\nl\qquad\cdot
\delta^{4|4}(e^{-i\varphi}\Lambda_3\sin\alpha+e^{i\vartheta}\Lambda'\cos\alpha-\bar\Lambda_{1})
\nl\qquad\cdot
\delta^{4|4}(e^{-i\vartheta}\Lambda_3\cos\alpha-e^{i\varphi}\Lambda'\sin\alpha-\bar\Lambda_{2})+
\begin{array}{c}\mbox{two cyclic}\\\mbox{images}\end{array}.
\>
The ranges for integration read $0\leq \alpha\leq \pi/2$
and $0\leq \varphi,\vartheta<2\pi$. The cyclic images account for the different combinations of
energy signatures as described in \appref{app:AnomalyVertex}.

For comparison, the two-vertices for the free generators $\gen{S},\bar{\gen{S}}$ read,
cf.\ \eqref{eq:StdRep}
\<
(S_2)_{\alpha B} (\Lambda_1,\Lambda_2)\eq \partial_{1,\alpha}\partial_{1,B}\delta^{4|4}(\Lambda_1-\bar\Lambda_2)\,
,
\nln
(\bar{S}_2)_{\dot \alpha}^B(\Lambda_1,\Lambda_2)
\eq \tilde{\partial}_{1,\dot \alpha} \eta^B_{1} \delta^{4|4}(\Lambda_1-\bar\Lambda_2)\,
.
\>

An important point is that the deformed tree-level symmetries relate
all tree amplitudes. As described above we can label the amplitudes by
the number of $\eta$'s, $A_{n,k}\sim\eta^{4k}$.
The correction to $\bar{\gen{S}}$ keeps $k$ fixed while increasing
$n$ by one whereas the correction to $\gen{S}$ increases both $k$
and $n$ by one. Thus by use of the generators we find relations between the
variations of all amplitudes.

It has recently been shown, \cite{Korchemsky:2009hm}, that  the
free symmetries (including dual superconformal symmetries)
alone are insufficient to uniquely fix the
tree amplitudes. They merely determine that the amplitudes be linear
combinations of dual superconformal invariants.  However,
the demand of  correct
collinear behaviour (\emph{or} the absence of so-called spurious poles) is, under
certain mild assumptions,
sufficient to fix all relative coefficients and so uniquely determine the
full tree-level amplitude \cite{Hodges:2009hk,Korchemsky:2009hm}.
Equivalently, demanding that the corrected generators are exact
symmetries, fixes all tree-level amplitudes.
Of course all tree-level amplitudes have already
been explicitly determined in \cite{Drummond:2008cr},
by use of the BCFW  recursion relations
\cite{Britto:2004ap,Britto:2005fq}, and their generalisations
\cite{Bianchi:2008pu,Brandhuber:2008pf,Elvang:2008na,ArkaniHamed:2008gz}.
However, the extent to which the symmetries fix the amplitudes is
an important question, particularly beyond tree level where we no
longer have such efficient methods as BCFW.

A related analysis of the symmetries of tree level amplitudes
was performed in \cite{Sever:2009aa}. This work made use of the CSW
\cite{Cachazo:2004kj} approach to constructing scattering amplitudes and,
with some assumptions regarding the regularisation of divergences,
could be generalised to loop level. We will comment on the
relation of our proposals to that of \cite{Sever:2009aa} in \secref{sec:comparison_prop}.

\section{Superconformal Symmetry at One Loop}
\label{sec:sym1loop}
We now wish to consider scattering amplitudes beyond tree level in order
to account for the new features to which loops give rise.
One important aspect of massless theories
is the existence of infra-red divergences which necessitate the
introduction of a regulator.
In part, these divergences originate from
virtual particles in loops becoming collinear with external legs.
These divergences cannot be removed,
but rather cancel only when calculating physical observables, and so the amplitudes will
explicitly depend on the regulator.

We will expand the amplitudes
using the loop counting parameter%
\footnote{We shall use a minimal subtraction scheme
without absorbing predictable numerical combinations
like $\gamma$ or $\log 4\pi$ into the coupling constant.
Instead we will carry them along in a
constant $c_\epsilon=1+\order{\epsilon}$, see \protect\eqref{eq:ceps}.}
\footnote{For general gauge groups we write the loop counting parameter in terms
of the quadratic Casimir. For $\grp{U}(N_c)$ gauge group and with normalisation
$\Tr(t^at^b)=\delta^{ab}$, $C\indup{A}=N\indup{c}$.}
\[\label{eq:looppardef}
g^2=\frac{g\indup{YM}^2C\indup{A}}{16\pi^2}\,,
\]
so that
\[
A^{a_1 \dots a_n}_n(1,\dots, n)=\sum_{\ell=0}^\infty g^{2\ell} \left(A^{(\ell)}\right)^{a_1 \dots a_n}_n(1,\dots, n)~.
\]
In principle one can further expand the amplitudes in an appropriate
basis of colour structures at each order. For example, at one loop
there are double traces in addition to the single traces
seen already at tree level. However we will for the most part
treat the general case and only simplify to specific colour structures
in considering the planar limit.

In explicit calculations of amplitudes it is common to make use of
dimensional regularisation%
\footnote{In order to maintain consistency with the supersymmetric Ward
identities a supersymmetric variant, such as dimensional reduction
\cite{Siegel:1979wq, Capper:1979ns},
should be chosen.}
 where $D=4-2\epsilon$ and for concrete calculations
this is the regularisation we will consider.
The amplitudes will have singularities
as $\epsilon\to 0$, typically $1/\epsilon^2$ per loop level for a conformal theory.
The structure of these divergences is well understood,
see e.g.\ \cite{Sen:1982bt, DelDuca:1989jt, Magnea:1990zb, Sterman:2002qn},
being determined by an evolution equation which follows from
gauge invariance and the factorisation of processes separated by
energy scales.

\begin{figure}
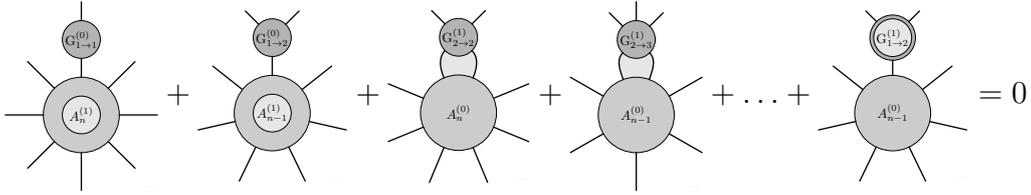
\centering
$\begin{array}{l}
\includegraphicsbox[scale=0.5]{FigAmp1G011.mps}
+\includegraphicsbox[scale=0.5]{FigAmp1G012.mps}
+\includegraphicsbox[scale=0.5]{FigAmp0G022.mps}
+\includegraphicsbox[scale=0.5]{FigAmp0G023.mps}
+
\ldots
+\includegraphicsbox[scale=0.5]{FigAmp0G112.mps}
=0
\end{array}$
\caption{Deformations of generators
necessary for invariance of scattering amplitudes at loop level.}
\label{fig:LoopInvariance}
\end{figure}

The introduction of the regulator breaks conformal symmetry and thus the
divergent parts of the amplitude will manifestly fail to be invariant.%
\footnote{In fact, the most divergent parts actually are invariant, but for the
subleading (including finite) parts invariance fails}
Generically even the finite parts of the loop-level amplitudes will not
be annihilated by the tree-level generators. However, by introducing further
deformations of the generators we can account for these effects and show that
the amplitudes are indeed invariant.
Said deformations will in general involve more external legs and are schematically of the form
\[
\gengen_{m \to n}\sim\int \Tr \bigbrk{\underbrace{J \dots J}_{n}\underbrace{\check{J} \dots \check{J} }_{m}}.
\]
Such an operator grabs $m$ legs of an amplitude and replaces them by $n$.
Acting on an $\ell$-loop amplitude with $p$ external legs
this could cancel a term arising from the free generator acting on
an $\ell+m-1$ loop amplitude with $p+n$ external legs.
In addition to creating loops by acting on multiple legs of an amplitude, deformations
can contain loops within themselves. The general structure of the deformations
necessary to annihilate
the one-loop amplitudes is shown in \figref{fig:LoopInvariance}. It is the goal
of the subsequent sections to find the explicit form of these deformations.

\subsection{General One-Loop Anomaly of Cuts}

Considered as functions of the kinematic invariants,
loop-level amplitudes have branch cuts,
in addition to collinear singularities and
multi-particle poles which appear already at tree level.
The form of the discontinuity  across a given cut is determined by unitarity
and at one loop can be expressed as a
phase space integral over products of tree-level amplitudes.
In fact, in supersymmetric theories one
can reconstruct the entire amplitude from its cuts.
Such unitarity methods, introduced in \cite{Bern:1994zx,Bern:1994cg} and
further developed in \cite{Bern:1997sc,Britto:2004nc},
have proved tremendously powerful calculational tools
and are a convenient method for uncovering
the structure of the symmetries at one loop
\cite{ Brandhuber:2009xz, Korchemsky:2009hm,Sever:2009aa, Brandhuber:2009kh}.

\begin{figure}
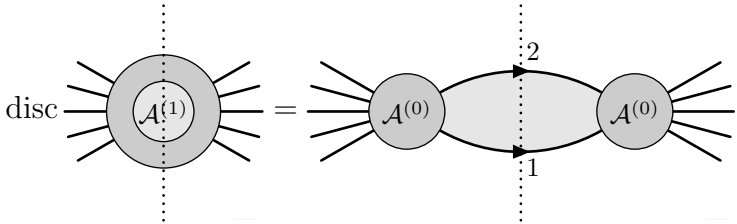
\centering
$\disc
\includegraphicsbox{FigCutA1.mps}
=
\includegraphicsbox{FigCutA0A0.mps}
$
\caption{Discontinuity of $\mathcal{A}^{(1)}$}
\label{fig:UnitarityCut}
\end{figure}

As was the case at tree level it is convenient to make use
of the generating functional language
as it naturally allows for the length-changing deformations and
includes the sum over all cuts.
We start with with the definition of the unitarity cut,
see \figref{fig:UnitarityCut},%
\footnote{In anticipation of taking the planar limit for the $\grp{U}(N\indup{c})$ case,
we have included a factor of the quadratic Casimir, $\cas$, in the loop
counting parameter $g^2$, see \protect\eqref{eq:looppardef},
and thus we need to cancel it here.}
\[
\disc \mathcal{A}^{(1)}=
-\quarter
\prefac
\int
(d^{4|4}\Lambda)^2
\Delta^\epsilon_{12}
(\check J^a_{\bar 1} \check J^b_{\bar 2} \mathcal{A}^{(0)})
(\check J^a_1 \check J^b_2 \mathcal{A}^{(0)}).
\]
The above unitarity relation is to be understood as follows:
The generating functionals $\mathcal{A}^{(0)}$ both represent
a tree level subamplitude with an indeterminate number of legs
(saturated with source fields $J$).
From each subamplitude we grab two legs by action with the
source variation $\check J$. We then perform an on-shell
integration over the momenta for each pair of legs.
The measure $\Delta^\epsilon_{12}$ is
non-zero only where the energies
of the two particles have equal signs.
Technically it is achieved by the following step function $\theta$
of the two-particle invariant $s_{j,k}$
\[
\Delta^\epsilon_{12}=\theta(-s_{12})+\order{\epsilon},
\qquad
s_{j,k}=(p_j+p_k)^2=\sprod{j}{k}\cprod{k}{j}.
\]
This step function also specifies which particular two-particle channel we
are talking about.
In some cases the above integral is divergent
so that the $\order{\epsilon}$ contributions to the
measure $\Delta^\epsilon_{12}$ become important and will serve as a regulator.
We will specify its precise form later where we need it.

\begin{figure}
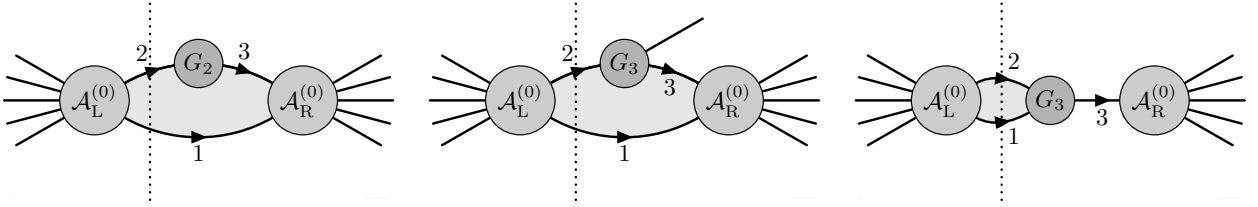
\centering
\includegraphics[width=0.3\textwidth]{FigCutAnom1.mps}\quad
\includegraphics[width=0.3\textwidth]{FigCutAnom2.mps}\quad
\includegraphics[width=0.3\textwidth]{FigCutAnom3.mps}
\caption{Three types of anomalies on the cut.}
\label{fig:CutAnomalies}
\end{figure}

The tree amplitude functionals $\mathcal{A}^{(0)}[J]$ in the cut are both invariant,
so the tree generator $\gengen^{(0)}$ will see only the source variations $\check J$
\[
\gengen^{(0)} \disc \mathcal{A}^{(1)}
=-
\half
\prefac
\int
(d^{4|4}\Lambda)^2
\Delta^\epsilon_{12}
(\check J^a_{\bar 1} \check J^b_{\bar 2} \mathcal{A}^{(0)})
\bigbrk{\comm{\gengen^{(0)}}{\check J^a_1 \check J^b_2} \mathcal{A}^{(0)}}.
\]
We can now substitute the definition of the deformed generic generator at tree level
$\gengen^{(0)}=\gengen^{(0)}_{1\to1}+\gengen^{(0)}_{1\to2}$ with
\[
\gengen^{(0)}_{1\to1}=
\int (d^{4|4}\Lambda)^2
\gengend_{1\bar 2}^{ab} J^a_1 \check J^b_2,
\qquad
\gengen^{(0)}_{1\to2}=
\half\int (d^{4|4}\Lambda)^3
\sign(E_1E_2)
\gengend^{abc}_{12\bar 3} J^a_1 J^b_2 \check J^c_3.
\]
Here, and subsequently,
we use an analogous notation for the kernels as was introduced for the
sources e.g.\ $\gengend_2^{ab}(\Lambda_1, \bar \Lambda_2)=\gengend_{1\bar2}^{ab}$.
The action of the generator
on the $\check J^a_{\bar 1} \check J^b_{\bar 2} \mathcal{A}^{(0)}$ subamplitudes yields
\<
\earel{}
\prefac
\int
(d^{4|4}\Lambda)^3
\Delta^\epsilon_{12}
\gengend_{2\bar 3}^{bc}
(\check J^a_{\bar 1} \check J^b_{\bar 2} \mathcal{A}^{(0)})
(\check J^a_1 \check J^c_3 \mathcal{A}^{(0)})
\nl
+
\prefac
\int
(d^{4|4}\Lambda)^4
\Delta^\epsilon_{12}
\sign(E_4E_2)
\gengend^{dbc}_{42\bar 3}
(\check J^a_{\bar 1} \check J^b_{\bar 2} \mathcal{A}^{(0)})
(J^d_4 \check J^a_1 \check J^c_3 \mathcal{A}^{(0)})
\nl
+
\half
\prefac
\int
(d^{4|4}\Lambda)^3
\Delta^\epsilon_{12}\sign(E_1E_2)
\gengend^{abc}_{12\bar 3}
(\check J^a_{\bar 1} \check J^b_{\bar 2} \mathcal{A}^{(0)})
(\check J^c_3 \mathcal{A}^{(0)}).
\>
The three types of contributions are depicted in
\figref{fig:CutAnomalies}, they correspond to

\begin{itemize}

\item For the generators we are interested in, for example
the dilatation generator $\gen{D}$, or the superconformal boost
$ {\bar{\gen{S}}} $, the first term
can be integrated by parts to write it as a commutator of the generator
with the measure.
For the case of the generators
without a holomorphic anomaly, e.g.\ the dilatation generator,
this is the only anomalous contribution.

\item The second term corresponds to the case where the anomaly sits partially inside the
loop integral. It occurs where one external and one internal leg become collinear.

\item The third term occurs when two internal legs become collinear; that is when
the anomaly vertex sits entirely inside the loop integral. As we will  see below this term
corresponds to a one-loop correction to the collinear limit and is only non-trivial
for the two particle cut.
\end{itemize}
Let us consider these various contributions in more detail.

\subsection{Anomaly of the Measure}
\label{subsec:Anomaly_measure}

We start with the first type of terms
where we substitute the definition of the
two-vertex $\gengend_2$ \eqref{eq:twovertex}
and write it in a symmetric fashion as
\<
\earel{}
\quarter
\prefac
\int
(d^{4|4}\Lambda)^2
\Delta^\epsilon_{12}
\bigbrk{(\gengen\supup{free}_{\bar 1}+\gengen\supup{free}_{\bar 2}) \check J^a_{\bar 1} \check J^b_{\bar 2} \mathcal{A}^{(0)}}
(\check J^a_1 \check J^b_2 \mathcal{A}^{(0)})
\nl+
\quarter
\prefac
\int
(d^{4|4}\Lambda)^2
\Delta^\epsilon_{12}
(\check J^a_{\bar 1} \check J^b_{\bar 2} \mathcal{A}^{(0)})
\bigbrk{(\gengen\supup{free}_{1}+\gengen\supup{free}_{2})\check J^a_1 \check J^b_2 \mathcal{A}^{(0)}}.
\>
Effectively this is the action of the free superconformal generator $\gen{G}$
on the internal particles of the cut integral.
If the integral had been finite and if there had been no measure factor,
the integral would have been perfectly superconformally invariant
and the above expression would have vanished.
Now one can convince oneself that
for all free generators in \eqref{eq:StdRep}
this is equivalent upon integration by parts
to the generator acting on the one-loop measure factor
\[
-\quarter
\prefac
\int
(d^{4|4}\Lambda)^2
(\check J^a_{\bar 1} \check J^b_{\bar 2} \mathcal{A}^{(0)})
\bigcomm{\gengen\supup{free}_{1}+\gengen\supup{free}_{2}}{\Delta^\epsilon_{12}}
(\check J^a_1 \check J^b_2 \mathcal{A}^{(0)})
.
\]
The derivation depends on the number of derivatives in the generator $\gengen$,
but the result is always the same.
In particular there is no anomaly for those generators
under which the one-loop measure is invariant, i.e.\
the super-Poincar\'e generators
$\gen{L}$, $\bar{\gen{L}}$, $\gen{R}$, $\gen{Q}$, $\bar{\gen{Q}}$ and $\gen{P}$.
The extra generators $\gen{D}$, $\gen{S}$, $\bar{\gen{S}}$ and $\gen{K}$
in the superconformal algebra are anomalous. The anomaly of $\gen{K}$
follows from the one of $\gen{S}$, $\bar{\gen{S}}$ plus the algebra so we will not
consider it separately.

The commutator gives rise to an overall factor of $\epsilon$ and so,
as the actions of the generators are finite,
we can focus on the IR-divergent part of the phase space integral.
Divergent contributions arise only if one of the two subamplitudes
has four legs. They originate within this subamplitude and they
are localised where the ingoing legs are both collinear with the outgoing ones.
For the purpose of computing the divergent part,
we can therefore replace the loop momenta in the second subamplitude
by the external momenta of the first
(for a discussion of this see e.g.\ \cite{Brandhuber:2009kh}).
Importantly, the $n$-point tree-level amplitude can be pulled out
of the integral so that we see that the action of the anomaly is diagonal which is to
say it takes two legs and gives two legs back.
Using the Schoutens identity and its cyclic symmetries the
full four leg subamplitude, as opposed to just a single colour ordering (see \eqref{eq:AmpPerm}),
can be written as
\[
\label{eq:tot4amp}
\mathcal{A}_4=-
\sfrac{1}{12}
f^{abe}f^{cde}
\int (d^{4|4}\Lambda)^4\,
J_1^aJ_2^bJ_3^cJ_4^d\,
A_{4},
\qquad
A_{4}=
\frac{\delta^4(P)\,\delta^8(Q)}{\sprods{1}{2}\sprods{2}{3}\sprods{3}{4}\sprods{4}{1}}~.
\]
We must at this point also address the definition of the measure factor $\Delta_{12}^\epsilon$ which
regulates the IR divergences. In the calculation of amplitudes it is common
to use dimensional regularisation, however it is difficult to define
the action of the super-conformal generators away from four dimensions. So
we choose a regulator which can be written in four dimensional
spinor variables but which reproduces
the answers of dimensional regularisation that is to say%
\footnote{This is essentially the same regulator, after using momentum
conservation across the two-particle cut, as was used in \cite{Bena:2004xu}.}
\[
\Delta^\epsilon_{12}=
c_\epsilon\lrbrk{\frac{\sprods{1}{4}\sprods{2}{3}\cprods{1}{2}}{\mu^2\sprods{3}{4}}}^{-\epsilon}\theta(-s_{12})~,
\]
where $1$ and $2$ label the internal momenta and $3$ and $4$ label the external legs.
The constant $c_\epsilon=1+\order{\epsilon}$
defined in \eqref{eq:ceps}
contains some unphysical artifacts of dimensional regularisation.
This factor vanishes when external leg $1$ becomes collinear with leg $4$ or leg $2$
becomes collinear with leg $3$ thus softening the divergence in the loop integral
that occurs for this configuration.
With this definition it can be shown that for the generators of interest,
\[
\delta^{(8)}(Q)\lrcomm{\gengen\supup{free}_{1}+\gengen\supup{free}_{2}}{\log {\frac{\sprods{1}{4}\sprods{2}{3}\cprods{1}{2}}{\mu^2\sprods{3}{4}}}} = \delta^{(8)}(Q)
\lrcomm{\gengen\supup{free}_{3}+\gengen\supup{free}_{4}}{\log \frac{s_{34}}{-\mu^2}}\,.
\]
This is obvious for the dilatation generator and requires only a little more effort for
the fermionic special conformal generators. Using this expression
 when one of the amplitudes has only four legs, and using \eqref{eq:tot4amp}
 such that
 \[
 \check{J}_{\bar 1}^a\check{J}_{\bar 2}^b {\cal A}_4
 =f^{ade}f^{bce} \int d^{4|4}\Lambda_3d^{4|4}\Lambda_4\, J_3^c J_4^d A_{\bar 1\bar 2 34}~,
 \]
 the anomaly can be written as
\<
\frac{ \epsilon }{4}
\prefac
\int& &\kern-15pt
d^{4|4}\Lambda_3d^{4|4}\Lambda_4\,\left(\int d^{4|4}\Lambda_1 d^{4|4}\Lambda_2
\Delta^\epsilon_{12}
A_{\bar 1\bar 234}\right)\nn\\
& &\times
\lrcomm{\gengen\supup{free}_{3}+\gengen\supup{free}_{4}}{\log \frac{s_{34}}{-\mu^2}}
f^{ace}f^{bde}(J^c_3J^d_4)(\check J^a_3 \check J^b_4 \mathcal{A}^{(0)}).
\>
 We can  use the fermionic delta-function to perform the Gra\ss{}mann integrations
 over legs $1$ and $2$ of the
 four point amplitude.  We can use the fact that
we have chosen our regulator  such that it
reproduces the answers of dimensional
regularisation to write
\<
\int d^{4}\lambda_1 d^{4}\lambda_2
\Delta^\epsilon_{12}\, \frac{\sprods{1}{2}^4\,\delta^4(P)}{\sprods{1}{2}\sprods{2}{3}\sprods{3}{4}\sprods{4}{1}}
\eq (2 \pi)^2 \int d^D \mathrm{LIPS}(\lambda_{1}, \lambda_{2} )\,
\frac{\sprods{1}{2}^4}{\sprods{1}{2}\sprods{2}{3}\sprods{3}{4}\sprods{4}{1}}\nn\\
\eq 2i (2\pi)^2  \disc\,
\frac{c_\epsilon}{\epsilon^2}
\lrbrk{\frac{s_{34}}{-\mu^2}}^{-\epsilon}.
\>
We substitute this into the above anomaly, relabel the indices and obtain
\[
\frac{-1}{2  \cas}
\int (d^{4|4}\Lambda)^2\,
\left(
\disc\,  \frac{c_\epsilon}{\epsilon}\lrbrk{\frac{s_{12}}{-\mu^2}}^{-\epsilon}\right)
\lrcomm{\gengen\supup{free}}{\log \frac{s_{12}}{-\mu^2}}
f^{ace}f^{bde}(J^c_1J^d_2)(\check J^a_1 \check J^b_2 \mathcal{A}^{(0)}).
\]

We have now obtained the anomaly of the cut arising from the measure factor.
It is finite as $\epsilon\to 0$ and rational.
However we are interested in the anomaly of the loop integral,
not just its cuts. In principle we should perform a dispersion integral
to obtain the loop anomaly, but here the result is obvious
due to finiteness and rationality.
For the generators of interest we have
\[
\left(\disc\,  \frac{c_\epsilon}{\epsilon}\left(\frac{s_{12}}{-\mu^2}\right)^{-\epsilon}\right)
\lrcomm{\gengen\supup{free}}{\log \frac{s_{12}}{-\mu^2}}
=
\disc\, \lrcomm{\gengen\supup{free}}{
-\frac{c_\epsilon}{\epsilon^2}\left(\frac{s_{12}}{-\mu^2}\right)^{-\epsilon}}
\]
and we then simply remove the discontinuity operator from inside
the anomaly. This is equivalent to multiplying by the logarithm and adding
a constant, though divergent term. The multiplication by the logarithm
clearly reproduces the correct discontinuity and the addition of the divergent
constant corresponds to the usual ambiguity involved in
reconstructing a function from its cuts. However as we shall see
in the comparison to the explicit answers for the amplitudes
in \secref{sec:MHVex}
this is the correct procedure.

Altogether the loop anomaly due to the measure reads
\[\label{eq:MeasureAnomaly}
-\frac{1}{2\cas}
\int (d^{4|4}\Lambda)^2\,
\lrcomm{\gengen\supup{free}}{\frac{c_\epsilon}{\epsilon^2}\lrbrk{\frac{s_{12}}{-\mu^2}}^{-\epsilon}}
f^{ace}f^{bde}J^c_1J^d_2\check J^a_1 \check J^b_2 \mathcal{A}^{(0)}.
\]
%

\subsection{Collinearities in Loops}
\label{sec:coll_in_loops}
We now turn to contributions of the second kind where
the anomaly vertex sits with two legs inside
the loop integral. Again we write it in a symmetric fashion
\<\earel{}
\half
\prefac
\int
(d^{4|4}\Lambda)^4
\Delta^\epsilon_{12}
\sign(E_4E_2)
J^d_4 \gengend^{dbc}_{42\bar 3}
(\check J^a_{\bar 1} \check J^b_{\bar 2} \mathcal{A}^{(0)})
(\check J^a_1 \check J^c_3 \mathcal{A}^{(0)})
\nl
+\half
\prefac
\int
(d^{4|4}\Lambda)^4
\Delta^\epsilon_{12}
\sign(E_4E_{\bar2})
J^d_4 \gengend^{dbc}_{4\bar 23}
(\check J^a_{\bar 1} \check J^c_{\bar 3} \mathcal{A}^{(0)})
(\check J^a_1 \check J^b_2 \mathcal{A}^{(0)}).
\>
We relabel particles $2$ and $3$ in the second term,
and the two terms combine
noting that on the cut the energies have equal signs
\[
\half
\prefac
\int
(d^{4|4}\Lambda)^4
\bigbrk{\Delta^\epsilon_{12}-\Delta^\epsilon_{13}}
\sign(s_{12}-s_{13})
J^d_4 \gengend^{dbc}_{42\bar 3}
(\check J^a_{\bar 1} \check J^b_{\bar 2} \mathcal{A}^{(0)})
(\check J^a_1 \check J^c_3 \mathcal{A}^{(0)}).
\]
We have furthermore used $\sign(E_1E_4)=\sign(-s_{14})=\sign(s_{12}-s_{13})$.
Counting the delta functions in the integrands --- in the amplitudes
and in the anomaly vertex \eqref{eq:AnomalyMain}  (see also \appref{app:AnomalyVertex})
--- we see that the three phase space integrals are completely localised:
The loop momentum yields four degrees of freedom while the on-shell
connections in the triangle (\figref{fig:CutAnomalies}) contribute one constraint each.
Collinearity in the anomaly vertex provides the final constraint
which localises the integral. Alternatively one can argue that the
anomaly vertex offers one degree of freedom corresponding
to the momentum fraction. It is used up by forcing the third side
of the triangle on shell.
Thus this cut anomaly is a finite and rational function of the kinematic variables.

Rationality and finiteness ensure
that discontinuities originate only from
the original cuts.
As we use dimensional regularisation
(represented through the measure $\Delta^{\epsilon}_{j,k}$ here),
it makes sense to consider $D$-dimensional cuts
with the discontinuity
\[
-2\pi i\,\Delta^{\epsilon}_{j,k}
=-\frac{c_\epsilon}{\epsilon} \disc \lrbrk{\frac{s_{j,k}}{-\mu^2}}^{-\epsilon}\,.
\]
Note that the factor $\sign(s_{12}-s_{13})$ does not lead to a discontinuity because it compensates
a sign originating from the on-shell integration over $\Lambda_{1,2,3}$.
Dropping the discontinuity operator
we find the following expression for the loop anomaly
\<\earel{}
\half
\lrbrk{\frac{1}{16\pi^3\cas}}
\int
(d^{4|4}\Lambda)^4
\sign(s_{12}-s_{13})
J^d_4 \gengend^{dbc}_{42\bar 3}
(\check J^a_{\bar 1} \check J^b_{\bar 2} \mathcal{A}^{(0)})
(\check J^a_1 \check J^c_3 \mathcal{A}^{(0)})
\nl\qquad\cdot
\frac{c_\epsilon}{\epsilon}
\lrbrk{
\lrbrk{\frac{s_{12}}{-\mu^2}}^{-\epsilon}
-\lrbrk{\frac{s_{13}}{-\mu^2}}^{-\epsilon}
}.
\>
As this expression is finite, we are free to expand
the bracket in $\epsilon$
\[\label{eq:CollLoopAnomaly}
-\lrbrk{\frac{1}{16\pi^3\cas}}
\int
(d^{4|4}\Lambda)^4
\sign(s_{12}-s_{13})
J^d_4 \gengend^{dbc}_{42\bar 3}
(\check J^a_{\bar 1} \check J^b_{\bar 2} \mathcal{A}^{(0)})
(\check J^a_1 \check J^c_3 \mathcal{A}^{(0)})
\log\frac{s_{12}}{s_{13}}\,.
\]
This result actually follows
directly by replacing $\Delta^\epsilon_{j,k}$
by step functions
originating from the discontinuity of a logarithm
\[
-2\pi i\,\theta(-s_{j,k})
=\disc\log\frac{s_{j,k}}{-\mu^2}\,.
\]
%

\subsection{One-Loop Splitting}
\label{sec:1loopsplit}

We now turn to the third type of anomaly in \figref{fig:CutAnomalies}, which occurs when the anomaly vertex sits
entirely inside the loop integral
\[
\quarter
\prefac
\int
(d^{4|4}\Lambda)^3
\Delta^\epsilon_{12}\sign(E_1E_2)
\gengend^{abc}_{12\bar 3} \Big[
(\check J^a_{\bar 1} \check J^b_{\bar 2} \mathcal{A}^{(0)})
(\check J^c_3 \mathcal{A}^{(0)})+(\check J^c_3 \mathcal{A}^{(0)})
(\check J^a_{\bar 1} \check J^b_{\bar 2} \mathcal{A}^{(0)})\Big]~.
\]
Recalling the structure of the anomaly vertex we can see that this
contributes when the internal momenta, labelled $1$ and $2$,  become collinear and
proportional to $3$. Let us consider the case
where we have a four point amplitude on the
one side of the cut. Using the expression for the four-point amplitude, \eqref{eq:tot4amp}, we can write
\<
\gengend^{abc}_{12\bar 3}
(\check J^a_{\bar 1} \check J^b_{\bar 2} \mathcal{A}^{(0)}_4)
=if^{abc}f^{adg}f^{beg}G_3\int d^{4|4}\Lambda_3 d^{4|4}\Lambda_4
J_3^e J_4^d A_{\bar 1\bar 2 34}~.
\>
Now, with the aid of the Jacobi identity and the relation between the dual Coxeter number,
$f_{abc} f_d{}^{bc}=C\indup{V}\delta_{ad}$, and the quadratic Casimir,
$\cas$, we see that the colour structures combine to produce an overall factor of  $\cas f^{ade}$
which cancels the $\cas$ in the prefactor.

As the two internal legs become collinear the remaining, external, two legs
of the four-point amplitude also become collinear.
We see that this contribution arises from the limit of the $n$-point amplitude as two external
legs become collinear and so is related to the splitting function. There are several subtleties involved in taking this limit; for example the kinematic invariant for this channel, $s_{12}$, is actually zero. It is therefore useful to  recall some salient
facts about the one-loop splitting function; we will closely follow the discussion in \cite{Kosower:1999xi}.

\begin{figure}
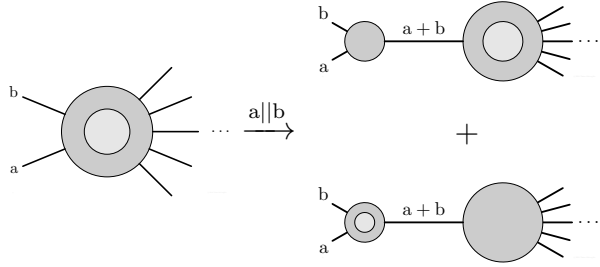
\centering
$\includegraphicsbox[scale=.6]{FigSplittingA1.mps}
\mathrel{\xrightarrow{\mathrm{a}||\mathrm{b}}}
\begin{array}{l}
\mathord{}
\includegraphicsbox[scale=.7]{FigSplittingA2.mps}
\\[4ex]\mathord{}
\hspace{1.8cm}{ +}
\\[2ex]\mathord{}
\includegraphicsbox[scale=.7]{FigSplittingA3.mps}
\end{array}
$
\caption{The two contributions to the collinear limit of the one-loop amplitude.}
\label{fig:one_loop_splitting}
\end{figure}

The one-loop amplitude, in the limit where two momenta become collinear
$p_{\rm a}\rightarrow z (p_{\rm a}+p_{\rm b})$, $p_{\rm b}\rightarrow(1- z) (p_{\rm a}+p_{\rm b})$,
 has two contributions, \figref{fig:one_loop_splitting}, which are given by
\<
A_n^{(1)}(1,\dots,{\rm a},{\rm b},\dots , n)& &\kern-10pt \xrightarrow{{\rm a}||{\rm b}}
{\rm Split^{(0)}}({\rm a},{\rm b}) A_{n-1}^{(1)}(1,\dots, ({\rm a}+{\rm b}), \dots , n)\nn\\
& &+{\rm Split}^{(1)}({\rm a},{\rm  b})A^{(0)}_{n-1}(1,\dots, ({\rm a}+{\rm b}),\dots, n).
\>

The first term is the tree-level splitting function which scales as $s_{\rm ab}^{-1/2}$ and the
effects of which have already been accounted for by the tree-level deformation of the
generators. In terms of cuts they are captured by $n$-particle cuts, $n>2$, where the
anomalous contribution forces two external legs to
become collinear. The second set of contributions correspond to the one loop splitting function
\[
\mathrm{Split}^{(1)}({\rm a},{\rm  b})=\mathrm{Split}^{(0)} ( {\rm a}, {\rm  b} ) r_S( z, s_{\rm ab} )
\]
where $ r\indup{S}( z, s_{\rm ab} )$ is independent of the flavour or helicity of the particles
labelled $\rm a$ and $\rm b$.

\begin{figure}
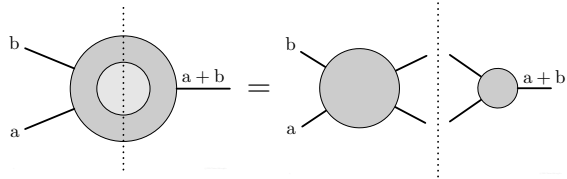
\centering
$
\includegraphicsbox[scale=.7]{FigSplittingCutA1.mps}\,
=\, \includegraphicsbox[scale=.7]{FigSplittingCutA2.mps}
$
\caption{Unitarity cut of the one-loop splitting function.}
\label{fig:one_loop_splitting_cut}
\end{figure}

This term is captured by the two-particle, ``singular'' channel, where on one side of the cut we have
a four point amplitude. As discussed in \cite{Kosower:1999xi}, the four point function
is singular in this limit, having a pole in $s_{\rm ab}$ rather than a square root
singularity. However as momentum conservation also
forces the loop momenta to become collinear we can use the factorisation of
the $n$-particle amplitude
on the other side of the cut to rewrite the expression as the cut of the
function $\mathrm{Split}^{(1)}({\rm a},{\rm  b})$, see
\figref{fig:one_loop_splitting_cut}.
By evaluating this cut or, alternatively, by taking the limit on the scalar box functions
which define the one-loop amplitude one can find an explicit
expression for $r\indup{S}(s_{\rm ab}, z)$,
\[
r\indup{S}(s_{\rm ab},z) =
-\left( \frac{ s_{\rm ab}}{ -\mu^2} \right)^{-\epsilon}\Big[ \frac{1}{\epsilon^2}- \frac{1}{\epsilon}\log z(1-z ) +\log^2\frac{1-z}{z} +\frac{\pi^2}{6}\Big]~.
\]
In this definition there is an ambiguity depending on the order in which one takes
the limits $\epsilon\rightarrow 0$ and
$s_{\rm ab}\rightarrow 0$. One prescription is to take the
singular momentum to zero before going to four dimensions this results in $r\indup{S}(0,z)=0$.
However one can
imagine different prescriptions such as setting
$s_{\rm ab}\rightarrow -\mu^2$ and
then taking $\epsilon\rightarrow 0$ defining $r\indup{S}(-\mu^2,z)$.

One can write the anomaly as a
 one-loop anomaly vertex, schematically, $G^{(1)}\sim r\indup{S} G^{(0)}$
 where $G^{(0)}$ denotes the tree-level anomaly vertex. More concretely
 for $\bar{\gen S}$,
 \<
 \left(\bar {\gen S}^{(1)}\right)_{123}^{abc}\eq
-2i f^{abc}\int d^{4|4}\Lambda'\,\delta^4(\lambda')\,
\varepsilon_{\dot\alpha\dot\gamma}\tilde\lambda_3^{\dot\gamma}\eta'^B
\int d\alpha\,d\varphi\,d\vartheta\,e^{i\varphi+i\vartheta}r\indup{S}(\alpha)
\nl\qquad\cdot
\delta^{4|4}(e^{-i\varphi}\Lambda_3\sin\alpha+e^{i\vartheta}\Lambda'\cos\alpha-\bar\Lambda_{1})
\nl\qquad\cdot
\delta^{4|4}(e^{-i\vartheta}\Lambda_3\cos\alpha-e^{i\varphi}\Lambda'\sin\alpha-\bar\Lambda_{2})+
\begin{array}{c}\mbox{two cyclic}\\\mbox{images}\end{array}~,
\>
where if we use the second prescription for the splitting function we have
($z=\sin\alpha$)
\[
r\indup{S}(\alpha)=
-\left[ \frac{1}{\epsilon^2}
  -\frac{2}{\epsilon}\log (\cos\alpha\sin\alpha)
  +4\log^2 \cot\alpha+\frac{\pi^2}{6}
\right]~.
\]
 Thus the anomalous contribution of the third kind
 is  explicitly
\[
\label{eq:SplitLoopAnomaly}
\half\int (d^{4|4}\Lambda)^3
\sign(E_1E_2)
(\gengend^{(1)})^{abc}_{12\bar 3} J^a_1 J^b_2  \check J^c_3~.
\]

In principle we  have to additionally consider the cases where
we have more than two external legs on both sides of the cut. These
terms correspond to one-loop corrections to multi-particle factorisation.
For example, the contribution of the cut with a five-point amplitude one side
is supported on the region of kinematical space where the sum
of three external particles becomes null. However for our definition of the
amplitudes, discussed further in \secref{sec:comparison_prop} but essentially
taking the principle part, there are no anomalous contributions to $\gen{S}$ or
$\bar{\gen S}$ on this support.

\subsection{Deformation of the Representation}

In summary of \eqref{eq:MeasureAnomaly},
\eqref{eq:CollLoopAnomaly} and
\eqref{eq:SplitLoopAnomaly}
we find the total one-loop anomaly
\<\label{eq:fullanomaly}
\gengen^{(0)}\mathcal{A}^{(1)}\eq
\frac{-1}{2\cas}
\int (d^{4|4}\Lambda)^2\,
\lrcomm{\gengen\supup{free}}{\frac{c_\epsilon}{\epsilon^2}\lrbrk{\frac{s_{12}}{-\mu^2}}^{-\epsilon}}
f^{ace}f^{bde}J^c_1J^d_2\check J^a_1 \check J^b_2 \mathcal{A}^{(0)}
\nl
-
\lrbrk{\frac{1}{16\pi^3\cas}}
\int
(d^{4|4}\Lambda)^4
\sign(s_{12}-s_{13})
J^d_4 \gengend^{dbc}_{42\bar 3}
(\check J^a_{\bar 1} \check J^b_{\bar 2} \mathcal{A}^{(0)})
(\check J^a_1 \check J^c_3 \mathcal{A}^{(0)})
\log\frac{s_{12}}{s_{13}}
\nl
-\half\int (d^{4|4}\Lambda)^3
\sign(E_1E_2)
\gengend^{(1)abc}_{12\bar 3} J^a_1 J^b_2 \check J^c_3 \mathcal{A}^{(0)}.
\>

Importantly all of these terms can be written as
some variation acting on the tree amplitude
$\gengen^{(0)}\mathcal{A}^{(1)}\sim \mathcal{A}^{(0)}$.
Thus we can cancel the anomaly easily
$\gengen^{(0)}\mathcal{A}^{(1)}+\gengen^{(1)}\mathcal{A}^{(0)}=0$
by introducing a corresponding
one-loop deformation of the superconformal representation
\[
\gengen^{(1)}=
\gengen^{(1)}_{2\to2}
+\sum_{k=3}^\infty
\gengen^{(1)}_{2\to k}
+
\gengen^{(1)}_{1\to 2}.
\]

The first term is meant to cancel the contribution due to
the measure anomaly. Here we first introduce an operator
$\hat{Z}^{(1)}_{2\to2}$
which captures the IR singularities in a one-loop amplitude
\[\label{eq:Zfactor}
\hat{Z}^{(1)}_{2\to2}=
-\frac{1}{2\cas}
\int (d^{4|4}\Lambda)^2\,
\frac{c_\epsilon}{\epsilon^2}\lrbrk{\frac{s_{12}}{-\mu^2}}^{-\epsilon}
f^{ace}f^{bde}J^c_1J^d_2\check J^a_1 \check J^b_2.
\]
Note that the momenta of the particles $1,2$ are not changed by this operator,
it merely acts non-trivially on the colour-structure and multiplies
by a divergent function of the two-particle invariant $s_{12}$.
In particular, this operator allows to split a one-loop amplitude
into IR-divergent contributions and a finite remainder $\tilde{\mathcal{A}}^{(1)}$
\cite{Mueller:1979ih, Collins:1980ih, Sen:1982bt, Korchemsky:1988pn, Korchemsky:1988hd, DelDuca:1989jt, Magnea:1990zb, Sterman:2002qn}
\[\label{eq:IRdivergentsplit}
\mathcal{A}^{(1)}=\hat{Z}^{(1)}_{2\to2}\mathcal{A}^{(0)}+\tilde{\mathcal{A}}^{(1)}.
\]
According to \eqref{eq:fullanomaly} the generator deformation
can now be written in the form of the commutator
\[\label{eq:G122comm}
\gengen^{(1)}_{2\to2}
=
\bigcomm{\hat{Z}^{(1)}_{2\to2}}{\gengen^{(0)}_{1\to1}}.
\]
It is obvious that this type of deformation respects the superconformal algebra
because it merely consists in a perturbative similarity transformation
of the free generators.

\begin{figure}
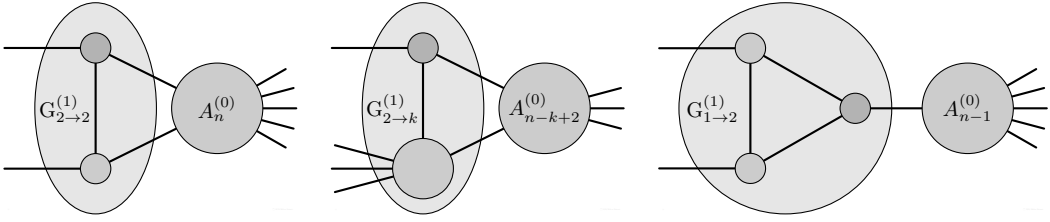
\centering
\includegraphicsbox[scale=0.8]{FigDeform22.mps}\quad
\includegraphicsbox[scale=0.8]{FigDeform2k.mps}\quad
\includegraphicsbox[scale=0.8]{FigDeform12.mps}
\caption{Structure of the deformations at one loop.}
\label{fig:DeformAll}
\end{figure}

The second term cancels the anomaly due to collinearities in the loop
\[\label{eq:G2kgen}
\gengen^{(1)}_{2\to k}
=
\lrbrk{\frac{1}{16\pi^3\cas}}
\int(d^{4|4}\Lambda)^4
\sign(s_{12}-s_{13})
J^d_4 \gengend^{dbc}_{42\bar 3}
(\check J^a_1 \check J^c_3 \mathcal{A}^{(0)}_{k+1})
\log\frac{s_{12}}{s_{13}}\,
\check J^a_{\bar 1} \check J^b_{\bar 2}.
\]
Note that this term is not uniquely determined
because we only know its action on tree amplitudes
$\mathcal{A}^{(0)}$ and not on generic functions.
The point is that the expression
already contains a tree amplitude $\mathcal{A}^{(0)}$
and thus when it acts on tree amplitudes it will
automatically symmetrise the two.
We could thus, alternatively, drop one of the terms of the
logarithm involving the invariants $s_{12}$ or $s_{13}$
in \eqref{eq:G2kgen} and multiply by two.
It has the same effect on tree amplitudes, but
it is a different deformation of the representation.
The third term removes the one-loop collinear anomaly
\[
\gengen^{(1)}_{1\to 2}
=
\half\int (d^{4|4}\Lambda)^3
\sign(E_1E_2)
\gengend^{(1)abc}_{12\bar 3} J^a_1 J^b_2 \check J^c_3.
\]
See \figref{fig:DeformAll} for an illustration of the
one-loop deformations $\gengen^{(1)}$ acting on an amplitude.

We will see in \secref{sec:MHVex} that exactly
these deformations are required
to make planar MHV amplitudes superconformally invariant.

\subsection{Planar Representation}

The above anomaly and the corresponding deformation
of the representation hold for arbitrary gauge groups of finite rank.
It is often convenient to restrict to the planar limit
in a $\grp{U}(N\indup{c})$ gauge group where
most expressions simplify.
Let us therefore formulate the deformation in the planar limit.
We act with the generators on a functional $\mathcal{X}[J]$
which is based on a colour-ordered function $X$
according to \eqref{eq:GenFuncOrdered}.
In the following we shall use a notation where the indices
in some quantity $X\rng{k}{j}$
denote a range of $k$ adjacent particles starting at particle $j$.
For example we introduce the partial momentum,
supermomentum and Lorentz invariants
\[\label{eq:sinv}
P\rng{k}{j}:=\sum_{i=0}^{k-1}p_{j+i},
\qquad
Q\rng{k}{j}:=\sum_{i=0}^{k-1}q_{j+i},
\qquad
t\rng{k}{j}:=(P\rng{k}{j})^2.
\]

The generator $\gengen^{(2)}_{2\to2}$
is given in terms of the IR-singularity operator
$\hat{Z}^{(1)}_{2\to2}$ in \eqref{eq:Zfactor}.
In the planar limit, the colour structure forces this operator
to act on two adjacent legs of the amplitude.
Hence we can write its action in the following form
\[\label{eq:Zfactorplanar}
\hat{Z}^{(1)}_{2\to 2}=
\sum_{i=1}^n
(\hat{Z}^{(1)}_{2\to 2})\rng{2}{i},
\qquad
(\hat{Z}^{(1)}_{2\to 2})\rng{2}{i}
=
-\frac{c_\epsilon}{\epsilon^2}
\lrbrk{\frac{\sinv{2}{i}}{-\mu^2} }^{-\epsilon}.
\]
Now the correction term $\gengen^{(1)}_{2\to 2}$ in \eqref{eq:G122comm} has a particularly simple
structure in the planar limit
\[\label{eq:planardiagonal}
\gengen^{(1)}_{2\to 2}
=
\bigcomm{\hat{Z}^{(1)}_{2\to2}}{ \gengen^{(0)}_{1\to 1}}
=\sum_{ i = 1 }^n
(\gengen^{(1)}_{2\to 2})\rng{2}{i},
\qquad
(\gengen^{(1)}_{2\to 2})\rng{2}{i}
=
\bigcomm{ (\hat{Z}^{(1)}_{2\to 2})\rng{2}{i} }{ \gengen\supup{free}_{i}+ \gengen\supup{free}_{i+1}}.
\]
This generator acts on two adjacent particles $i,i+1$.
As noted above this type of deformation applies to the generators
$\gen{D}$, $\gen{S}$, $\bar{\gen{S}}$ and $\gen{K}$.
The latter can always be expressed through commutators and
hence we list an explicit result only for the first three
\<\label{eq:planardiagonalDSSb}
(\gen{D}^{(1)}_{2\to 2})\rng{2}{i}
\eq
-\frac{2c_\epsilon}{\epsilon}\lrbrk{\frac{\sinv{2}{i}}{-\mu^2} }^{-\epsilon},
\nln
((\gen{S}^{(1)}_{2\to 2})\rng{2}{i})_{\alpha B}
\eq
-\frac{c_\epsilon}{\epsilon}
\lrbrk{\frac{\sinv{2}{i}}{-\mu^2}}^{-\epsilon}
\frac{\varepsilon_{\alpha\gamma}}{\sprod{i}{i+1}}
\lrbrk{\lambda_{i+1}^{\gamma}\partial_{i,B}-\lambda_i^{\gamma}\partial_{i+1,B}},
\nln
((\bar{\gen{S}}^{(1)}_{2\to2})\rng{2}{i})^B_{\dot\alpha}
\eq
-
\frac{c_\epsilon}{\epsilon}
\lrbrk{\frac{\sinv{2}{i}}{-\mu^2}}^{-\epsilon}
\frac{\varepsilon_{\dot\alpha\dot\gamma}}{\cprod{i}{i+1}}\,
\bigbrk{\tilde\lambda_{i+1}^{\dot\gamma}\eta_{i}^B-\tilde\lambda_{i}^{\dot\gamma} \eta_{i+1}^B}
\,.
\>

Let us now act on some colour-ordered amplitude function $X_n$
with the generator $\gengen^{(1)}_{2\to k}$.
Due to the inhomogeneous nature of the generator $\gengen^{(1)}_{2\to k}$
it will return some amplitude function $Y_{n+k-2}$ with $(n+k-2)$ legs
\[
\gengen^{(1)}_{2\to k} X_n = Y_{n+k-2}.
\]
As before the two variations in \eqref{eq:G2kgen} must hit adjacent sources, so we can
introduce a generator $(\gengen^{(1)}_{2\to k})\rng{2}{j}$
which acts on legs $j,j+1$ of $X_n$ and replaces them by $(k-2)$ new legs
\[
\gengen^{(1)}_{2\to k}=\sum_{j=1}^{n+k-2}(\gengen^{(1)}_{2\to k})\rng{k}{j},
\qquad
Y_{n+k-2}
=\sum_{j=1}^{n+k-2}(Y_{n+k-2})\rng{k}{j}
=\sum_{j=1}^{n+k-2}(\gengen^{(1)}_{2\to k})\rng{k}{j}X_n
.
\]
The range of the sum may seem surprising at first sight,
but it is the only way the resulting expression can make sense:
Both $X_n$ and $Y_{n+k-2}$ must be cyclic functions.
Consequently, it does not matter which pair of legs of $X_n$ is chosen.
The problem is that the contribution $(Y_{n+k-2})\rng{k}{j}$ is not cyclic.
Cyclicity of $Y_{n+k-2}$ is only restored
in a sum over all cyclic permutations of its $(n+k-2)$ legs.
Evaluating the colour structures in \eqref{eq:G2kgen}
we can write the planar action as
\<\label{eq:G2kplanar}
(\gengen^{(1)}_{2\to k})\rng{k}{j}X_n\eq
-
\frac{1}{16\pi^3}
\int
d^{4|4}\Lambda_{\mathrm{a}}\,
d^{4|4}\Lambda_{\mathrm{b}}\,
d^{4|4}\Lambda_{\mathrm{c}}\,
X_n(1,\ldots,j-1,\mathrm{b},\mathrm{c},j+k,\ldots,n+k-2)
\nl
\cdot\bigg[
\sign(\sinv{k}{j}-\sinv{k-1}{j+1})
\gengend_3(\mathrm{\bar a},\mathrm{\bar b},j)
A^{(0)}_k(\mathrm{a},j+1,\ldots,j+k-1,\mathrm{\bar c})
\log\frac{\sinv{k}{j}}{\sinv{k-1}{j+1}}
\nl\quad
-
\sign(\sinv{k}{j}-\sinv{k-1}{j})
A^{(0)}_k(\mathrm{\bar b},j,\ldots,j+k-2,\mathrm{a})
\gengend_3(\mathrm{\bar c},\mathrm{\bar a},j+k-1)
\log\frac{\sinv{k}{j}}{\sinv{k-1}{j}}
\bigg]
.
\nl
\>

Similarly, the planar action of the third generator reads
\<
(\gengen^{(1)}_{1\to 2})\rng{2}{j}X_n\eq
\int d^{4|4}\Lambda_{\mathrm{a}}\,
\sign(E_jE_{j+1})\,
\gengend^{(1)}_3(j,j+1,\mathrm{\bar a})
\nl\quad\cdot
X_n(1,\ldots,j-1,\mathrm{a},j+2,\ldots,n+1)
.
\>
%

\section{Superconformal Symmetry of MHV Amplitudes}
\label{sec:MHVex}

Having established the general framework for superconformal
symmetry of one-loop amplitudes, we will confirm it using
the simple set of planar MHV amplitudes $A\indup{MHV}$.
To avoid clutter, we shall drop the label MHV
from the amplitude functions $A$ and functionals $\mathcal{A}$
throughout this section.

\subsection{One-Loop Correction}

We summarise the construction and the properties of one-loop MHV amplitudes
in \appref{sec:MHVapp} in order to focus on the one-loop anomalies here.
For MHV amplitudes the helicity-dependence
is fully constrained by the symmetry.
It forces the exact amplitude to equal
the tree result times a function of the
particle momenta
\[
A_n^{(\ell)}
=
A_n^{(0)}
M_n^{(\ell)},
\qquad
A_n^{(0)}=\frac{\delta^4(P)\,\delta^8(Q)}{\sprods{1}{2}\ldots\sprods{n}{1}}\,,\quad
M_n^{(0)}=1.
\]
The one-loop amplitude in dimensional reduction reads
\cite{Bern:1994zx}
\<\label{eq:Aloop}
M^{(1)}_n\eq
-
\sum_{j=1}^n
\frac{c_\epsilon}{\epsilon^2}
\lrbrk{\frac{\sinv{2}{j}}{-\mu^2}}^{-\epsilon}
+\sfrac{1}{6}n\pi^2
-
\half\sum_{k=3}^{n-3}\sum_{j=1}^n
\Li_2\lrbrk{1-\frac{\sinv{k-1}{j+1}\sinv{k+1}{j}}{\sinv{k}{j}\sinv{k}{j+1}}}
\nl
-
\half\sum_{k=2}^{n-3}\sum_{j=1}^n
\log^2\frac{\sinv{k}{j}}{\sinv{k+1}{j}}
+
\quarter\sum_{k=2}^{n-2}\sum_{j=1}^n\log^2\frac{\sinv{k}{j}}{\sinv{k}{j+1}}\,.
\>
The loop function depends on invariants $\sinv{k}{j}$ associated
to the overall momentum of $k$ consecutive particles starting at particle $j$
introduced in \eqref{eq:sinv}.
Furthermore $c_\epsilon=1+\order{\epsilon}$
is some function of the dimensional reduction
parameter $\epsilon$ and $\mu$ is the regularisation scale. It is perhaps
worth nothing that this formula is chosen to reproduce only the
most ``complicated" part of the loop integral. That is to say, it does
not reproduce the imaginary parts of the logarithm and dilogarithms.
In order to  define the function one must specify the appropriate Riemann
sheet for all values of the kinematic variables.

\subsection{Measure Anomaly}

First we will consider symmetry generators
$\gengen$ which are anomaly-free at tree level,
i.e.\ they act as in the free theory $\gengen^{(0)}=\gengen^{(0)}_{1\to1}=\gengen\supup{free}$.
In particular, the dilatation generator $\gen{D}$
and effectively also the superconformal boost $\gen{S}$ when acting on MHV amplitudes
are of this form.
The proposed one-loop deformation \eqref{eq:planardiagonal} is simple:
It acts on nearest neighbouring particles only
with a simple commutator form for the pairwise action
in terms of the free generator $\gengen\supup{free}$
\[
\gengen^{(1)}=\gengen^{(1)}_{2\to 2}=\sum_{j=1}^n
(\gengen^{(1)}_{2\to2})\rng{2}{j},
\qquad
(\gengen^{(1)}_{2\to2})\rng{2}{j}
=
\lrcomm{\gengen\supup{free}}
{\frac{c_\epsilon}{\epsilon^2}\lrbrk{\frac{\sinv{2}{j}}{-\mu^2}}^{-\epsilon}}
.
\]

The simplest non-trivial anomaly is the one of the generator
of scale transformations $\gen{D}$.
The free representation \eqref{eq:StdRep}
and the one-loop deformation \eqref{eq:planardiagonalDSSb} read
\[
\label{44}
\gen{D}\supup{free}_{j}=
1+
\half\lambda^\alpha_j\partial_{j,\alpha}+
\half\tilde\lambda^{\dot\alpha}_j\tilde\partial_{j,\dot\alpha}
,
\qquad
(\gen{D}^{(1)}_{2\to2})\rng{2}{j}
=
-\frac{2c_\epsilon}{\epsilon}\lrbrk{\frac{\sinv{2}{j}}{-\mu^2}}^{-\epsilon}
.
\]
The only term in \eqref{eq:Aloop} violating
scaling invariance is the one containing the
regularisation scale $\mu$. The scaling anomaly reads
\[
\gen{D}^{(0)}_{1\to1}A^{(1)}_n=
A^{(0)}_{n}
\sum_{j=1}^n
\frac{2c_\epsilon}{\epsilon}
\lrbrk{\frac{\sinv{2}{j}}{-\mu^2}}^{-\epsilon}.
\]

As anticipated the anomaly depends on the momenta
of two adjacent particles only.
Obviously, the one-loop deformation cancels precisely the anomaly
and makes the one-loop amplitude
exactly invariant under (deformed) scaling transformations
\[
\gen{D}^{(0)}_{1\to1}A^{(1)}_n+
\gen{D}^{(1)}_{2\to2}A^{(0)}_n=0.
\]

Next we consider the superconformal boost generator $\gen{S}$
given in \eqref{eq:StdRep} and \eqref{eq:planardiagonalDSSb}
\[
(\gen{S}\supup{free}_{j})_{\alpha B}=\partial_{j,B}\partial_{j,\alpha},
\qquad
((\gen{S}^{(1)}_{2\to2})\rng{2}{j})_{\alpha B}
=-
\frac{c_\epsilon}{\epsilon}
\lrbrk{\frac{\sinv{2}{j}}{-\mu^2}}^{-\epsilon}
\frac{\varepsilon_{\alpha\delta}}{\sprod{j}{j+1}}
\lrbrk{\lambda_{j+1}^{\delta}\partial_{j,B}-\lambda_j^{\delta}\partial_{j+1,B}}
.
\]
In applying the free generator to $A^{(1)}_n$, the fermionic derivative will
act on the $\delta^8(Q)$ in $A^{(0)}_n$ because it is the only
piece depending on the $\eta$'s.
The bosonic derivative must act on the loop function $M^{(1)}_n$
because the tree-level amplitude $A^{(0)}_n$ is invariant
\[\label{eq:S0onA1}
(\gen{S}^{(0)}_{1\to1})_{\alpha B}A^{(1)}_n
=
\sum_{j=1}^n
(\partial_{j,B}A^{(0)}_n)\,(\partial_{j,\alpha} M^{(1)}_n)
=
\sum_{j=1}^n
\frac{\partial A^{(0)}_n}{\partial Q^{\gamma B}}\,
\lambda^\gamma_j\partial_{j,\alpha}M^{(1)}_n.
\]
The second form uses the identity $\partial_{j,B} Q^{\alpha C}=\delta^C_B\lambda_j^\alpha$,
cf.\ \eqref{eq:supermom}.
The combination $\lambda^\gamma_j\partial_{j,\alpha}$, summed over all sites,
equals the Lorentz generator $\gen{L}^\gamma{}_\alpha$ up to its trace.
The function $M^{(1)}_n$ is a Lorentz invariant, and hence it is annihilated
by $\gen{L}^{\gamma}{}_\alpha$. Furthermore the trace contribution
measures the weight in $\lambda$'s
which is the same as the scaling weight for the invariants $\sinv{k}{j}$.
The superconformal boost anomaly is thus very similar to the scaling anomaly:
\[
(\gen{S}^{(0)}_{1\to1})_{\alpha B}A^{(1)}_n
=
\frac{\partial A^{(0)}_n}{\partial Q^{\gamma B}}\,
\lrbrk{
\gen{L}^{\gamma}{}_\alpha
+\half\delta^\gamma_{\alpha}
\sum_{j=1}^n
\lambda^\delta_j\partial_{j,\delta}
}
M^{(1)}_n
=
\frac{\partial A^{(0)}_n}{\partial Q^{\alpha B}}\,
\sum_{j=1}^n
\frac{c_\epsilon}{\epsilon}
\lrbrk{\frac{\sinv{2}{j}}{-\mu^2}}^{-\epsilon}.
\]

We should compare this expression to the one-loop deformation.
As above we make use of the fact that the fermionic derivative $\partial_{j,B}$
only hits the fermionic delta function $\delta^8(Q)$
and that $\partial_{j,B} Q^{\alpha C}=\delta^C_B\lambda_j^\alpha$
\<
(\gen{S}^{(0)}_{2\to2})_{\alpha B}A^{(0)}_n
\eq
-\sum_{j=1}^n
\frac{c_\epsilon}{\epsilon}
\lrbrk{\frac{\sinv{2}{j}}{-\mu^2}}^{-\epsilon}
\frac{\varepsilon_{\alpha\delta}}{\sprod{j}{j+1}}
\lrbrk{\lambda_{j+1}^{\delta}\partial_{j,B}-\lambda_j^{\delta}\partial_{j+1,B}}
A^{(0)}_n
\,
\nln
\eq
-\sum_{j=1}^n
\frac{c_\epsilon}{\epsilon}
\lrbrk{\frac{\sinv{2}{j}}{-\mu^2}}^{-\epsilon}
\varepsilon_{\alpha\delta}\frac{\lambda_j^{\gamma}\lambda_{j+1}^{\delta}-\lambda_j^{\delta}\lambda_{j+1}^{\gamma}}{\sprod{j}{j+1}}
\,\frac{\partial A^{(0)}_n}{\partial Q^{\gamma B}}
\nln
\eq
-\sum_{j=1}^n
\frac{c_\epsilon}{\epsilon}
\lrbrk{\frac{\sinv{2}{j}}{-\mu^2}}^{-\epsilon}
\frac{\partial A^{(0)}_n}{\partial Q^{\alpha B}}\,
.
\>
Altogether we obtain the invariance condition
\[
\gen{S}^{(0)}_{1\to1}A^{(1)}_{n}+
\gen{S}^{(1)}_{2\to2}A^{(0)}_{n}=0.
\]
Note that $\gen{S}$ is not anomaly-free at tree level \cite{Bargheer:2009qu}.
In general one therefore expects the correction term
$\gen{S}^{(0)}_{1\to2}$ at tree level \cite{Bargheer:2009qu}
and further corrections $\gen{S}^{(1)}_{2\to k}$ loop level.
This anomaly however does not apply to MHV amplitudes
which is why the treatment of $\gen{S}$ was relatively simple.

\subsection{Collinearities in Loops}
\label{sec:Inhom_corr}

For generators $\gengen$ which are anomalous at tree level,
$\gengen^{(0)}=\gengen^{(0)}_{1\to 1}+\gengen^{(0)}_{1\to2}$,
we have to work harder.
The prototype example when acting on MHV amplitudes is $\bar{\gen{S}}$.
In addition to the homogeneous $\gengen^{(1)}_{2\to 2}$ corrections,
there are inhomogeneous terms $\gengen^{(1)}_{2\to k}$, $k\geq 3$
\[
\gengen^{(1)}=
\gengen^{(1)}_{2\to 2}+
\sum_{k=3}^{n-2} \gengen^{(1)}_{2\to k},
\qquad
\gengen^{(1)}_{2\to k}=\sum_{j=1}^n
(\gengen^{(1)}_{2\to k})\rng{2}{j}
\]
When acting on the $n-k+2$-particle amplitude $A^{(0)}_{n-k+2}$ they yield
an $n$-particle function to cancel the 	anomaly.
The action of $\gengen^{(1)}_{2\to k}$ on particles $j,j+1$ of $A^{(0)}_{n-k+2}$
defined in \eqref{eq:G2kplanar} uses the anomaly three-vertex $\gengend_3$
in \eqref{eq:AnomalyMain}.

In principle there can be loop corrections $\gengen^{(1)}_{1\to2}$
to the collinear anomaly $\gengen^{(0)}_{1\to2}$ itself.
In this section we assume for convenience that the
particle momenta are in a general position
and pairwise linearly independent.
The case of collinear external momenta will be considered
in the following section.

We now consider the conjugate superconformal generator $\bar{\gen{S}}$.
We first act on $A^{(1)}_{n}$ with the free generator
$(\bar{\gen{S}}\supup{free}_j)_{\dot\alpha}^B=
\eta_j^B\tilde\partial_{j,\dot\alpha}$
\eqref{eq:StdRep}.
The straight-forward variations will produce a lot of terms.
Let us therefore first consider the general variation of
the loop function $M^{(1)}_n$ under shifts of the invariants $\sinv{k}{j}$,
and simplify it as far as possible.
A very convenient expression for the variation
of \eqref{eq:Aloop} reads
(see \appref{sec:Mvar} for some intermediate expressions)
\<\label{eq:Mvar}
\delta M^{(1)}_n\eq
\sum_{j=1}^n
\lrbrk{\delta\log\frac{\sprod{j-1}{j}\cprod{j}{j+1}\sprod{j+1}{j+2}}{-\mu^2 \sprod{j}{j+1}}}
\frac{c_\epsilon}{\epsilon}\lrbrk{\frac{\sinv{2}{j}}{-\mu^2}}^{-\epsilon}
\nl
-
\sum_{k=2}^{n-3}\sum_{j=1}^n
\lrbrk{\delta\log\frac{\Upsilon\rng{k+1}{j-1}}{\Upsilon\rng{k}{j}}}
\log \frac{\sinv{k+1}{j}}{\sinv{k}{j}}\,.
\>
The symbol $\Upsilon\rng{k}{j}$ is defined as the following
Lorentz invariant combination
(see \eqref{eq:sinv} for the
definition of the fractional momentum $P\rng{k}{j}$)
\[
\Upsilon\rng{k}{j}=\sbra{j}P\rng{k-1}{j+1}\cket{j+k}
=\varepsilon_{\beta\delta}\varepsilon_{\dot\alpha\dot\gamma}\lambda_j^\beta (P\rng{k-1}{j+1})^{\delta\dot\alpha} \tilde\lambda_{j+k}^{\dot\gamma},
\]
and it originates from the following combination
occurring frequently in $\delta M^{(1)}_n$, see
\appref{sec:Mvar},
\[
\sinv{k}{j}\sinv{k}{j+1}-\sinv{k-1}{j+1}\sinv{k+1}{j}=-\Upsilon\rng{k}{j}\Upsilon\rng{n-k}{j+k}.
\]
Note that this identity makes use
of momentum conservation $P\rng{k}{j}=-P\rng{n-k}{j+k}$
and the fact that $D=4$.
In particular the latter is interesting because
the above factorisation is crucial for conformal symmetry
which is special to four dimensions.

For simplifying the anomaly arising from the first line in
\eqref{eq:Mvar} we note
that $\bar{\gen{S}}$ acts on conjugate spinors only.
Thus for the purpose of $\bar{\gen{S}}^{(0)}$ we
can replace the argument of the logarithm by $\sinv{2}{j}
=\sprod{j}{j+1}\cprod{j+1}{j}$.
This yields
\[
(\gen{S}^{(0)}_{1\to 1})^B_{\dot\alpha}
\log\frac{\sinv{2}{j}}{-\mu^2}
=
\frac{\varepsilon_{\dot\alpha\dot\gamma}
\bigbrk{\eta_j^B\tilde\lambda^{\dot\gamma}_{j+1}-\eta_{j+1}^B\tilde\lambda^{\dot\gamma}_{j}}}{\cprod{j}{j+1}}\,.
\]
The action in the brackets of the second line in \eqref{eq:Mvar}
can be evaluated and simplified using spinor algebra:
\[\label{eq:SbarUpsilon}
(\bar{\gen{S}}^{(0)}_{1\to1})^B_{\dot\alpha}
\log\frac{\Upsilon\rng{k+1}{j-1}}{\Upsilon\rng{k}{j}}=
\varepsilon_{\dot\alpha\dot\gamma}\tilde\lambda_{j+k}^{\dot\gamma}
\frac{\sprod{j-1}{j}}{\Upsilon\rng{k}{j}\Upsilon\rng{k+1}{j-1}}
\lrbrk{
\tilde\lambda_{j+k}^{\dot\epsilon}\varepsilon_{\dot\epsilon\dot\kappa}
(P\rng{k}{j})^{\lambda\dot\kappa}\varepsilon_{\lambda\delta}(Q\rng{k}{j})^{\delta B}
-\sinv{k}{j} \eta_{j+k}^{B}}.
\]
Here $Q\rng{k}{j}$ is a fractional supermomentum defined in \eqref{eq:sinv}.
Altogether the conjugate superconformal boost anomaly reads%
\footnote{Essentially identical formulae were found by \cite{Korchemsky:2009hm}
and \cite{Sever:2009aa} in their analysis of the ${\bar {\gen S}}$
(or correspondingly dual $\bar {\gen Q}$)
anomaly of the cuts of MHV and NMHV amplitudes.}
\<
(\bar{\gen{S}}^{(0)}_{1\to1})^B_{\dot\alpha} A^{(1)}_n\eq
A^{(0)}_n
\sum_{j=1}^n
\lrbrk{(\bar{\gen{S}}^{(0)}_{1\to1})^B_{\dot\alpha}\log\frac{\sinv{2}{j}}{-\mu^2}}
\frac{c_\epsilon}{\epsilon}\lrbrk{\frac{\sinv{2}{j}}{-\mu^2}}^{-\epsilon}
\nl
-
A^{(0)}_n
\sum_{k=2}^{n-3}\sum_{j=1}^n
\lrbrk{(\bar{\gen{S}}^{(0)}_{1\to 1})^B_{\dot\alpha}\log\frac{\Upsilon\rng{k+1}{j-1}}{\Upsilon\rng{k}{j}}}
\log \frac{\sinv{k+1}{j}}{\sinv{k}{j}}\,.
\>

The anomaly on the first line is obviously cancelled by
the $\bar{\gen{S}}^{(1)}_{2\to2}$ correction \eqref{eq:planardiagonal}.
The remaining anomaly from the terms on the second line
should be cancelled by terms from $\gen{S}^{(1)}_{2\to k}$ in
\eqref{eq:G2kplanar}.
\begin{figure}\centering
\includegraphics{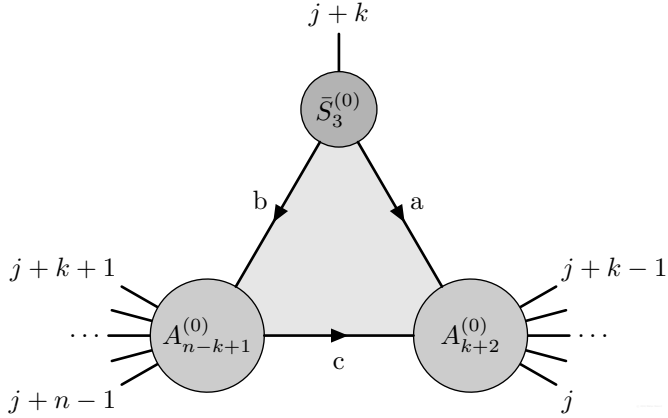}
\caption{$\bar{\gen{S}}$ cut anomaly $T\rng{k}{j}$.}
\label{fig:SbarTriangle}
\end{figure}
We write this as
\[
((\bar{\gen{S}}^{(1)}_{2\to n-k})\rng{k}{j})^B_{\dot\alpha} A^{(0)}_{k+2}
=
-
\half (T\rng{k}{j-k})^B_{\dot\alpha}
\log\frac{\sinv{n-k}{j}}{\sinv{n-k-1}{j+1}}
+
\half (T\rng{n-k-1}{j})^B_{\dot\alpha}
\log\frac{\sinv{n-k}{j}}{\sinv{n-k-1}{j}}
\]
where $T\rng{k}{j}$ is the on-shell triangle integral,
see \figref{fig:SbarTriangle} for the configuration of momenta,
\<\label{eq:Bdef}
(T\rng{k}{j})^B_{\dot\alpha}\eq
\frac{1}{8\pi^3} \int
d^{4|4}\Lambda\indup{a}\,
d^{4|4}\Lambda\indup{b}\,
d^{4|4}\Lambda\indup{c}\,
\sign(\sinv{k}{j}-\sinv{k+1}{j})\,
\bar S_3(\mathrm{\bar b},\mathrm{\bar a},j+k)^B_{\dot\alpha}
\nl\cdot
A^{(0)}_{k+2}(\mathrm{a},\mathrm{c},j,\ldots,j+k-1)
A^{(0)}_{n-k+1}(\mathrm{\bar c},\mathrm{b},j+k+1,\ldots,j+n-1).
\>
We carry out the lengthy calculation for $T$ in \appref{app:Triangle}, the final result
can be related to the quantity in \eqref{eq:SbarUpsilon}
\<
(T\rng{k}{j})^B_{\dot\alpha}\eq
A^{(0)}_{n}
\varepsilon_{\dot\alpha\dot\gamma}\tilde\lambda_{j+k}^{\dot\gamma}
\bigbrk{\tilde\lambda_{j+k}^{\dot\kappa} \varepsilon_{\dot\kappa\dot\lambda}
(P\rng{k}{j})^{\delta\dot\lambda} \varepsilon_{\delta\epsilon}(Q\rng{k}{j})^{\epsilon B}
-\sinv{k}{j}\eta_{j+k}^B}
\frac{\sprod{j-1}{j}}{\Upsilon\rng{k}{j}\Upsilon\rng{k+1}{j-1}}
\nln
\eq
A^{(0)}_{n}
(\bar{\gen{S}}^{(0)}_{1\to 1})^B_{\dot\alpha}\log\frac{\Upsilon\rng{k}{j}}{\Upsilon\rng{k+1}{j-1}}\,.
\>
Summing over all contributions, expanding the logarithms
and reordering some of the sums,
we find the contribution of the remaining deformation
\<
\sum_{k=2}^{n-3}
(\bar{\gen{S}}^{(1)}_{2\to n-k})^B_{\dot\alpha} A^{(0)}_{k+2}
\eq
+\sum_{k=2}^{n-3}\sum_{j=1}^n
(T\rng{k}{j})^B_{\dot\alpha}
\log\frac{\sinv{k+1}{j}}{\sinv{k}{j}}
\nln\eq
A^{(0)}_{n}
\sum_{k=2}^{n-3}\sum_{j=1}^n
\lrbrk{(\bar{\gen{S}}^{(0)}_{1\to 1})^B_{\dot\alpha}\log\frac{\Upsilon\rng{k}{j}}{\Upsilon\rng{k+1}{j-1}}}
\log\frac{\sinv{k+1}{j}}{\sinv{k}{j}}\,.
\>
By comparison to the above results we find proper invariance under deformed conjugate superconformal boosts
\[
\bar{\gen{S}}^{(0)}_{1\to 1} A^{(1)}_n+\bar{\gen{S}}^{(1)}_{2\to2} A^{(0)}_n+
\sum_{k=2}^{n-3}\bar{\gen{S}}^{(1)}_{2\to n-k} A^{(0)}_{k+2}
=0.
\]
%

\subsection{Splitting Anomaly}
\label{sec:splitting_amp}
Finally we consider the terms arising when the generator acts
on the tree-level prefactor.
As described previously, when the conjugate superconformal generator is applied
to the  tree-level MHV amplitude, it will see the holomorphic
poles as anomalies.
This produces a delta-function which forces certain momenta in the loop integral
to become collinear, effectively setting some
$\sinv{2}{k}=(p_k+p_{k+1})^2$ to zero.
This in turn corresponds to considering
the collinear limits of the one-loop amplitude
which is known to be governed by the one-loop splitting function, $r\indup{S}$
\cite{Bern:1994zx},
\<
\bar{\gen{S}}^{(0)} A_n^{(1)}\eq\left(\bar{\gen{S}}^{(0)} A_n^{(0)}\right)M_n^{(1)}
+\dots\nn\\
\earel{\propto} \delta^{(2)}(\sprod{k}{k+1}) A_{n-1}^{(0)}M_n^{(1)}+\dots\nn\\
\eq \delta^{(2)}(\sprod{k}{k+1}) A_{n-1}^{(0)}\left(M_{n-1}^{(1)}+r\indup{S}\right)+\dots~.
\>
%
For the appropriate definitions of the collinear
limit $r\indup{S}$ is  the same function as was found in the analysis
of the unitarity cuts, \secref{sec:1loopsplit}. The first term is simply cancelled
by the tree level deformation $ \bar{\gen{S}}^{(0)}_{1\to 2}$ while the second
are cancelled by $\bar{\gen{S}}^{(1)}_{1\to 2}$.
Combining this with the previous terms we find the full deformation of the
conjugate superconformal generator at one loop
\[
\bar{\gen{S}}^{(0)}_{1\to 1} A^{(1)}_n+\bar{\gen{S}}^{(0)}_{1\to 2} A^{(1)}_{n-1}+\bar{\gen{S}}^{(1)}_{1\to 2} A^{(0)}_{n-1}+
\bar{\gen{S}}^{(1)}_{2\to2} A^{(0)}_n+
\sum_{k=2}^{n-3}\bar{\gen{S}}^{(1)}_{2\to n-k} A^{(0)}_{k+2}
=0.
\]

As was discussed in  \secref{sec:1loopsplit}, and further in \appref{app:app_collinear}, the exact
definition of the collinear limit is subtle. There is an inherent ambiguity related to the
order in which one takes the $\epsilon\rightarrow 0$ and collinear limits. Different prescriptions
will give different results, however as long as we are consistent this is accounted for
by the appropriate definition of $r\indup{S}$  that appears in both the deformed generator and the collinear
limit of the amplitude.

\section{Invariance of the Six-Point NMHV Amplitude}
\label{sec:6NMHVex}

The invariance of one-loop amplitudes
with respect to the deformed generators, as described in \secref{sec:sym1loop},
is designed to apply to generic amplitudes not just MHV.
To check that this is indeed the case and to test the specifics of
the  proposal
we examine the
simplest non-trivial NMHV amplitude i.e.\ the six-point
one-loop NMHV amplitude, $A\indup{6;NMHV}^{(1)}$.

A convenient, manifestly supersymmetric, expression for this
amplitude  was given in \cite{Drummond:2008vq}
(see \cite{Bern:1994cg} for earlier calculations of the component
 amplitudes). It is written in terms of the
dual-superconformal invariants, $R_{rst}$, which are polynomials
in the $\eta$'s of order four,
\[
A^{(1)}\indup{6;NMHV}=A^{(0)}\indup{6;MHV}\Big[R_{146}F_6^{[1]}+\mathrm{cyclic}\Big]~.
\]
The functions $F^{[i]}_{6}$ depend on the kinematic invariants $t_j^k$ and
are combinations of two-mass hard and one-mass scalar
box functions. We give explicit expression for the $R_{rst}$'s and $F^{[i]}_6$'s in
\appref{sec:NMHVapp}.

For simplicity we will focus on the variation of this amplitude with
respect to the generator $\bar{\gen{S}}$;
for 6 legs the action of $\gen{S}$ follows by conjugation.
As in previous sections we will
show that the non-trivial action of the tree level $\bar{\gen{S}}^{(0)}_{1\to1}$
is cancelled by the deformations constructed previously. That is to say,
\[
\bar{\gen{S}}^{(0)}_{1\to1}A^{(1)}\indup{6;NMHV}+\bar{\gen{S}}^{(1)}_{2\to2}A^{(0)}\indup{6;NMHV}
+\sum_{k=2}^3\bar{\gen{S}}^{(1)}_{2\to n-k}A^{(0)}_{k+2;{\rm NMHV}}=0~.
\]
In this equation we are ignoring the anomaly in the tree-level amplitudes;
these terms are cancelled by the deformations, $\bar{\gen{S}}^{(0)}_{1\to2}$ and
$\bar{\gen{S}}^{(1)}_{1\to2}$, corresponding to the splitting function. We will thus focus
on the variation of the loop integral portion of the amplitude.

\subsection{Variation}
Using the
 explicit expressions for the functions $F_6^{[i]}$, \eqref{eq:boxfn},
one can
straightforwardly calculate their variation. Let us first consider
the terms in the variation with log cuts in the three-particle channels, say
$t_1^3$. These terms occur only multiplying the $R_{146}$ and $R_{413}$
structures, for example,
\[
\label{eq:x14vartot}
 {\bar{\gen{S}}^{(0)}_{1\to1}} A^{(1)}\indup{6;NMHV}=A^{(0)}\indup{6;MHV}R_{146} \log t_1^3
 \left(  {\bar{\gen{S}}^{(0)}_{1\to1}}  \log\Big[\frac{t_1^2 t_2^2}{t_3^2t_6^2}\Big]\right)+\dots~.
\]
In principle one could completely expand in powers of the $\eta$'s and
consider the various terms independently. It turns out to be more convenient
to leave the $R$'s intact and to consider the coefficients
of the various $\eta_A R_{rst}$ terms.
However, the $R$'s are not independent but rather satisfy various identities.
For example, for the six-point amplitude one can use the identity \eqref{eq:threetermid}
to remove one of the six $R$'s and so one could expect cancellations
between different terms.%
\footnote{In fact, if we use the overall supermomentum
delta-function and the fermionic delta-function
in the $R$ to remove three of the $\eta_A$'s we have fifteen
terms, $\eta_A R_{rst}$, in the variation.
If we alternatively use only the supermomentum condition we
get four $\eta$'s times six $R$'s. Now for each $R$ we have an additional
constraint between the $\eta$'s -- six constraints -- while for
each $\eta$ we have a relation
between the various $R$'s -- four constraints -- giving
a total of ten constraints and hence only fourteen independent
terms. Thus if these constraints are not degenerate we find that there
must be at least one additional relation between the various terms.}
Nonetheless, through a judicious choice we will be able to
consider the coefficients of the $\eta_A R_{rst}$ terms separately.
In fact, for the case of the three-particle channel,
the coefficients of $\eta_A R_{146}$ and $\eta_A R_{413}$ are
independent as can seen by noting that the first term will give rise
to terms, once we include the overall fermionic supermomentum delta-function, such
as $\eta_A\eta_4^4\eta_5^4\eta_6^4$ which cannot arise from the second term.
Concretely, we focus on the coefficient of $R_{146}$ and
using the fermionic delta-functions to remove $\eta_1$, $\eta_3$ and $\eta_5$.
We find
\<
\label{eq:x14var}
A^{(0)}\indup{6;MHV} R_{146}\bar{\gen{S}}^{(0)}_{1\to1}\log\Big[\frac{t_2^2}{t_3^2}\Big]
\eq A^{(0)}\indup{6;MHV}  R_{146}
\frac{{\bar \lambda}_3\langle 1|P_1^3|4]}{\sprods{1}{3}[23][34][46]}
\left(\eta_2[46]
+\eta_4[62]
+\eta_6[24]\right),\nn\\
A^{(0)}\indup{6;MHV} R_{146}\bar{\gen{S}}^{(0)}_{1\to1}\log\Big[\frac{t_1^2}{t_6^2}\Big]
\eq A^{(0)}\indup{6;MHV} R_{146}
\frac{{\bar \lambda}_1\langle 3|P_3^3|6]}{\sprods{1}{3}[12][46][61]}
\left(\eta_2[46]+\eta_4[62]
+\eta_6[24]\right)~.
\nl
\>
Combining these terms we find the complete variation in the three-particle channel
$t_1^3$ with coefficient $R_{146}$.
It is interesting to note that this calculation is essentially the same
as that performed in \cite{Cachazo:2004dr} which used the holomorphic
anomaly of the collinearity operator to fix the box coefficients of the split-helicity
NMHV amplitudes. Indeed the coefficient of $\eta_2$ in $\bar{\gen{S}}$
is exactly the collinearity operator for particles $1$, $2$ and $3$ used
in \cite{Cachazo:2004dr}. All the other three particle cuts can be found using
cyclicity.

Turning to the two-particle channel cuts of the variation, for example cuts
in the variable $t_1^2$,
we choose to remove $R_{413}$  using \eqref{eq:threetermid}. The resulting expression
for the variation of the amplitude in terms of the remaining $R$'s is
\<
\label{eq:x13var}
\gen{{\bar S}}_{1\to 1}^{(0)}A^{(1)}\indup{6;NMHV}\eq A^{(0)}\indup{6;MHV}
 \left(
\left(R_{146}+R_{362}+ R_{524}\right)\,
\gen{{\bar S}}_{1\to 1}^{(0)}
\Big[-\frac{c_{\epsilon}}{\epsilon^2}\left( \frac{ t_1^2}{-\mu^2}\right)^{-\epsilon}\Big]\right.
\nl
\left.
-\, \log\frac{ t_1^2}{-\mu^2} R_{146}\, \gen{{\bar S}}_{1\to 1}^{(0)} \log\Big[\frac{t_2^2  }{t_3^2}\Big]
 -\log\frac{ t_1^2}{-\mu^2} R_{635}\, \gen{{\bar S}}_{1\to 1}^{(0)}\log\Big[\frac{t_6^2 }{t_5^2}\Big]^{}\right)
 +\dots
\nln\eq
A\indup{6;NMHV}^{(0)}
\gen{{\bar S}}_{1\to 1}^{(0)}
\Big[-\frac{c_{\epsilon}}{\epsilon^2}\left( \frac{ t_1^2}{-\mu^2}\right)^{-\epsilon}\Big]
\nl
 -\,A\indup{6;MHV}^{(0)}\log\frac{ t_1^2}{-\mu^2}
\left(
R_{146}\, \gen{{\bar S}}_{1\to 1}^{(0)} \log\Big[\frac{t_2^2  }{t_3^2}\Big]
 +R_{635}\, \gen{{\bar S}}_{1\to 1}^{(0)} \log\Big[\frac{t_6^2 }{t_5^2}\Big]\right)+\dots~.
\nl
\>
Looking at this term we can immediately see that the first term is the same divergent structure
as appeared for the MHV amplitudes. As was the case there, this term corresponds
to the $\bar{\gen{S}}_{2\to2}^{(1)}$ deformation due to the measure correction. The
subsequent terms are given by \eqref{eq:x14var} and its cyclic
permutation. We expect these terms to be cancelled by $\bar{\gen{S}}_{2\to k}^{(1)}$
terms arising from collinear anomalies in  loops and, as we will see,  indeed this is the case.

\subsection{Deformations}

 \begin{figure}
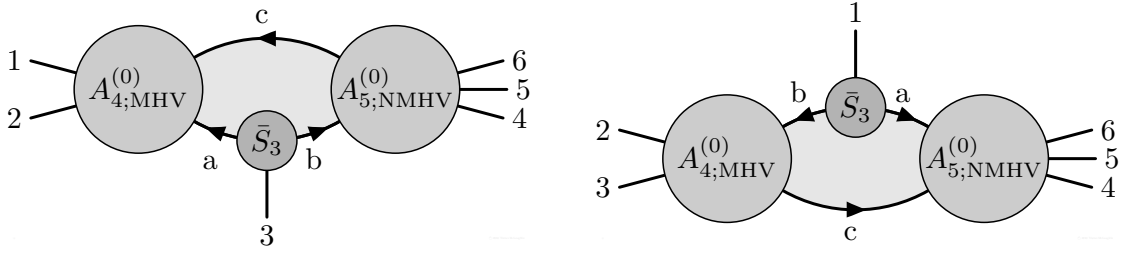
\centering
\includegraphics[width=0.4\textwidth]{FigCutAnomNMHV1.mps}\qquad
\includegraphics[width=0.4\textwidth]{FigCutAnomNMHV2.mps}
\caption{Anomaly for six-point NMHV with coefficient $R_{146}$.}
\label{fig:NMHVtriangle1}
\end{figure}
\begin{figure}\centering
\includegraphics[width=0.4\textwidth]{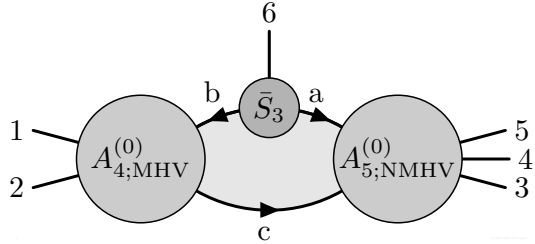}
\caption{Anomaly for six-point NMHV with coefficient $R_{635}$.}
\label{fig:NMHVtriangle2}
\end{figure}

The deformation due to the measure factor, $\bar{\gen{S}}_{2\to2}^{(1)}$,
is straightforwardly seen to be the same for NMHV amplitudes as for
the MHV ones. It cancels the divergent terms (and the finite parts grouped
with them). Similarly, the contributions for the splitting function, at tree and one-loop level,
are the same due to the universality of the splitting function and it is well known that
the amplitudes have the correct collinear behaviour.
Thus we move to consider the
contributions from triangle diagrams where the anomaly sits inside
the loop, the analogues of \figref{fig:SbarTriangle}. These arise
from the deformation
\[
\sum_{k=2}^3\bar{\gen{S}}_{2 \to n-k}^{(1)}A_{k;{\rm MHV}}^{(0)}
=\sum_{k=2}^3\sum_{j=1}^6T_{j}^k\log \frac{t_j^k}{t_j^{k+1}}~,
\]
where, $T_j^k$,  are on-shell triangle integrals. Let us first calculate
the coefficient of the three-particle channel cuts in the variable $t_1^3$
 which are proportional to  $R_{146}$ i.e.\ the terms needed to cancel
 \eqref{eq:x14vartot}. That is we wish to evaluate
 \[
 (T_1^2-T^3_1) \log t^3_1
 \]
where the triangle integrals are represented in  \figref{fig:NMHVtriangle1}
and, for example,
\[
T_1^2=\frac{1}{(2\pi)^3}\int (d^{4|4}\Lambda)^3 \sign(t^2_1-t_1^3){\bar S}_3(\mathrm{\bar b}, \mathrm{\bar a}, 3)
A^{(0)}\indup{4;MHV}(\mathrm{a},\mathrm{c},1,2)A^{(0)}\indup{5;NMHV}(\mathrm{\bar c},\mathrm{b},4,5,6).
\]
As for the MHV case the delta-functions enable one to trivially
evaluate these expressions and indeed the calculations are quite
similar. It can then be shown that
\<
\label{eq:NMHVtrianglesfin}
T_1^2\eq A\indup{6;MHV}^{(0)}R_{146}\frac{\langle 1|P_1^3|4]{\bar \lambda}_3}{\sprods{1}{3}[23][34][46]}
\left(\eta_2[46]+\eta_4[26]+\eta_6[24]\right),\nn\\
-T_4^3\eq A\indup{6;MHV}^{(0)}R_{146}\frac{\langle 3|P_3^3|6]{\bar \lambda}_1}{\sprods{1}{3}[12][46][61]}
\left(\eta_2[46]+\eta_4[26]+\eta_6[24]\right),
\>
which  are identical to \eqref{eq:x14var} and so cancel the variation \eqref{eq:x14vartot}.

Turning to the two-particle channels in the variable $t_1^2$ we see that they can only arise
with coefficients $R_{146}$, from the first triangle diagram in \figref{fig:NMHVtriangle1},
and $R_{635}$, from the triangle diagram \figref{fig:NMHVtriangle2}. From
\eqref{eq:NMHVtrianglesfin}, and its appropriate permutation,  it
 is clear that these terms will indeed cancel the
relevant terms in the variation \eqref{eq:x13var}. The remaining
two-particle channels similarly follow by use of cyclicity.
Thus we see that the deformations
constructed previously also annihilate the six-point NMHV amplitude.

\section{Yangian Symmetry at One Loop}
\label{sec:Yangian}
In addition to the standard super-conformal symmetries it has recently
become clear that planar scattering amplitudes in $\mathcal{N}=4$
transform covariantly under a ``dual'' superconformal algebra
\cite{Drummond:2008vq}. First hints towards this symmetry appeared at the level
of loop integrals contributing to the amplitudes
\cite{Drummond:2006rz,Bern:2006ew,Bern:2007ct}.
This surprising additional symmetry of the planar theory has its origin in the
ordinary conformal symmetry of the dual Wilson loop description of MHV amplitudes \cite{Drummond:2007cf,Drummond:2007au}.
Tree amplitudes have been proven to be covariant with respect to dual superconformal
transformations \cite{Brandhuber:2008pf,Drummond:2008cr}.
At loop level the dual conformal boost
$\genD{K}^{\alpha\dot\alpha}$ is naively broken and picks up an
anomaly term whose form, however, was
conjectured to be under control to all loop orders in \cite{Drummond:2007au,Drummond:2008vq}.
Recently it was shown that
all one-loop $\mathcal{N}=4$ SYM scattering amplitudes indeed obey the dual conformal
anomaly relation \cite{Brandhuber:2009kh}, following earlier results on MHV and NMHV
amplitudes
\cite{Drummond:2008vq,Drummond:2008bq,Brandhuber:2008pf,Brandhuber:2009xz,Elvang:2009ya}.

\begin{figure}
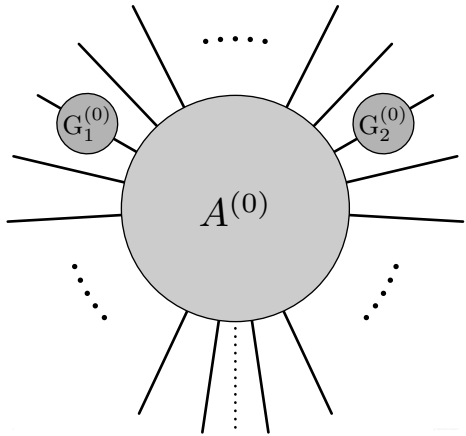
\centering
\includegraphicsbox{FigYang0.mps}
\caption{Action of the Yangian generator $\gengenY$
on the colour-ordered amplitude $A$ at tree level.
The action is defined as the bi-local insertion
of two superconformal generators $\gengen_1$ and $\gengen_2$,
the former acting to the left of the latter.
To that end one has to define an origin
for the colour-ordered amplitude (dotted line).}
\label{fig:treeyang}
\end{figure}

The dual superconformal symmetry can be made
manifest by introducing new dual coordinates on which the dual symmetry generators act
locally
\[
\label{eq:px}
\lambda_i^{\beta}\tilde\lambda_{i}^{ \dot \alpha}=x_i^{\beta \dot\alpha}-x_{i+1}^{\beta \dot\alpha}~,\,~~
\, ~~\lambda^{\beta}_i\eta_i^A=\theta^{\beta A}_i-\theta^{\beta A}_{i+1}
\]
and one makes the identifications $x_{n+1}:=x_1$, $\theta_{n+1}:=\theta_1$,
\cite{Drummond:2008vq}. At tree level it was shown in  \cite{Drummond:2009fd}
that this additional symmetry can be understood to lift the conventional superconformal
symmetry algebra $\alg{psu}(2,2|4)$ to a Yangian symmetry $\grp{Y}[\alg{psu}(2,2|4)]$.
The level-one Yangian generators $\gengenY$
are given by the standard coproduct rule for evaluation representations
with homogeneous evaluation parameters, cf.\ \figref{fig:treeyang},
\[
\label{eq:Yangian_def}
\gengenY_M=
f_M^{KL}\sum_{1\leq k<\ell\leq n}
\gengen_{k,K}\gengen_{\ell,L}~.
\]
The level-one generators $\gengenY$ thereby satisfy
\[
\label{eq:Yangian_com}
\gcomm{\gengen_K}{\gengenY_L} = f^{M}_{KL}\, \gengenY_{M}
~.
\]
Making use of the invariance of tree-amplitudes with respect to the locally acting
central charge, $\gen{C}_{i}$, it can be shown that this representation is compatible with
the cyclicity of the amplitudes \cite{Drummond:2009fd}. Furthermore by means of the Serre relations
and the covariance of
the tree-level amplitudes under the ``dual'' conformal generators
one can show that the
amplitudes are indeed invariant under the full Yangian algebra \cite{Drummond:2009fd}.

In the following we determine the one-loop deformation of the Yangian generators
$\genY{P}^{\alpha\dot\alpha}$ and  $\genY{Q}^{\alpha A}$.

\subsection{Dual Conformal Boost alias Level-One Momentum}

Indeed by virtue of \eqref{eq:Yangian_com} it suffices to construct the one-loop
deformation of $\genY{P}_{\alpha\dot\alpha}$ as all the other level-one Yangian
generators follow by commutation with level-zero ones.

To begin with, let us then quantify the relation between dual conformal boost
\[
\label{eq:KKtilde}
\genD{K}^{\alpha\dot \alpha} = \sum_{i=1}^{n} \left [ x_{i}^{\alpha\dot\beta}
x_{i}^{\dot\alpha\beta}\, \frac{\partial}{\partial x_{i}^{\beta\dot\beta}}
+ x_{i}^{\dot\alpha\beta}
\theta_{i}^{\alpha B}\, \frac{\partial}{\partial \theta_{i}^{\beta\dot\beta}}
+ x_{i}^{\alpha\dot\alpha }
\right]
\]
and the level-one Yangian
%
\[\label{eq:PhatTree}
(\genY{P}^{(0)})^{\alpha\dot\alpha}=
\sum_{1\leq j<i\leq n}  \left\lbrack
\left( \gen{L}_{j}^{\alpha}{}_{ \gamma} \,\delta^{\dot{\alpha}}_{\dot{\gamma}}
+ \bar{\gen{L}}_{j}^{\dot{\alpha}}{}_{ \dot{\gamma}} \, \delta^{\alpha}_{\gamma}
+ \gen{D}^{(0)}_{j} \delta^{\alpha}_{\gamma} \delta^{\dot{\alpha}}_{\dot{\gamma}} \right)
\gen{P}_j^{ \gamma \dot{\gamma}} + \gen{Q}_i^{\alpha C}   \bar{\gen{Q}}^{\dot{\alpha}}_{j C }
- ( i \leftrightarrow j ) \right\rbrack
\]
at tree-level \cite{Drummond:2009fd}.
Note that both $\genD{K}^{\alpha\dot \alpha}$ and
$\genY{P}^{\alpha\dot\alpha}$ annihilate $A_{n}^{(0)}$. Following \cite{Drummond:2009fd}
one may solve the new dual coordinates in terms of the original ones via \eqref{eq:px}
\[
x_{i}^{\alpha\dot\alpha}=x_{1}^{\alpha\dot\alpha} -\sum_{1\leq j<i} \lambda_{j}^{\alpha}\tilde{\lambda}_{j}^{\dot\alpha}\, , \qquad
\theta_{i}^{\alpha A}=\theta_{1}^{\alpha A} -\sum_{1\leq j<i} \lambda_{j}^{\alpha}\eta_{j}^{A}\, ,
\]
and rewrite the dual conformal boost \eqref{eq:KKtilde} as a differential operator
acting in the original on-shell superspace coordinates
$\{\lambda_{i},\tilde\lambda_{i},\eta_{i}\}$ to find
\<
\genD{K}^{\alpha\dot \alpha} \eq \half \genY{P}^{\alpha\dot\alpha}
+\sum_{i}\left ( \,\gen{P}_i^{\alpha\dot\alpha}\, \gen{C}_{i}\, \right ) -\half \gen{P}^{\alpha\dot\alpha}
\nl
- \half \gen{P}^{\gamma\dot\gamma}\, \left (\gen{L}^{\alpha}{}_{\gamma}\, \delta^{\dot{\alpha}}_{\dot{\gamma}}
+ \bar{\gen{L}}^{\dot{\alpha}}{}_{\dot{\gamma}} \, \delta^{\alpha}_{\gamma}
+ \gen{D}\, \delta^{\alpha}_{\gamma} \delta^{\dot{\alpha}}_{\dot{\gamma}} \right)  -
\half \gen{Q}^{ \alpha B}\bar{\gen{Q}}^{ \dot{\alpha} }_{B}
\nl
+ x_{1}^{\gamma\dot\gamma}\,  \left (\gen{L}^{\alpha}{}_{\gamma}\, \delta^{\dot{\alpha}}_{\dot{\gamma}}
+ \bar{\gen{L}}^{\dot{\alpha}}{}_{\dot{\gamma}} \, \delta^{\alpha}_{\gamma}
+ \gen{D}\, \delta^{\alpha}_{\gamma} \delta^{\dot{\alpha}}_{\dot{\gamma}} \right)
\nl
+\theta^{1B}_{\alpha}\, \bar{\gen{Q}}^{\dot\alpha}_{ B} \, ,
\label{eq:66}
\>
where the only non-local structure on the right-hand side resides in the level-one
Yangian momentum generator. We now make use of the full one-loop anomaly relation
\cite{Brandhuber:2009kh} in our conventions
\<
\label{eqn:Kanomaly}
\genD{K}^{\alpha\dot \alpha} A_{n}^{(1)} \eq 2 \, A_{n}^{(0)}\,
\sum_{i=1}^n x_{i+1}^{\alpha\dot\alpha}\, \frac{c_{\epsilon}}{\epsilon}
\left(\frac{x_{i,i+2}^{2}}{-\mu^{2}}\right)^{-\epsilon} \nln
\eq-2 \, A_{n}^{(0)}\, \biggl [
\sum_{1\leq j<i\leq n} p_j^{\alpha\dot\alpha}\, \frac{c_{\epsilon}}{\epsilon}
\left(\frac{t_{i}^{2}}{-\mu^{2}}\right)^{-\epsilon}
+ \sum_{i=1}^n p_i^{\alpha\dot\alpha}\, \frac{c_{\epsilon}}{\epsilon}
\left(\frac{t_{i}^{2}}{-\mu^{2}}\right)^{-\epsilon}
\nl\qquad\qquad
-x_1^{\alpha\dot\alpha} \sum_{i=1}^n \frac{c_{\epsilon}}{\epsilon}
\left (\frac{t_{i}^{2}}{-\mu^{2}}\right)^{-\epsilon} \biggr ]
\>
where $x_{i,j}:=x_{i}-x_{j}$. Now acting with \eqref{eq:66} on the  one-loop amplitude,
taking into account that the tree-level generators $\{\gen{L}^{\alpha}{}_{\beta},
\bar{\gen{L}}^{\dot\alpha}{}_{\dot\beta},\gen{Q}^{\alpha B},\bar{\gen{Q}}^{\dot\alpha}_{B}\}$
all annihilate $A_{n}^{(1)}$ as well as the \emph{local} generator
$\gen{C}_{i}\,\genD{K}^{\alpha\dot \alpha}=0$ and that
\eqref{eq:planardiagonalDSSb}
\[
\gen{D}^{(0)}\, A_{n}^{(1)} = 2A_{n}^{(0)}\, \sum_{j=1}^n\frac{c_{\epsilon}}{\epsilon}
\left(\frac{t_{j}^{2}}{-\mu^{2}}\right)^{-\epsilon}
\]
one easily computes the action of the tree-level Yangian generator on the one-loop
amplitude
\[
(\genY{P}^{(0)})^{\alpha\dot\alpha} A_{n}^{(1)} = -2
\left [ \sum_{1\leq j<i\leq n} \left ( p_{j}^{\alpha\dot\alpha}  \frac{c_{\epsilon}}{\epsilon}
\left(\frac{t_{i}^{2}}{-\mu^{2}}\right)^{-\epsilon} - (i\leftrightarrow j) \,\right )+
\sum_{i=1}^n p_{i}^{\alpha\dot\alpha}\frac{c_{\epsilon}}{\epsilon}
\left(\frac{t_{i}^{2}}{-\mu^{2}}\right)^{-\epsilon} \right ] A_{n}^{(0)}
\, ,
\]
remarkably recovering a bi-local structure again. We hence conclude that the one-loop
deformation of $\genY{P}^{\alpha\dot\alpha}$ takes the simple form
\[
(\genY{P}^{(1)})^{\alpha\dot\alpha} = -
\left [ \sum_{1\leq j<i\leq n} \left ( \gen{P}_{j}^{\alpha\dot\alpha}\,
(\gen{D}^{(1)}_{2\to2})^{2}_{i} -  (i\leftrightarrow j) \,\right )+
\sum_{i=1}^n\gen{P}_{i}^{\alpha\dot\alpha}\,
(\gen{D}^{(1)}_{2\to2})^{2}_{i}\, \right ]\, .
\]
where we have inserted the the one-loop deformation of the dilatation operator
of \eqref{eq:planardiagonalDSSb}.
 Rewriting $\genY{P}_{\alpha\dot\alpha}^{(1)}$ in a fashion of
keeping the bi-local terms free of contact terms we have
\[\label{eq:P1contact}
(\genY{P}^{(1)})^{\alpha\dot\alpha} = -
\left [ \sum_{1\leq j<i \leq n} \left ( \gen{P}_{j}^{\alpha\dot\alpha}\,
(\gen{D}^{(1)}_{2\to2})^{2}_{i} -  \gen{P}_{i}^{\alpha\dot\alpha}\,
(\gen{D}^{(1)}_{2\to2})^{2}_{j-1}\,\right )+
\sum_{i=1}^n(\gen{P}_{i}^{\alpha\dot\alpha}-\gen{P}_{i+1}^{\alpha\dot\alpha})\,
(\gen{D}^{(1)}_{2\to2})^{2}_{i}\, \right ]\, .
\]
Now the following curious picture emerges: Recalling the split
of the one-loop amplitude into a finite and a divergent
piece as done in \eqref{eq:IRdivergentsplit} but here restricted to
the planar limit \eqref{eq:Zfactorplanar}
\[
A_{n}^{(1)}=\hat{Z}_{2\to 2}^{(1)}A_{n}^{(0)}+\tilde A_{n}^{(1)}\, \, ,
\quad \text{with} \quad
\hat{Z}_{2 \to 2}^{(1)}=-\sum_{j=1}^{n}\frac{c_{\epsilon}}{\epsilon^{2}}
\left(\frac{t_{j}^{2}}{-\mu^{2}}\right)^{-\epsilon} \, ,
\]
we can identify the bi-local deformation of $\genY{P}_{\alpha\dot\alpha}^{(1)}$
in \eqref{eq:P1contact} as arising from the
action of the tree-level generator on $\hat{Z}_{2 \to 2}^{(1)}$
%
\<
\label{eins}
\bigcomm{(\genY{P}^{(0)})^{\alpha\dot\alpha}}{\hat{Z}_{2\to 2}^{(1)}}
\eq \sum_{i=2}^{n-1}
\left [\sum_{1\leq j<i }\gen{P}_j\, (\gen{D}^{(1)}_{2\to 2})^2_i- \sum_{i< j\leq n }\gen{P}_j\, (\gen{D}^{(1)}_{2\to 2})^2_{i-1}\right ] -
\left ( \gen{P}_{n}^{\alpha\dot\alpha} -  \gen{P}_{1}^{\alpha\dot\alpha} \right ) (\gen{D}^{(1)}_{2\to2})^{2}_{n}\nn\\
\eq\sum_{i=1}^{n}
\left [\sum_{1\leq j<i }\gen{P}_j\, (\gen{D}^{(1)}_{2\to 2})^2_i- \sum_{i< j\leq n }\gen{P}_j\, (\gen{D}^{(1)}_{2\to 2})^2_{i-1}\right ] ~.
\>
Conversely, the last local term in \eqref{eq:P1contact}
is nothing but the anomaly of $\tilde{A}^{(1)}_n$
once spelled out in the dual coordinates
\[
(\genY{P}^{(0)})^{\alpha\dot\alpha}\, \tilde{A}^{(1)}_n
= 2\, \sum_{i=1}^nx_{i,i+1}^{\alpha\dot\alpha}\, \log \left(\frac{x^{2}_{i,i+2}}{x^{2}_{i-1,i+1}} \right)A^{(0)}_n\,
=-(\genY{P}^{(1)}_{2\to2})^{\alpha\dot\alpha}A^{(0)}_n
,
\]
with
\[
(\genY{P}^{(1)}_{2\to2})^{\alpha\dot\alpha} =
-\sum_{i=1}^n(\gen{P}_{i}^{\alpha\dot\alpha}-\gen{P}_{i+1}^{\alpha\dot\alpha})\,
(\gen{D}^{(1)}_{2\to2})^{2}_{i}.
\]
In conclusion we can cleanly separate the deformation from the divergent measure
and a genuine one-loop anomaly
%
\[\label{eq:Phat1}
(\genY{P}^{(1)})^{\alpha\dot\alpha} =
-\bigcomm{(\genY{P}^{(0)})^{\alpha\dot\alpha}}{\hat{Z}_{2\to 2}^{(1)}}
+(\genY{P}^{(1)}_{2\to2})^{\alpha\dot\alpha}\, .
\]
While perhaps a mere curiosity it is worth noting that one can find the anomaly from the action
of a bi-local operator on ${\hat{Z}_{2\to 2}^{(1)}}$,
%
\[
(\genY{P}^{(1)}_{2\to2})^{\alpha\dot\alpha}=
 \sum_{1\leq j<i\leq n}  \left [2
\left( \gen{L}_{j}^{\alpha}{}_{\gamma}\,\delta^{\dot{\alpha}}_{\dot{\gamma}}
+ \bar{\gen{L}}_{j }^{\dot{\alpha}}{}_{\dot{\gamma}} \, \delta^{\alpha}_{\gamma}
\right)
\gen{P}_i^{ \gamma \dot{\gamma}} -(i\leftrightarrow j), {\hat{Z}_{2\to 2}^{(1)}}\right ].
\]
Effectively the first term in \eqref{eq:Phat1} is sufficient
to reproduce the complete $\genY{P}^{(1)}$
if the sign of $\gen{L}$ and $\bar{\gen{L}}$ in $\genY{P}^{(0)}$, cf.\ \eqref{eq:PhatTree},
is flipped.

\subsection{The All-Loop Form of \texorpdfstring{$\gen{D}$}{D} and
\texorpdfstring{$\genY{P}$}{Phat}}
\label{sec:allloopDPhat}

This insight together with the all-loop conjecture for the form of the dual conformal anomaly
of Drummond, Henn, Korchemsky and Sokatchev \cite{Drummond:2008vq}
then leads to a transparent conjecture for the all-loop form of the dilatation and Yangian level-one
momentum generators. Due to the exponentiation of IR singularities the all-loop expression
for the planar $n$-point amplitudes takes the form
\[
A_{n}(g^2)= \exp \bigbrk{\hat{Z}_{2 \to 2}(g^2,\epsilon)}\, \tilde A_{n}(g^2)\, ,
\]
with $\tilde A_n$ being the all-loop finite piece,
and ${\hat{Z}}_{2\to 2}$ is the logarithm of the all-loop IR divergences.
The form of this function is well studied with the leading term for planar amplitudes being given by%
\footnote{In the following we shall use a sloppy notation
where we disregard the effect of the loop order
on the $\epsilon$-dependence.
This dependence is very systematic and can be implemented
easily by the rule $\epsilon\to\ell\epsilon$ at $\ell$ loops. For a review of
the all-loop structure of ${\cal N}=4$ SYM amplitudes including their
IR divergences see e.g.\ \cite{Alday:2008yw} and the
many papers referenced therein.}
\[
\hat{Z}_{2\to 2}(g^2,\epsilon)=
\quarter\Gamma(g^{2},\epsilon)\, {\hat{Z}}^{(1)}_{2\to 2}(\epsilon)+\order{\epsilon^0}~,
\]
where $\Gamma(g^2,\epsilon)$ contains the cusp dimension
$\Gamma\indup{cusp}(g^{2})=g^{2}+\order{g^{4}}$
as well as the collinear dimension
$\Gamma\indup{coll}(g^{2})=\order{g^{4}}$
\[
\Gamma(g^2,\epsilon)=\Gamma\indup{cusp}(g^{2})+\epsilon \Gamma\indup{coll}(g^2)+\order{\epsilon^2}.
\]
The subleading in $\epsilon$ terms are scheme dependent.
Acting on this with tree-level $\gen{D}$ yields
\[
\gen{D}^{(0)}\, A_{n}= \quarter\Gamma(g^{2},\epsilon) \bigcomm{\gen{D}^{(0)}}{\hat{Z}^{(1)}_{2 \to 2}} A_{n}=
-\quarter\Gamma(g^{2},\epsilon)\, \gen{D}^{(1)}\, A_{n}
\]
hence the all-loop \emph{planar} dilatation operator is simply
\[
\label{aldilop}
\gen{D}(g^2)=\gen{D}^{(0)} +\quarter\Gamma(g^{2},\epsilon)\,\gen{D}^{(1)} ,
\]
such that $\gen{D}\, A_{n}=0$.

Similarly, the all-loop form of the level-one Yangian generator $\genY{P}^{\alpha\dot\alpha}$
can be established. For this we observe that the action of the tree-level
$(\genY{P}^{(0)})^{\alpha\dot\alpha}$ results in
\[
\label{drei}
(\genY{P}^{(0)})^{\alpha\dot\alpha}\, A_{n} =
 \quarter\Gamma(g^{2},\epsilon)\,
\bigcomm{(\genY{P}^{(0)})^{\alpha\dot\alpha}}{\hat{Z}^{(1)}_{2\to 2}}
A_{n}
+
\exp \bigbrk{\hat{Z}_{2 \to 2}}
(\genY{P}^{(0)})^{\alpha\dot\alpha}\, \tilde A_{n}\,
.
\]
Now by virtue of the conjectured form \cite{Drummond:2008vq} of the
all-loop dual-conformal anomaly we note
\<
(\genY{P}^{(0)})^{\alpha\dot\alpha}\, \tilde A_n \eq 2\, \genD{K}^{\alpha\dot \alpha}\, \tilde A_n=
\half\Gamma(g^{2},\epsilon)\,
\sum_{i=1}^n x_{i,i+1}^{\alpha\dot\alpha}\, \log \left(\frac{x^{2}_{i,i+2}}{x^{2}_{i-1,i+1}} \right)
\tilde A_n
\nln\eq
-\quarter\Gamma(g^{2},\epsilon)\, (\genY{P}^{(1)}_{2\to2})^{\alpha\dot\alpha}\, \tilde A_n\, ,
\>
where we have made use of the fact that the discrepancy terms between
$(\genY{P}^{(0)})^{\alpha\dot\alpha}$ and $2\genD{K}^{\alpha\dot \alpha}$ in
\eqref{eq:66} are inactive when acting on $\tilde A$. Hence we know all terms on the
right-hand-side of \eqref{drei}. This enables us to also conjecture the all-loop form
of the level-one Yangian generator $\genY{P}^{\alpha\dot\alpha}$
\[
\genY{P}(g^2)^{\alpha\dot\alpha}= (\genY{P}^{(0)}
)^{\alpha\dot\alpha}+\quarter \Gamma(g^{2},\epsilon)\,
(\genY{P}^{(1)}) ^{\alpha\dot\alpha},
\]
whose structure is dictated by the one-loop deformation, as is also the case for the
dilatation operator \eqref{aldilop}.

\subsection{Dual Superconformal Boosts and Bi-Local Generators}

Invariance of the amplitude under the dual superconformal boosts
$\genD{S}$ alias the level-one supersymmetry generators $\genY{Q}$
follows straight-forwardly from invariance under
$\genY{P}$ and $\bar{\gen{S}}$ using \eqref{eq:Yangian_com}
\[\label{eq:Qhatcomm}
\bigcomm {\bar{\gen{S}}^A_{\dot\beta}}{\genY{P}^{\gamma\dot\delta}}
=-\delta^{\dot\delta}_{\dot\beta}\genY{Q}^{\gamma A}.
\]
Nevertheless, it is instructive to see how the one-loop
deformation of $\genY{Q}$ acts qualitatively.
In particular, we have seen in \secref{sec:allloopDPhat} that $\genY{P}$ receives
only relatively simple corrections, but here length-changing interactions
as discussed in \secref{sec:coll_in_loops} enter. This follows directly
from the above commutator which introduces them via $\bar{\gen{S}}$.

\begin{figure}
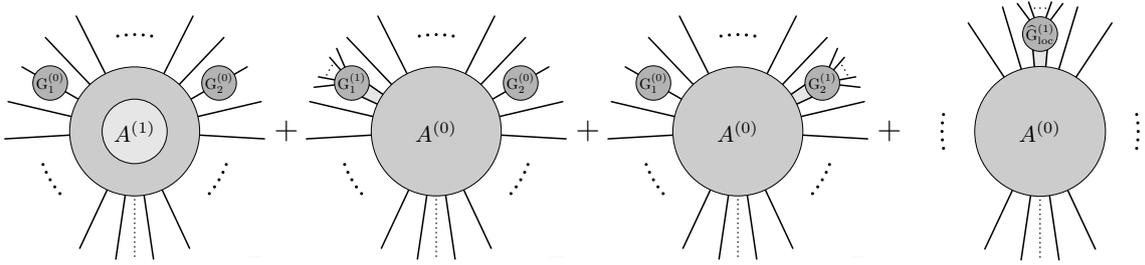
\centering
$\includegraphicsbox[width=0.2\textwidth]{FigYang1a.mps}
+\includegraphicsbox[width=0.2\textwidth]{FigYang1b.mps}
+\includegraphicsbox[width=0.2\textwidth]{FigYang1c.mps}
+\includegraphicsbox[width=0.2\textwidth]{FigYang1d.mps}$
\caption{Action of the Yangian generator $\gengenY$
on the colour-ordered planar amplitude $A$ at one loop.
The standard contribution of $\gengenY^{(1)}$ is the same as at tree level
where one constituent superconformal generators $\gengen_1,\gengen_2$
is replaced by its one-loop deformation.
Furthermore $\gengenY^{(1)}\indup{loc}$ represents the action when the
constituent generators overlap.
Here, one may have to specify separately how $\gengenY^{(1)}\indup{loc}$
acts on the two particles separated by the origin.}
\label{fig:loopyang}
\end{figure}

Let us compare the structure of the one-loop deformation
of $\genY{P}$ in \eqref{eq:P1contact}
with the tree-level generator in \eqref{eq:PhatTree}:
For the bi-local contribution in \eqref{eq:P1contact}
one promotes all instances of $\gen{D}$ in \eqref{eq:PhatTree}
to its one-loop correction and drops the other terms
which involve only super-Poincar\'e generators
$\gen{L},\bar{\gen{L}},\gen{P},\gen{Q},\bar{\gen{Q}}$.
In other words, the bi-local contribution is obtained
from a perturbative generalisation of \eqref{eq:Yangian_def}
\[\label{eq:yangbiall}
\gengenY_M=
f_M^{KL}
\gengen_{K}\otimes\gengen_{L}+
\gengenY_{\mathrm{loc},M}
:=
f_M^{KL}\sum_{k\ll \ell}
\gengen_{k,K}\,\gengen_{\ell,L}+
\sum_k\gengenY_{\mathrm{loc},k,M}
~,
\]
where $\gengen_K$ are the perturbative generalisations of the
superconformal generators.
The local contributions $\gengenY\indup{loc}$ are needed to
specify the action of the bi-local generators when the
constituent operators overlap.
When expanded to one loop one obtains, cf.\ \figref{fig:loopyang},
\[\label{eq:yangbione}
\gengenY^{(1)}_M=
f_M^{KL}\gengen^{(1)}_{K}\otimes\gengen^{(0)}_{L}
+f_M^{KL}\gengen^{(0)}_{K}\otimes\gengen^{(1)}_{L}
+\gengenY^{(1)}_{\mathrm{loc},M}.
\]
In the case of $\genY{P}$ the only bi-local deformation originates from the
dilatation generator $\gen{D}$, and it is precisely of this form.
Moreover, the anomaly $\genY{P}^{(1)}_{2\to2}$ equals the local term
$\genY{P}\indup{loc}^{(1)}$.

The structure \eqref{eq:yangbione,eq:yangbiall} is
well-known for perturbative Yangians,
cf.\ \cite{Serban:2004jf,Zwiebel:2006cb,Beisert:2007jv,Bargheer:2008jt,Bargheer:2009xy}.
Importantly, the bi-local term
is stable under the adjoint action \eqref{eq:Yangian_com}
provided that the superconformal algebra is satisfied.
The local terms serve as a regularisation and are more sensitive to the
details of the algebra and its deformation.

For the correction to the dual superconformal boost $\genY{Q}$
these considerations imply the structure
\[
(\genY{Q}^{(1)})^{A\beta}=
\gen{P}^{\beta\dot\gamma}\wedge(\bar{\gen{S}}^{(1)})^A_{\dot\gamma}
+\half\gen{Q}^{A\beta}\wedge\gen{D}^{(1)}
+\genY{Q}^{A\beta}\indup{loc},
\]
where $A\wedge B:=A\otimes B-B\otimes A$.
This expression has length-preserving contributions
from $\bar{\gen{S}}^{(1)}_{2\to2}$,
$\gen{D}^{(1)}_{2\to2}$ and $\genY{Q}^{A\beta}\indup{loc,2\to2}$,
but there are also length-changing contributions
from $\bar{\gen{S}}^{(1)}_{2\to k}$ and $\genY{Q}^{A\beta}_{\mathrm{loc},2\to k}$,
cf.\ \secref{sec:sym1loop}.
The bi-local contributions follow directly from the
correction to the superconformal generators,
whereas the local terms follow from the commutator \eqref{eq:Qhatcomm}.
We refrain from presenting the results of such a computation
as the result is guaranteed to annihilate all amplitudes anyway.

Similar considerations apply for $\genY{\bar{Q}}$ which has
a structure analogous to $\genY{Q}$ but does not derive from one of the dual
superconformal generators.
The remaining level-one generators follow the same pattern,
but they all involve $\gen{K}$ which has a yet more
complicated structure than $\gen{S}$ and $\bar{\gen{S}}$.

We have not yet addressed the issues regarding the algebra of the
deformed generators and while we postpone a detailed consideration
to future work a few comments are in order. Due to the choice
of regulator all the deformations respect the manifest super-Poincar\'e algebra.
For the special conformal and fermionic conformal generators the
situation is much less clear. Of course, the algebra is
trivially satisfied on the space of amplitudes  however on
larger spaces it is remains non-trivial to demonstrate closure.
Already at tree-level the algebra is seen to close
only up to field dependent gauge transformations
\cite{Bargheer:2009qu} and
demonstrating closure at one-loop remains an open problem.
Furthermore, to establish the existence of
a Yangian algebra one must additionally show that the
deformed generators satisfy the Serre relations. It is not clear that
Yangian algebra will be satisfied,
and it is very possible that this is a subtle issue.
Note that some evidence in favour of this point of view
was already found in a study of the perturbative Yangian \cite{Zwiebel:2006cb}.
As a point of comparison, at strong coupling
the scattering amplitudes are described by open strings
in the AdS geometry which, being essentially a
coset sigma model, are known to possess an infinite family of
non-local charges \cite{Bena:2003wd}.
While the full quantum algebra of these charges is
not currently known we point out that even at the classical level there
is likely an inherent ambiguity in their algebra. That this is so
was pointed out for the Yangian symmetries of the non-linear
sigma model by \cite{Luscher:1977rq} and discussed by e.g.\
\cite{deVega:1983gy,Maillet:1985fn, MacKay:1992rc}.
Essentially, due to the non-ultralocal terms in
the current algebra, one must define a regularisation for the
charges as integrals over densities, taking particular care
with the end points, and different regularisations can
lead to different terms on the r.h.s.\ of \eqref{eq:Yangian_com}.

\section{Propagator Prescriptions}
\label{sec:comparison_prop}

In this section we comment on the exact definition of the amplitudes
we have analysed. This will allow us to clarify the relationship to
a different proposal \cite{Sever:2009aa}
of how to deform the symmetry generators at loop level.
As it appears to be different already at tree level,
we shall start there and later continue at loops.

For simplicity let us consider a generator $\gengen$ which
has only deformations of the type $1\to2$ at tree level.
In practice this can be one of the two superconformal boost generators
$\gen{S}$ or $\bar{\gen{S}}$, but not the conformal boost $\gen{K}$.

\subsection{Tree Level}
\label{sec:CompareBBGLM.SV}

In \cite{Bargheer:2009qu} a deformation of the free
representation was proposed
in order to make all tree-level amplitudes exactly invariant.
In \cite{Sever:2009aa} it was subsequently shown that additional
terms are needed for exact invariance of the tree-level S-matrix.
Here we shall illustrate
the differences between the two proposals
and show that they are in fact compatible.

\begin{figure}
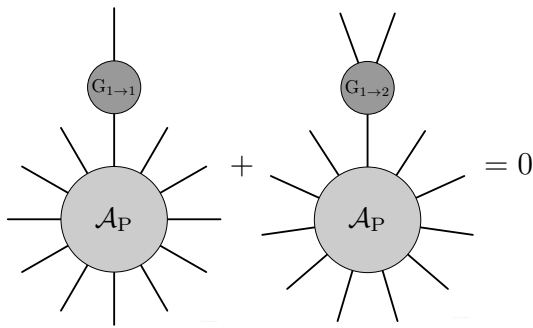
\centering
$\includegraphicsbox[scale=0.7]{FigInvBBGLM1.mps}
+\includegraphicsbox[scale=0.7]{FigInvBBGLM2.mps}
=0$
\caption{Invariance condition for tree amplitudes
with principal part prescription.}
\label{fig:InvBBGLM}
\end{figure}

In \cite{Bargheer:2009qu} it was shown that
the amplitude generating functional $\mathcal{A}\indup{P}$
is annihilated by the deformed
representation of the generator $\gengen=\gengen_{1\to1}+\gengen_{1\to2}$
(see \figref{fig:InvBBGLM})
\[\label{eq:InvBBGLM}
(\gengen_{1\to1}+\gengen_{1\to2})\mathcal{A}\indup{P}=0.\]

\begin{figure}
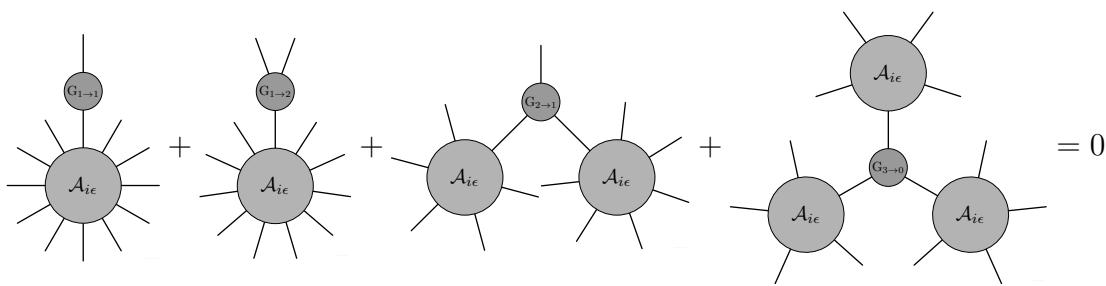
\centering
$\includegraphicsbox[scale=0.5]{FigInvSV1.mps}
+\includegraphicsbox[scale=0.5]{FigInvSV2.mps}
+\includegraphicsbox[scale=0.5]{FigInvSV3.mps}
+\includegraphicsbox[scale=0.5]{FigInvSV4.mps}
=0$
\caption{Invariance condition for tree amplitudes
with $i\epsilon$ prescription.}
\label{fig:InvSV}
\end{figure}

In \cite{Sever:2009aa} it was shown that
additional deformations are needed, namely
$\gengen=\gengen_{1\to1}+\gengen_{1\to2}+\gengen_{2\to1}+\gengen_{3\to0}$.
The deformations
$\gengen_{2\to1}$ and $\gengen_{3\to0}$ are almost the same as $\gengen_{1\to2}$,
but they have a different distribution of in and out legs.
Moreover, the connected amplitude $\mathcal{A}_{i\epsilon}$ itself
is not invariant, but only its exponential
\[\label{eq:InvSV}
(\gengen_{1\to1}+\gengen_{1\to2}+\gengen_{2\to1}+\gengen_{3\to0})\exp(i\mathcal{A}_{i\epsilon})
=0.\]
While $\mathcal{A}_{i\epsilon}$ is a connected amplitude,
the expansion of the exponential yields disconnected graphs.
The additional generators $\gengen_{2\to1}$ and $\gengen_{3\to0}$
can now connect two or three subgraphs into a single component.
The invariance equation is depicted in \figref{fig:InvSV}.

It appears that the two invariance equations differ by terms
which are non-zero in general and thus only one
of them could be true. However, one has to be careful about
the precise definition of the amplitudes considered in each case.
The first equation \eqref{eq:InvBBGLM} assumes
a principal value prescription for all internal propagators
of the tree amplitude $\mathcal{A}\indup{P}$.
Conversely, the second equation \eqref{eq:InvSV} requires
use of an $i\epsilon$ prescription for the internal propagators
of $\mathcal{A}_{i\epsilon}$.
Indeed, this minor change is what accounts for the
difference in the two equations.

\begin{figure}
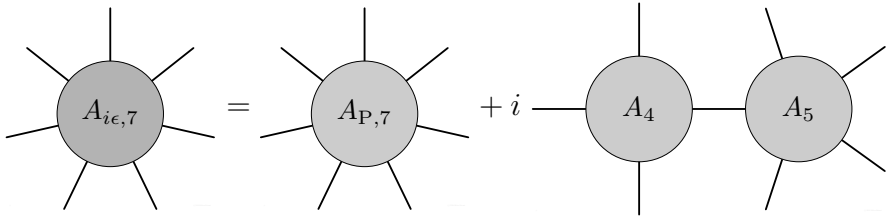
\centering
$\includegraphicsbox[scale=0.7]{FigUniTree1.mps}
=\includegraphicsbox[scale=0.7]{FigUniTree2.mps}
+i\ \includegraphicsbox[scale=0.7]{FigUniTree3.mps}$
\caption{Reduction of
tree amplitude with $i\epsilon$ prescription
to combinations of amplitudes with
principal value prescription.}
\label{fig:UniTree}
\end{figure}

Consider, e.g., an amplitude with $7$ legs:
It has poles corresponding to an internal propagator going on shell.
The residue is given by the product of two subamplitudes
with $4$ and $5$ legs, respectively. Now, the difference
between a principal value and an $i\epsilon$ prescription
is given by the same residue supported by a delta function
forcing the internal particle on shell
\[\label{eq:ieps}
\frac{1}{p^2\mp i\epsilon} = \frac{1}{p^2}\pm i\pi\delta(p^2).
\]
Consequently there is the relation
$A_{i\epsilon,7}=A\indup{P,7}+iA_4\mbox{---}A_5$
which is illustrated in \figref{fig:UniTree}.
The point is that the $A_{\mathrm{P},7}$ is exactly
annihilated by $\gengen_{1\to1}+\gengen_{1\to2}$,
but $A_4\mbox{---}A_5$ requires an extra
contribution from $\gengen_{2\to1}$
acting on $\ihalf (A_4)^2$,
i.e.\ the third term in \figref{fig:InvSV}.

How do the extra anomalies in $\mathcal{A}_{i\epsilon}$ arise in practice?
When the free representation $\gengen_{1\to1}$
acts on a rational function,
there can be delta-function contributions localised at the poles.
Collinear singularities yield such anomalies which are subsequently
cancelled by $\gengen_{1\to2}$.
Secondly, there are multi-particle singularities,
but these do not cause anomalies,
whether they are evaluated in principal value or in $i\epsilon$ prescriptions.
The third type of singularity can cause anomalies, but it is spurious and
cancels in the complete amplitude because the residues cancel.
Almost! In the $i\epsilon$ prescription there are some left-over terms
with delta function support. These are the anomalies to be cancelled
by $\gengen_{2\to1}$ and $\gengen_{3\to0}$ acting on disconnected amplitudes.

It turns out that the two invariance equations
are perfectly compatible, and they have the same physical
and mathematical implications.
Is one of the two preferable over the other?
One the one hand, the amplitude $\mathcal{A}_{i\epsilon}$
with $i\epsilon$ prescription
may appear to be a more natural and more physical object
than the amplitude $\mathcal{A}\indup{P}$
with principal value prescription.
On the other hand,
the symmetry relations for $\mathcal{A}\indup{P}$ are simpler.

\begin{figure}\centering
$\displaystyle
\includegraphicsbox[scale=0.8]{FigAmpE1.mps}
=
\includegraphicsbox[scale=0.8]{FigAmpP1.mps}
+\frac{i}{2}\ \includegraphicsbox[scale=0.8]{FigAmpP2.mps}
-\frac{1}{2}\ \includegraphicsbox[scale=0.8]{FigAmpP3.mps}
-\frac{i}{2}\ \includegraphicsbox[scale=0.8]{FigAmpP4a.mps}
-\frac{i}{6}\ \includegraphicsbox[scale=0.8]{FigAmpP4b.mps}
+\ldots$
\caption{Relation between amplitudes with $i\epsilon$ prescription and
amplitudes with principal part prescription at tree level.}
\label{fig:Aieps}
\end{figure}

In the remainder of this section we will explore the formal relationship
between the various quantities in order to gain a clearer
understanding of amplitudes and their invariance conditions.
Using the above relationship \eqref{eq:ieps} between
propagators it is clear that amplitudes
with $i\epsilon$ prescription can be expanded in terms
on-shell connections of amplitudes with principal part prescription,
see \figref{fig:Aieps}.
The prefactors are the natural symmetry factors associated to the graphs.
Note that we have terminated the expansion at four powers of $\mathcal{A}\indup{P}$
and at tree level.
This relation can be formalised as follows
\[
\mathcal{S}[J]=
\exp\bigbrk{i\mathcal{A}_{i\epsilon}[J]}
=\exp(\hat{C})
\exp\bigbrk{i\mathcal{A}\indup{P}[J]},
\qquad
\hat{C}=\frac{1}{4}\int d^{4|4}\Lambda\, \check J^a(\Lambda)\check J^a(\bar\Lambda).
\]
In words it says that the S-matrix is given by a collection of arbitrarily many
amplitudes $\mathcal{A}_{i\epsilon}$. Equivalently,
it is given by a collection of arbitrarily many amplitudes
$\mathcal{A}\indup{P}$ with arbitrarily many on-shell connections $\hat{C}$ between the legs.

We start with the statement of superconformal symmetry at tree level \cite{Bargheer:2009qu}%
\footnote{Note that $\mathcal{A}\indup{P}[J]$ uses principal part propagators.
As loop corrections require at least one on-shell propagator,
$\mathcal{A}\indup{P}[J]$ terminates at tree level.
Thus the following argument
(which is analogous to the one used in \cite{Sever:2009aa})
\emph{formally} applies to all loops
when neglecting effects of regularisation.}
\[
\brk{\gengen_{1\to1}+\gengen_{1\to2}}\mathcal{A}\indup{P}[J]=0.
\]
Now we would like to transform this equation to a statement for
$\mathcal{S}$. The generators $\gengen_{1\to1}$ and $\gengen_{1\to2}$
are linear in derivatives and therefore
$\exp(i\mathcal{A}\indup{P})=\exp(-\hat{C})\mathcal{S}$ is equally invariant.
We conclude that $\mathcal{S}$ is invariant under
the same generator conjugated by $\exp(\hat{C})$
\[
\exp(\hat{C})\bigbrk{\gengen_{1\to1}+\gengen_{1\to2}}\exp(-\hat{C})\,
\mathcal{S}[J]=0.
\]
This is precisely the claim of \cite{Sever:2009aa}.
Namely, one can easily confirm that
the free generator is invariant under conjugation
\[\label{eq:G2conj}
\gengen_{1\to1}=\exp(\hat{C})\gengen_{1\to1}\exp(-\hat{C}).
\]
This translates to the statement that on-shell contractions respect
free superconformal symmetry. Furthermore the deformations obey
\[\label{eq:G3conj}
\gengen_{1\to2}+\gengen_{2\to1}+\gengen_{3\to0}
=
\exp(\hat{C})
\gengen_{1\to2}
\exp(-\hat{C}).
\]
In other words $\gengen_{2\to 1}=\comm{\hat{C}}{\gengen_{1\to 2}}$ and
$\gengen_{3\to 0}=\half\comm{\hat{C}}{\gengen_{2\to 1}}$ which is essentially
how the additional deformations were derived in \cite{Sever:2009aa}.
Starting with $\gengen_{1\to2}$ in \eqref{eq:twovertex} we find
in agreement with \cite{Sever:2009aa}
\<\label{eq:G21G30}
\gengen_{1\to 2} \eq \frac{1}{2}\int (d^{4|4}\Lambda)^3 \, \sign(E_1E_2)\, \gengend^{abc}_{123}
J^a_1 J^b_2 \check J^c_{\bar3},
\nln
\gengen_{2\to1}\eq
\bigcomm{\hat{C}}{\gengen_{1\to2}}
=
\frac{1}{2}
\int (d^{4|4}\Lambda)^3 \, \sign(E_1E_2)\, \gengend^{abc}_{123}
J^a_1 \check J^b_{\bar2}\check J^c_{\bar3}
\nln\eq
-\frac{1}{2}
\int (d^{4|4}\Lambda)^3 \, \theta(E_2E_3)\, \gengend^{abc}_{123}
J^a_1 \check J^b_{\bar2}\check J^c_{\bar3},
\nln
\gengen_{3\to0}\eq
\half \comm{\hat{C}}{\gengen_{2\to1}}
=
-\frac{1}{8}
\int (d^{4|4}\Lambda)^3 \, \theta(E_2E_3)\, \gengend^{abc}_{123}
\check J^a_{\bar1} \check J^b_{\bar2}\check J^c_{\bar3}
\nln\eq
-\frac{1}{24}
\int (d^{4|4}\Lambda)^3 \, \gengend^{abc}_{123}
\check J^a_{\bar 1} \check J^b_{\bar 2}\check J^c_{\bar 3}.
\>
The transformations between the sign and step factors
makes use of permutation symmetries and momentum conservation.
Thus the two proposals for superconformal symmetry agree.

Let us conclude with some remarks.
One may wonder about the different structures
of $\gengen_{1\to2}$, $\gengen_{2\to1}$ and $\gengen_{3\to0}$
concerning the signs of the particle energies \cite{Sever:2009aa}.
For instance, all signatures compatible with energy-momentum conservation
are permitted in $\gengen_{1\to2}$ and $\gengen_{3\to0}$.
The difference is that $\gengen_{1\to2}$ makes explicit reference
to the sign of energies while $\gengen_{3\to0}$ does not.
Conversely, $\gengen_{2\to1}$ requires the two in-going particles
to have equal signs.
In a canonical quantisation framework this distinction between
$\gengen_{1\to2}$ and $\gengen_{2\to1}$ actually makes sense.
In such a picture, positive energy states are represented by
creation operators and negative energy states by annihilation operators.
The deformed symmetry generator would thus consist of two creation
and one annihilation operator or vice versa.
Now invariance of an operator means that it commutes with a symmetry generator.
Commuting the deformed generator with some operator can connect the two objects
by one (on-shell) Wick contraction (corresponding to $\gengen_{1\to2}$) or by two
(corresponding to $\gengen_{2\to1}$).
In the case of two contractions, the energies automatically align
in agreement with the structure of $\gengen_{2\to1}$.%
\footnote{One may wonder what about $\gengen_{3\to0}$
which apparently has no representation in a canonical quantisation framework.
Consequently the S-matrix operator does not seem to commute with
the deformed generator. Nevertheless there appears to exist
a slightly deformed version of the S-matrix operator
which does commute and which is unitary.}

We would also like to remark that the structures
$\gengen_{1\to2}$ and $\gengen_{2\to1}$ in \eqref{eq:G3conj}
are reminiscent of the brackets and cobrackets in a
classical double of a Lie bialgebra
(cf.\ \cite{Beisert:2007ty} for some
explicit expressions).
In this analogy,
positive and negative  energy states correspond to the two copies
of the original bialgebra. Brackets are defined between elements
of both copies, while the cobracket remains confined
to each subalgebra.
The deeper meaning of this observation remains obscure,
but it may help to obtain a better understanding
of the deformation.

\subsection{One Loop}

Having convinced ourselves of the equivalence of the
two proposals \cite{Bargheer:2009qu} and \cite{Sever:2009aa}
at tree level, we should now compare our one-loop results
to \cite{Sever:2009aa}.

The most obvious difference is that while \cite{Sever:2009aa}
calculate the action of generators on amplitudes
in the final analysis they are
only concerned with the invariance of  so-called
IR-finite observables such as inclusive cross-sections, see for example \cite{Bork:2009ps,Bork:2009nc}.
These are derived from cross sections made from scattering amplitudes.
Conversely, we formulate an invariance condition
for the scattering amplitudes themselves.
Therefore anomaly contributions due to the integration measure
could be discarded in \cite{Sever:2009aa}
while we have to take them into account.
In particular, \eqref{eq:G2conj} does not hold
strictly which eventually leads to the deformation
$\gengen^{(1)}_{2\to2}$ derived in \secref{subsec:Anomaly_measure}.

One could now wonder whether our remaining deformations
$\gengen^{(1)}_{2\to k}$, $k\geq 3$, match with
the contribution from $\gengen_{2\to 1}$ in the framework of \cite{Sever:2009aa}.
This expectation is reasonable because the contributions have a similar structure,
and in the case of MHV amplitudes they actually yield coincident contributions.
There is however one conceptual problem with applying
the generator $\gengen_{2\to 1}$ to some tree amplitude
$\mathcal{A}_{i\epsilon,n+1}$:
The generator forces two legs
of the amplitude to be strictly collinear.
On the one hand, the amplitude diverges near collinear configurations.
On the other hand, there are cancellations in the numerator
which compensate the divergence.
We give an explicit example in \appref{sec:5NMHV}.
In practice the amplitude $\mathcal{A}_{i\epsilon,n+1}$
is split up into two subamplitudes
$\mathcal{A}_{k+1}$ and $\mathcal{A}_{n-k+2}$ in \cite{Sever:2009aa}.
The split is performed using the CSW rules, \cite{Cachazo:2004kj},
which require one to go off shell or to violate momentum conservation
by a tiny amount. Then the calculation can be completed
in terms of the subamplitudes and one ends up with finite
contributions to $\gengen_{2\to 1}\mathcal{A}_{i\epsilon,n+1}$
(unless one of the two subamplitudes has only three legs).

In our approach we substitute the deformation
$\gengen_{2\to 1}$ by the set of generators $\gengen^{(1)}_{2\to k}$.
While $\gengen_{2\to 1}$ is an extremely simple generator,
its action involves computing loop integrals which are
complicated and potentially divergent.
For our generators we have essentially already performed
the regularised loop integrals.
Consequently the generators are somewhat more complicated,
but their action on amplitudes is straight-forward,
and the IR divergences are manifest.


\section{Conclusions}

In this work we have analysed, in the context of $\mathcal{N}=4$ SYM,
 the fate of the superconformal symmetries
of generic scattering amplitudes
and of the Yangian symmetries of planar amplitudes
 once radiative
corrections are taken into account. Our central message is that the IR singularities
pose no serious threat, as the symmetry generators can be deformed in such a
fashion as to render
the amplitudes, defined in a
dimensional regularisation scheme, invariant
to one-loop order. The key input was the inclusion of the deformation
of the tree-level generators due to collinear terms which give rise to leg
changing effects. Here it proved
very advantageous to represent these terms arising from the holomorphic anomaly
as an on-shell triangle graph.
Acting with the tree level generators on the branch
cuts of the one-loop scattering amplitudes then led to a natural one-loop
deformation of the representation which moreover could be lifted off the cut
by virtue of the cut-constructibility of the theory. Specialising to the planar theory,
an analogous result was obtained for the level-one Yangian generator
of momentum, which, along with the dilatation generator, could even be deformed
to the all-loop level.

Importantly, we were able to represent a universal form
of the deformation of any generator by putting the on-shell triangle anomaly
inside the loop in all possible ways. This construction turned out to
localise all integrals, so that effectively no loop integration had to be performed
and only tree-level data, consisting of vertices and amplitudes, entered the construction.

It would be interesting to repeat the study of deformations with an alternative regulator.
In particular the recently introduced Higgs IR regulator of \cite{Alday:2009zm,Henn:2010bk}
comes to mind, where the all-loop deformation of the dilatation and dual conformal
generator are simple and have a natural five-dimensional holographic
interpretation.

The most pressing question left open in our work is the closure of the algebra of the
one-loop deformed generators. While it is obvious that this will be the case for the
super-Poincar\'e and R-symmetry generators (by virtue of the regularisation procedure)
this is not at all so for the super-conformal part. Acting on the amplitudes the algebra
is of course trivially obeyed, but one would like to know if it closes for the generators.
Note that this will depend on the functional space on which the generators are allowed to act:
For instance, even at tree level the algebra closes
only on gauge-invariant functions \cite{Bargheer:2009qu}.
At one loop, extra constraints may become necessary,
such as, perhaps, (super) Poincar\'e-invariance or cyclicity.
The question of the algebra is also intimately connected
with the existence of a deformed Yangian symmetry at loop level, as the algebra
along with the explicit form of the level-one momentum generator gives rise to all
further level-one generators. To establish the complete Yangian symmetry a check
of the super-Serre relations for the deformed generators is also needed. However,
it is possible that while one can find an infinite tower of charges that annihilate amplitudes
the issue regarding their algebra may be subtle as is the case for non-ultralocal two-dimensional
integrable field theories. We intend to address these questions in the future.

Recently, remarkable formulae have been proposed based on an integral
over a certain Gra\ss{}mannian with manifest superconformal invariance,
which reproduce the $\superN=4$ SYM tree-level amplitudes
and even integral coefficients of higher-loop integral topologies \cite{ArkaniHamed:2009dn,Mason:2009qx,ArkaniHamed:2009vw}
thereby respecting Yangian symmetry \cite{Drummond:2010qh}.
It would be interesting to clarify the relation to our approach.

A further important question is whether the loop-deformed symmetries are constructive
in the sense of determining the loop-level amplitudes completely. This was argued
 to be the case for the tree-level amplitudes upon the incorporation of
the collinear terms in \cite{Korchemsky:2009hm, Bargheer:2009qu}.
We would certainly expect this to remain
true at the loop level. One immediate question is
what can the symmetries tell us about the remainder function, that is the difference between
the ABDK/BDS ansatz of \cite{Anastasiou:2003kj,Bern:2005iz}
and the true finite part of the amplitude,
starting at the two-loop order?
It is known that the naive dual conformal symmetries alone
are insufficient to fix this function and,
while there has already been significant
numerical and analytic work
\cite{Drummond:2008aq, Bern:2008ap, DelDuca:2009au, Brandhuber:2009da,Alday:2009dv},
determining its exact form remains a challenging open problem.

\subsection*{Acknowledgements}

We are most grateful to James Drummond for collaboration
in the early stages of this work, and we
thank him for useful comments throughout the work.
It is a pleasure to thank Lance Dixon, Gregory Korchemsky, Marc Magro, Radu Roiban,
Amit Sever, Emery Sokatchev, David Skinner and Pedro Vieira for discussions.
This work was supported in part by the Volkswagen Foundation.

\appendix

\section{Anomaly as a Three-Vertex}
\label{app:AnomalyVertex}

In this appendix we derive the form the anomaly vertex for superconformal
boosts by acting on a 3-vertex by a superconformal boost generator.
The result is expected to reproduce the superconformal boost deformation
found in \cite{Bargheer:2009qu}.

\subsection{Three-Vertices}

Consider the MHV 3-vertex
\[
A_3=\frac{\delta^4(P)\,\delta^8(Q)}{\sprod{1}{2}\sprod{2}{3}\sprod{3}{1}}\,.
\]
Momentum conservation does not allow for a proper phase space in $(3,1)$ signature.
Let us therefore continue in $(2,2)$ signature where the full phase space exists.
In this form, however, the anomaly cannot be seen easily,
and we first recast the vertex into a different form.
The two-component spinors $\lambda$ and $\tilde\lambda$ are now independent and real.
The amplitude can be represented in an alternative form which will
be useful for further considerations:
We note identities which allow to express spinors $\lambda$ or $\tilde\lambda$
in a given basis of
two different spinors $\mu,\mu'$ or $\tilde\mu,\tilde\mu'$, respectively
\<\label{eq:spinorbasis}
1\eq\bigabs{\sprod{\mu}{\mu'}}\int dx\,dx'\,
\delta^2(\lambda-x\mu-x'\mu'),
\nln
1\eq\bigabs{\cprod{\tilde\mu}{\tilde\mu'}}\int d\tilde x\,d\tilde x'\,
\delta^2(\tilde\lambda-\tilde x\tilde\mu-\tilde x'\tilde\mu').
\>
We use the latter to express $\tilde\lambda_1$ and $\tilde\lambda_2$ in a
basis of $\tilde\lambda_3$ and some reference spinor $\tilde\mu$.
\<
A_3\eq
\frac{\cprod{3}{\tilde\mu}^2}{\sprod{1}{2}\sprod{2}{3}\sprod{3}{1}}
\int
d\tilde x_1\,d\tilde x'_1\,d\tilde x_2\,d\tilde x_2'\,
\delta^2(\tilde\lambda_1-\tilde x_1\tilde\lambda_3-\tilde x_1'\tilde\mu)\,
\delta^2(\tilde\lambda_2-\tilde x_2\tilde\lambda_3-\tilde x_2'\tilde\mu)
\nl\quad\cdot
\delta^4\bigbrk{(\tilde x_1\lambda_1+\tilde x_2\lambda_2+\lambda_3)\tilde\lambda_3+(\tilde x_1'\lambda_1+\tilde x_2'\lambda_2)\tilde\mu}
\,
\delta^8\bigbrk{\lambda_1\eta_1+\lambda_2\eta_2+\lambda_3\eta_3}.
\nl
\>
A further identity converts a spinorial delta function into
a product of two regular delta functions
\[\label{eq:deltasplit}
\delta^2(a\lambda+b\mu)
=\frac{\delta(a)\,\delta(b)}{\bigabs{\sprod{\lambda}{\mu}}}\,,
\qquad
\delta^2(a\tilde\lambda+b\tilde\mu)
=\frac{\delta(a)\,\delta(b)}{\bigabs{\cprod{\tilde\lambda}{\tilde\mu}}}\,.
\]
The momentum delta function implies
$\lambda_3=-\tilde x_1\lambda_1-\tilde x_2\lambda_2$
which we can use to convert the supermomentum delta function to
$\delta^8\brk{\lambda_1(\eta_1-\tilde x_1\eta_3)+\lambda_2(\eta_2-\tilde x_2\eta_3)}$.
Again we can use the identity \eqref{eq:deltasplit},
but now the argument is fermionic and the measure factor
$\abs{\sprod{\lambda_1}{\lambda_2}}$ must be in the numerator instead of the denominator.
We end up with $A_3$ given through a set of delta functions
and a signum function
\<
A_3\eq
\sign\bigbrk{\sprod{1}{2}}
\int \frac{d\tilde x_1}{\tilde x_1}\,\frac{d\tilde x_2}{\tilde x_2}\,
\delta^2(\tilde x_1\lambda_1+\tilde x_2\lambda_2+\lambda_3)
\nl\qquad\cdot
\delta^2(\tilde\lambda_1-\tilde x_1\tilde\lambda_3)\,
\delta^2(\tilde\lambda_2-\tilde x_2\tilde\lambda_3)\,
\delta^4(\eta_1-\tilde x_1\eta_3)\,
\delta^4(\eta_2-\tilde x_2\eta_3).
\>
Similarly the conjugate MHV 3-vertex reads
\<
\bar A_3\eq
\sign\bigbrk{\cprod{1}{2}}
\int \frac{dx_1}{x_1}\,\frac{dx_2}{x_2}\,
\delta^2(\lambda_1-x_1\lambda_3)\,
\delta^2(\lambda_2-x_2\lambda_3)
\nl\qquad\cdot
\delta^2(x_1\tilde\lambda_1+x_2\tilde\lambda_2+\tilde\lambda_3)
\delta^4(x_1\eta_1+x_2\eta_2+\eta_3).
\>
%

\subsection{Anomaly Three-Vertices}

The collection of delta functions is superconformal
and hence it is annihilated by the superconformal boosts
$\gen{S}$ and $\bar{\gen{S}}$.
Only the signum factor violates invariance under
the conjugate superconformal boost $\bar{\gen{S}}\supup{free}$.
According to the identity $d\sign(x)=2dx\,\delta(x)$
the derivative in the generator converts it to a delta function
forcing $\tilde\lambda_1$ and $\tilde\lambda_2$ to be collinear.
Expressing $\tilde\lambda_1$ and $\tilde\lambda_2$ in a basis
of $\tilde\lambda_3$ and a reference spinor $\tilde\mu$ we arrive at
\<\label{eq:firstSbA3}
(\bar{\gen{S}}\supup{free})^B_{\dot\alpha}\bar A_3
\eq
2\varepsilon_{\dot\alpha\dot\gamma}
(\tilde\lambda_2^{\dot\gamma}\eta^B_1-\tilde\lambda_1^{\dot\gamma}\eta^B_2)
\int \frac{dx_1}{x_1}\,\frac{dx_2}{x_2}\,d\tilde x_1\,d\tilde x_2\,
\delta(1+x_1\tilde x_1+x_2\tilde x_2)
\nl\qquad\cdot
\delta^2(\lambda_1-x_1\lambda_3)\,
\delta^2(\lambda_2-x_2\lambda_3)
\nl\qquad\cdot
\delta^2(\tilde\lambda_1-\tilde x_1\tilde\lambda_3)\,
\delta^2(\tilde\lambda_2-\tilde x_2\tilde\lambda_3)\,
\delta^4(x_1\eta_1+x_2\eta_2+\eta_3).
\>

We can recast this expression into a different
form which may be more convenient for some purposes.
To that end we insert $1=\int d^{4|4}\Lambda'
\delta^2(\lambda')\delta^2(\tilde\lambda')
\delta^4(\eta'-\tilde x_2\eta_1+\tilde x_1\eta_2)$
and use an identity which holds when
$1+x_1\tilde x_1+x_2\tilde x_2=0$
\[
\delta^4(\eta'-\tilde x_2\eta_1+\tilde x_1\eta_2)\,
\delta^4(x_1\eta_1+x_2\eta_2+\eta_3)=
\delta^4(\eta_1-\tilde x_1\eta_3+x_2\eta')\,
\delta^4(\eta_2-\tilde x_2\eta_3-x_1\eta').
\]
Next we supplement $d^4\eta'$ by
$d^4\lambda'=d^2\lambda'd^2\tilde\lambda'$
and the corresponding delta function $\delta^4(\lambda')=
\delta^2(\lambda')\delta^2(\tilde\lambda')$
to $d^{4|4}\Lambda'\delta^4(\lambda')$. Subsequently we can add terms $\lambda',\tilde\lambda'$
to the delta function to make them appear more symmetric
\<
(\bar{\gen{S}}\supup{free})^B_{\dot\alpha}\bar A_3
\eq
2\int d^{4|4}\Lambda'\,\delta^4(\lambda')\,
\varepsilon_{\dot\alpha\dot\gamma}\tilde\lambda_3^{\dot\gamma}\eta'^B
\int \frac{dx_1}{x_1}\,\frac{dx_2}{x_2}\,d\tilde x_1\,d\tilde x_2\,
\delta(1+x_1\tilde x_1+x_2\tilde x_2)
\nl\qquad\cdot
\delta^2(\lambda_1-x_1\lambda_3+\tilde x_2\lambda')\,
\delta^2(\lambda_2-x_2\lambda_3-\tilde x_1\lambda')\,
\nl\qquad\cdot
\delta^2(\tilde\lambda_1-\tilde x_1\tilde\lambda_3+x_2\tilde\lambda')\,
\delta^2(\tilde\lambda_2-\tilde x_2\tilde\lambda_3-x_1\tilde\lambda')
\nl\qquad\cdot
\delta^4(\eta_1-\tilde x_1\eta_3+x_2\eta')\,
\delta^4(\eta_2-\tilde x_2\eta_3-x_1\eta').
\>

In order to convert the expression to the physical $(3,1)$ spacetime signature
we perform a change of variables such that $\tilde x_{1,2}=\pm\bar x_{1,2}$.
Here we must distinguish three different cases depending on
the energy signatures of the particles:
$(\pm\pm\mp)$, $(\mp\pm\pm)$ and $(\pm\mp\pm)$.
They are achieved by the substitutions
($0\leq \alpha,\beta\leq \half\pi$, $0\leq \varphi,\vartheta<2\pi$)
\[
\begin{array}[b]{rclrclrclrcl}
x_1\eq e^{-i\varphi}\sin\alpha,&
\tilde x_1\eq-e^{i\varphi}\sin\alpha,&
x_2\eq e^{-i\vartheta}\cos\beta,&
\tilde x_2\eq-e^{i\vartheta}\cos\beta,
\\[1ex]
x_1\eq e^{-i\varphi+i\vartheta}\tan\alpha,&
\tilde x_1\eq +e^{i\varphi-i\vartheta}\tan\alpha,&
x_2\eq e^{i\vartheta}\sec\beta,&
\tilde x_2\eq -e^{-i\vartheta}\sec\beta,
\\[1ex]
x_1\eq e^{i\varphi}\sec\alpha,&
\tilde x_1\eq -e^{-i\varphi}\sec\alpha,&
x_2\eq e^{-i\vartheta+i\varphi}\tan\beta,&
\tilde x_2\eq+e^{i\vartheta-i\varphi}\tan\beta.
\end{array}
\]
The delta function for the $x$'s leads to $\beta=\alpha$.
We combine the delta functions
$\delta^{4|4}(\Lambda)=\delta^4(\lambda)\delta^4(\eta)
=\delta^2(\lambda)\delta^2(\tilde\lambda)\delta^4(\eta)$
and multiply with a suitable prefactor of $-1/2$ to obtain the anomaly vertex
\<\label{eq:AnomVertexSb}
(\bar S_3)^B_{\dot\alpha}
\eq
-\half(\bar{\gen{S}}\supup{free})^B_{\dot\alpha}\bar A_3
\nln\eq
-2\int d^{4|4}\Lambda'\,\delta^4(\lambda')\,
\varepsilon_{\dot\alpha\dot\gamma}\tilde\lambda_3^{\dot\gamma}\eta'^B
\int d\alpha\,d\varphi\,d\vartheta\,e^{i\varphi+i\vartheta}
\nl\qquad\cdot
\delta^{4|4}(e^{-i\varphi}\bar\Lambda_3\sin\alpha+e^{i\vartheta}\bar\Lambda'\cos\alpha-\Lambda_1)
\nl\qquad\cdot
\delta^{4|4}(e^{-i\vartheta}\bar\Lambda_3\cos\alpha-e^{i\varphi}\bar\Lambda'\sin\alpha-\Lambda_2)
\nl+\mbox{2 cyclic images}.
\>

An analogous construction leads to the anomaly vertex
for the superconformal boost
\<
(\gen{S}\supup{free})_{\alpha B}A_3\eq
2\varepsilon_{\alpha\gamma}
(\lambda_2^\gamma \partial_{1,B}-\lambda_1^\gamma\partial_{2,B})
\int \frac{d\tilde x_1}{\tilde x_1}\,\frac{d\tilde x_2}{\tilde x_2}\,dx_1\,dx_2\,
\delta(1+x_1\tilde x_1+x_2\tilde x_2)
\nl\quad\cdot
\delta^2(\lambda_1-x_1\lambda_3)\,
\delta^2(\lambda_2-x_2\lambda_3)
\nl\quad\cdot
\delta^2(\tilde\lambda_1-\tilde x_1\tilde\lambda_3)\,
\delta^2(\tilde\lambda_2-\tilde x_2\tilde\lambda_3)
\nl\quad\cdot
\delta^4(\eta_1-\tilde x_1\eta_3)\,
\delta^4(\eta_2-\tilde x_2\eta_3).
\>
Here we insert $1=\int d^{4|4}\Lambda' \delta^{4|4}(\Lambda')$,
expand some of the delta function by terms in $\Lambda'$
and finally make the above replacements to
convert to $(3,1)$ signature
\<
(S_3)_{\alpha B}\eq
-\half
(\gen{S}\supup{free})_{\alpha B}A_3
\nln\eq
-2\int d^{4|4}\Lambda'\, \delta^{4|4}(\Lambda')\,
\varepsilon_{\alpha\gamma}\lambda_3^\gamma \partial'_{B}
\int d\alpha\,d\varphi\,d\vartheta\,e^{-i\varphi-i\vartheta}
\nl\qquad\cdot
\delta^{4|4}(e^{-i\varphi}\bar\Lambda_3\sin\alpha+e^{i\vartheta}\bar\Lambda'\cos\alpha-\Lambda_{1})
\nl\qquad\cdot
\delta^{4|4}(e^{-i\vartheta}\bar\Lambda_3\cos\alpha-e^{i\varphi}\bar\Lambda'\sin\alpha-\Lambda_{2})
\nl+\mbox{2 cyclic images}.
\>

\subsection{Tree-Level Superconformal Anomaly}

The above expressions agree (up to a conventional overall factor and phase redefinitions)
with the superconformal boost deformations
found in \cite{Bargheer:2009qu}.
Let us repeat the calculation for the colour-ordered planar MHV
amplitudes in order to fix the overall factors.

Consider the holomorphic anomaly
for spinor variables \eqref{eq:spinoranomaly}.
First we resolve the delta function in terms of an explicit
relation between the spinors using an identity analogous to \eqref{eq:spinorbasis}
\[
\delta^2\bigbrk{\sprod{\lambda}{\mu}}
=
\int_0^\infty 2r\, dr\,\int_0^{2\pi} d\varphi \,
\delta^2(\lambda-re^{i\varphi}\mu)
\bigbrk{\delta^2(\tilde\lambda-re^{-i\varphi}\tilde\mu)
+\delta^2(\tilde\lambda+re^{-i\varphi}\tilde\mu)}.
\]
Now we can compute the anomaly of MHV amplitudes
\<
(\bar{\gen{S}}\supup{free})^B_{\dot\alpha}
A_{n}
\eq
4\pi
\varepsilon_{\dot\alpha\dot\gamma}
\sum_{k=1}^n
\int_0^\infty r\, dr\,\int_0^{2\pi} d\varphi \,
\frac{\bigbrk{\tilde\lambda^{\dot\gamma}_{k+1}\eta_k^B -\tilde\lambda^{\dot\gamma}_{k}\eta_{k+1}^B}\,\delta^4(P)\,\delta^8(Q)}{\sprods{1}{2}\ldots\sprod{k}{k+1}^0 \ldots\sprods{n}{1}}
\nl\qquad\cdot
\delta^2(\lambda_k-re^{i\varphi}\lambda_{k+1})\,
\bigbrk
{
\delta^2(\tilde\lambda_k-re^{-i\varphi}\tilde\lambda_{k+1})
-\delta^2(\tilde\lambda_k+re^{-i\varphi}\tilde\lambda_{k+1})
}.
\nl
\>

We compare this to the deformation $\bar{\gen{S}}_{1\to 2}$
of the representation which consists
in inserting the anomaly three-vertex into the amplitude
\<
(\bar{\gen{S}}_{1\to 2})^B_{\dot\alpha}
A_{n-1}\eq
\sum_{k=1}^n
\int d^{4|4}\Lambda_{\mathrm{a}}\,
\sign(E_{k}E_{k+1})
\nl\qquad\cdot
\bar S_3(k,k+1,\mathrm{\bar a})^B_{\dot\alpha}
A_{n-1}(1,\ldots,k-1,\mathrm{a},k+2,\ldots,n).
\>
For the anomaly vertex we use the above result \eqref{eq:AnomVertexSb}
where the prefactor was already chosen correctly.
The deformation of the representation yields the contribution
\<
(\bar{\gen{S}}_{1\to 2})^B_{\dot\alpha}
A_{n-1}\eq
4\pi
\sum_{k=1}^n
\varepsilon_{\dot\alpha\dot\gamma}
\frac{\bigbrk{\tilde\lambda_{k+1}^{\dot\gamma}\eta^B_{k}-\tilde\lambda_k^{\dot\gamma}\eta^B_{k+1}}\,\delta^4(P)\,\delta^8(Q)}{\sprods{1}{2}\ldots \sprod{k}{k+1}^0\ldots\sprods{n}{1}}
\nl\quad\cdot
\bigg[
-\int_0^\infty r\,dr\,d\varphi\,
\delta^{2}(\lambda_k-re^{i\varphi}\lambda_{k+1})\,
\delta^{2}(\tilde\lambda_{k}-re^{-i\varphi}\tilde\lambda_{k+1})
\nl\qquad
+\int_0^1 r\,dr\,d\varphi\,
\delta^{2}(\lambda_k-re^{i\varphi}\lambda_{k+1})\,
\delta^{2}(\tilde\lambda_{k}+re^{-i\varphi}\tilde\lambda_{k+1})
\nl\qquad
+\int_{1}^{\infty} r\,dr\,d\varphi\,
\delta^{2}(\lambda_k-re^{i\varphi}\lambda_{k+1})\,
\delta^{2}(\tilde\lambda_k+re^{-i\varphi}\tilde\lambda_{k+1})
\bigg],
\nl
\>
where each of the three term originates from the
above three components of \eqref{eq:AnomVertexSb}.
This shows that the prefactors for the three terms have to be
chosen as in \eqref{eq:AnomVertexSb} in order
for the anomaly to be cancelled.


\section{One-Loop MHV Amplitude}
\label{sec:MHVapp}

In this appendix we collect results and identities for the (planar)
one-loop MHV amplitude and the underlying ``2-mass easy'' box integrals.
The n-point MHV amplitude in ${\cal N}=4$ SYM was found in
\cite{Bern:1994zx} and the derivations of many of these results can be
found there.

\subsection{Box Integrals}

The one-loop correction to any amplitude can be obtained
as a linear combination of scalar box integrals
\[
I^{\square}=-i\int \frac{\mu^{2\epsilon}d^{4-2\epsilon}\ell}{(2\pi)^{4-2\epsilon}}\,
\frac{1}{\ell^2(\ell+p_1)^2(\ell+p_1+p_2)^2(\ell-p_4)^2}\,.
\]
For MHV amplitudes the only contributions come from
special ``2-mass easy'' box integrals
with light-like momentum inflow at two opposite corners, $p_2^2=p_4^2=0$.
It makes sense to split the integral $I^\square$
up into a dimensionless loop function $F$
and a rational prefactor
\[
I^\square=\frac{F}{16\pi^2\Delta}\,,
\qquad
\Delta=-\half(st-uv).
\]
where the invariants $s,t,u,v$ are defined as
\[
s=(p_1+p_2)^2,\quad
t=(p_1+p_4)^2,\quad
u=p_1^2,\quad
v=p_3^2.
\]
In dimensional regularisation the $F$ takes the following form%
\footnote{Actually all terms --- not just the first four, divergent ones ---
should be proportional to $\mu^{2\epsilon}$.
For the finite terms this plays almost no role
and hence such minute factors can be safely discarded.
The only place where it does matter is in collinear limits.}
\<\label{eq:F2me}
F\eq
-\frac{c_\epsilon}{\epsilon^2}\lrbrk{\frac{s}{-\mu^2}}^{-\epsilon}
-\frac{c_\epsilon}{\epsilon^2}\lrbrk{\frac{t}{-\mu^2}}^{-\epsilon}
+\frac{c_\epsilon}{\epsilon^2}\lrbrk{\frac{u}{-\mu^2}}^{-\epsilon}
+\frac{c_\epsilon}{\epsilon^2}\lrbrk{\frac{v}{-\mu^2}}^{-\epsilon}
+\half\log^2\frac{s}{t}
\nl
+\Li_2\lrbrk{1-\frac{u}{s}}
+\Li_2\lrbrk{1-\frac{u}{t}}
+\Li_2\lrbrk{1-\frac{v}{s}}
+\Li_2\lrbrk{1-\frac{v}{t}}
-\Li_2\lrbrk{1-\frac{uv}{st}}.
\>
It has been normalised such that the coefficients
of the resulting $\Li_2$ and $\log^2$ terms are $\pm 1$ and $\pm\half$,
respectively.
Here $c_\epsilon$ is a frequently occurring function of dimensional
regularisation parameter $\epsilon$
\[
c_\epsilon=(4\pi)^{\epsilon}\frac{\gammafn(1+\epsilon)\,\gammafn(1-\epsilon)^2}{\gammafn(1-2\epsilon)}
=1+\order{\epsilon}.
\]

The one-mass and massless box functions
can be viewed as a special case with $u=0$ or/and $v=0$.
Here the third or/and fourth terms in \eqref{eq:F2me} are singular
and the correct prescription in dimensional regularisation is to drop them altogether.

Note that the above expression is not meant to reproduce the
physically correct imaginary part in all cases.
One would have to pick the applicable Riemann sheet of the function
for each physical situation.
Here we have written it
such that the function is real when all invariants $s,t,u,v$ are negative.

\subsection{BCF Construction}

\begin{figure}\centering
\includegraphics{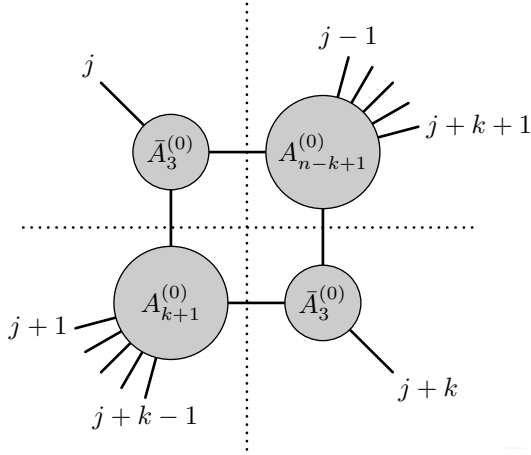}
\caption{A box integral contribution to the one-loop MHV amplitude}
\label{fig:BCFBox}
\end{figure}

The coefficients of the scalar box integrals for any one-loop amplitude in
${\cal N}=4$ SYM
are determined through quadruple cuts \cite{Britto:2004nc}.
Due to the tight supersymmetry constraints on MHV amplitudes,
all coefficients must equal the tree-level amplitude.
One can thus write
\[
A^{(1)}_n=A^{(0)}_n\,M^{(1)}_n.
\]
where $M^{(1)}$ is a sum over normalised 2-mass easy scalar box integrals,
cf.\ \figref{fig:BCFBox},%
\footnote{In this sum every term effectively appears twice,
hence a factor of $\half$.
This way of writing the sum has the benefit that no
distinction has to be made between even and odd $n$.}
\[\label{eq:MHVSum}
M^{(1)}_n=\half\sum_{k=2}^{n-2} \sum_{j=1}^n F(\sinv{k}{j},\sinv{k}{j+1},\sinv{k-1}{j+1},\sinv{k+1}{j}).
\]
Note that the boundary terms $k=2$ and $k=n-2$
in the sum a third leg of the box is light-like
and the loop function is actually a one-mass box.
As described above, in these cases one has to carefully drop singular terms
from $F$ in \eqref{eq:F2me}.
Furthermore, many of the individual terms in $F\supup{\square2m}$ cancel in the sum $M^{(1)}_{n}$
for which we obtain the convenient final result
\<\label{eq:AloopApp}
M^{(1)}_n\eq
-
\sum_{j=1}^n
\frac{c_\epsilon}{\epsilon^2}
\lrbrk{\frac{\sinv{2}{j}}{-\mu^2}}^{-\epsilon}
+\sfrac{1}{6}n\pi^2
-
\half\sum_{k=3}^{n-3}\sum_{j=1}^n
\Li_2\lrbrk{1-\frac{\sinv{k-1}{j+1}\sinv{k+1}{j}}{\sinv{k}{j}\sinv{k}{j+1}}}
\nl
-
\half\sum_{k=2}^{n-3}\sum_{j=1}^n
\log^2\frac{\sinv{k}{j}}{\sinv{k+1}{j}}
+
\quarter\sum_{k=2}^{n-2}\sum_{j=1}^n\log^2\frac{\sinv{k}{j}}{\sinv{k}{j+1}}\,.
\>
%

\subsection{Variations}
\label{sec:Mvar}

For acting with superconformal symmetries we must take derivatives
of loop function $M^{(1)}$ with respect to the external momenta.
As these appear only within the Mandelstam invariants $\sinv{k}{j}$
it suffices to compute the variation w.r.t.\ them
\<\label{eq:Fvar}
\delta M^{(1)}_n\eq
+
\half\sum_{j=1}^n
\lrbrk{\delta\log\frac{\sinv{2}{j-1}\sinv{2}{j+1}}{\mu^4}}
\frac{c_\epsilon}{\epsilon}\lrbrk{\frac{\sinv{2}{j}}{-\mu^2}}^{-\epsilon}
\nl
+
\half\sum_{k=2}^{n-3}\sum_{j=1}^n
\lrbrk{\delta\log\frac{\Delta\rng{k}{j}}{\Delta\rng{k+1}{j-1}}}
\log \frac{\sinv{k+1}{j}}{\sinv{k}{j}}\,.
\>
Here $\Delta\rng{k}{j}$ is the following combination of invariants
which arises from derivatives of the dilogs in \eqref{eq:F2me}
\[
\Delta\rng{k}{j}=-\half(\sinv{k}{j}\sinv{k}{j+1}-\sinv{k-1}{j+1}\sinv{k+1}{j}).
\]
Reducing all invariants to $\sinv{k-1}{j+1}$ plus extra terms, this equals
\[
\Delta\rng{k}{j}=
(p_k\cdot p_{j+k})(P\rng{k-1}{j+1}\cdot P\rng{k-1}{j+1})
-2(p_j\cdot P\rng{k-1}{j+1})(p_{j+k}\cdot P\rng{k-1}{j+1})
.
\]
In four dimensions one can furthermore
use spinor helicity variables
to write $p_j\cdot p_k=\half \sprod{j}{k}\cprod{k}{j}$.
Then $\Delta\rng{k}{j}$ magically factorises into
two terms $\Upsilon\rng{k}{j}=\sbra{j}P\rng{k-1}{j+1}\cket{j+k}$
\<\label{eq:DeltaFac}
\Delta\rng{k}{j}\eq
\quarter\sum_{m,n=j+1}^{j+k-1}\lrbrk{
(p_k\cdot p_{j+k})(p_{m}\cdot p_{n})
-(p_j\cdot p_m)(p_{j+k}\cdot p_n)
-(p_j\cdot p_n)(p_{j+k}\cdot p_m)
}
\nln\eq
-\half\sum_{m,n=j+1}^{j+k-1}
\sprod{j}{m}\cprod{m}{j+k}\sprod{j+k}{n}\cprod{n}{j}
\nln\eq
-\half\sbra{j}P\rng{k-1}{j+1}\cket{j+k}\sbra{j+k}P\rng{k-1}{j+1}\cket{j}
=\half\Upsilon\rng{k}{j}\Upsilon\rng{n-k}{j+k}.
\>
After substituting $\Delta\rng{k}{j}$ and using the identities
\<
\Upsilon\rng{2}{j-1}\eq\sprod{j-1}{j}\cprod{j}{j+1},
\nln
\Upsilon\rng{n-2}{j+2}\eq-\sprod{j+1}{j+2}\cprod{j}{j+1},
\nln
\sinv{2}{j}\eq-\sprod{j}{j+1}\cprod{j}{j+1}
\>
we end up with a convenient form for the variation of the loop function
\<
\label{eq:MHVvar}
\delta M^{(1)}_n\eq
\sum_{j=1}^n
\lrbrk{\delta\log\frac{\sprod{j-1}{j}\cprod{j}{j+1}\sprod{j+1}{j+2}}{-\mu^2 \sprod{j}{j+1}}}
\frac{c_\epsilon}{\epsilon}\lrbrk{\frac{\sinv{2}{j}}{-\mu^2}}^{-\epsilon}
\nl
-
\sum_{k=2}^{n-3}\sum_{j=1}^n
\lrbrk{\delta\log\frac{\Upsilon\rng{k+1}{j-1}}{\Upsilon\rng{k}{j}}}
\log \frac{\sinv{k+1}{j}}{\sinv{k}{j}}\,.
\>

\subsection{Collinear Configurations}
\label{app:app_collinear}
Finally we would like to address the question
what happens when two adjacent legs, say $n-1$ and $n$,
are strictly collinear
while evaluating the loop integral $M^{(1)}_{n}$.%
\footnote{Note that one has to distinguish between
the collinear limit of the loop function
and its value when two momenta are collinear.
This does not mean that the limit is not smooth,
but it apparently does not commute with removing the regulator.}
Most terms of the sum \eqref{eq:MHVSum}
reduce to terms of $M^{(1)}_{n-1}$
when combining the two collinear momenta
into one $p_{n-1}+p_n\to p_{n-1}$.
This is because the function $M^{(1)}_{n}$ depends
only on ranges of momenta $P\rng{k}{j}$ in $\sinv{k}{j}$.
The only exceptions arise when the range begins or ends between
the collinear momenta.

Let us therefore analyse more carefully the differences between
$M^{(1)}_{n}$ and $M^{(1)}_{n-1}$. Assume that the collinear momenta
obey
\[
p_{n-1}\to zp_{n-1},\quad
p_n\to \bar zp_{n-1},\qquad
z+\bar z=1.
\]
The Mandelstam invariants then reduce according to
\[
\sinv{k}{j}\to
\begin{cases}
\sinv{k}{j}                       &\mbox{when }j<n-k,\\
z\sinv{k}{j}+\bar z\sinv{k-1}{j}  &\mbox{when }j=n-k,\\
\sinv{k-1}{j}                     &\mbox{when }n-k<j<n,\\
\bar z\sinv{k}{j-1}+z\sinv{k-1}{j}&\mbox{when }j=n,\\
\sinv{k}{j-1}                     &\mbox{when }n<j,
\end{cases}
\]
To account for these different cases, we should split up the
sum over $j$ in \eqref{eq:MHVSum} into the
ranges $\set{1,\ldots,n-k-2}$, $\set{n-k+1,\ldots, n-2}$ and treat
the four remaining values separately.
It turns out that almost all terms combine as follows
(see e.g.\ \cite{Brandhuber:2004yw})
\[
M^{(1)}_n\to
M^{(1)}_{n-1}
+F(0,\bar z\sinv{2}{n-1},\sinv{2}{n-1},0)
+F(z\sinv{2}{n-2},0,0,\sinv{2}{n-2}).
\]
In combining some terms we made use of a splitting identity
for the box function
\[
F(s,t,u,v)
=
F(s,zt+\bar zv,\bar zs+zu,v)
+F(\bar zs+zu,t,u,zt+\bar zv).
\]
It follows from two dilog identities ($x=u/s$, $y=v/t$)
\<
0\eq
\Li_2(1-z)+\Li_2\lrbrk{1-\frac{1}{z}}+\half\log^2 z,
\nln
0\eq
+\Li_2\lrbrk{1-x}
+\Li_2\lrbrk{1-y}
-\Li_2(1-xy)
\nl
-\Li_2\bigbrk{z(1-x)}
-\Li_2\bigbrk{\bar z(1-y)}
-\log(\bar z+zx)\log(z+\bar zy)
\nl
-\Li_2\frac{z(1-y)}{z+\bar zy}
-\Li_2\frac{\bar z(1-x)}{\bar z+zx}
+\Li_2\frac{z(1-xy)}{z+\bar zy}
+\Li_2\frac{\bar z(1-xy)}{\bar z+zx}\,.
\>

It is tricky to determine the value of $F(0,t,u,0)$.
It originates form a one-mass box integral evaluated at $s=0$.
Unfortunately, the expression \eqref{eq:F2me} is very singular at this point.
One way to obtain a value is to
consider a particular configuration of invariants
and show that
$F(zu,t,u,zt)=0$.
Then the limit $z\to 0$ suggests that
$F(0,t,u,0)=0$,
but it is certainly not a smooth limit in general.
Another indication in favour of this result is
that the original box integral
$I^\square(0,t,u,0)$ is finite.
Furthermore, the multiplicative factor $\Delta=0$
and thus $F(0,t,u,0)=0$.
So we are led to the conclusion that the MHV loop factor
with two collinear momenta reduces exactly to
the loop factor with the two collinear momenta replaced by their sum
\[
\bigeval{M^{(1)}_n}_{n-1\parallel n}= M^{(1)}_{n-1}.
\]
However, this is not the only way to define this limit
and in general
\[
\bigeval{M^{(1)}_n}_{n-1\parallel n}= M^{(1)}_{n-1}+r\indup{S}~,
\]
with the function $r\indup{S}$ being non-trivial. For
example, in the prescription given
by \cite{Kosower:1999xi} and used widely in
literature this function is at one-loop
that given in \secref{sec:1loopsplit}.

\section{Computation of the On-Shell Triangle Anomaly}
\label{app:Triangle}

In this appendix we compute the on-shell triangle integral
\eqref{eq:Bdef}
\<
(T\rng{k}{j})^B_{\dot\alpha}\eq
\frac{1}{8\pi^3} \int
d^{4|4}\Lambda\indup{a}\,
d^{4|4}\Lambda\indup{b}\,
d^{4|4}\Lambda\indup{c}\,
\sign(\sinv{k}{j}-\sinv{k+1}{j})\,
\bar S_3(\mathrm{\bar b},\mathrm{\bar a},j+k)^B_{\dot\alpha}
\nl\cdot
A^{(0)}_{k+2}(\mathrm{a},\mathrm{c},j,\ldots,j+k-1)
A^{(0)}_{n-k+1}(\mathrm{\bar c},\mathrm{b},j+k+1,\ldots,j+n-1).
\>
representing the superconformal anomaly.
Substituting the anomaly vertex \eqref{eq:AnomalyMain},
performing the trivial phase integrals
over $\varphi$ and $\vartheta$
and pulling out an overall tree-level MHV amplitude
we arrive at
\<\label{eq:SbarTriangle}
(T\rng{k}{j})^B_{\dot\alpha}\eq
\frac{1}{\pi} A^{(0)}_{n}
\varepsilon_{\dot\alpha\dot\gamma}\tilde\lambda_{j+k}^{\dot\gamma}
\sign(\sinv{k+1}{j}-\sinv{k}{j})
\int d\alpha\,d^4\lambda\indup{c}\,
\delta^4(p\indup{a}+p\indup{c}+P\rng{k}{j})
\nl\quad
\int d^4\eta\indup{c}\,d^4\eta'\, \eta'^B\,\delta^8(q\indup{a}+q\indup{c}+Q\rng{k}{j})
\nl\quad
\frac{\sprod{j-1}{j}}{\sprod{j-1}{\mathrm{c}}\sprod{\mathrm{c}}{j}}\,
\frac{\sprod{j+k-1}{j+k}\sprod{j+k}{j+k+1}}
{\sprod{j+k-1}{\mathrm{a}}\sprod{\mathrm{a}}{\mathrm{c}}\sprod{\mathrm{c}}{\mathrm{b}}\sprod{\mathrm{b}}{j+k+1}}\,,
\>
where the spinor helicity variables of two intermediate
particles are determined through the momentum fraction angle $\alpha$
\[
\begin{array}{rclcrclcrcl}
\lambda\indup{a}\eq\lambda_{j+k}\cos\alpha,
&\quad&
\tilde\lambda\indup{a}\eq\tilde\lambda_{j+k}\cos\alpha,
&\quad&
\eta\indup{a}\eq\eta_{j+k}\cos\alpha-\eta'\sin\alpha,
\\[1ex]
\lambda\indup{b}\eq\lambda_{j+k}\sin\alpha,
&\quad&
\tilde\lambda\indup{b}\eq\tilde\lambda_{j+k}\sin\alpha,
&\quad&
\eta\indup{b}\eq\eta_{j+k}\sin\alpha+\eta'\cos\alpha.
\end{array}
\]
We now evaluate the three lines of the above expression in parts.
The bosonic integral on the first line of \eqref{eq:SbarTriangle}
is of the form%
\footnote{We use a delta function for momenta $P$ in spinor notation
$\delta^4(P^{\beta\dot\alpha})$ rather than in vector notation
$\delta^4(P^\mu)=4\delta^4(P^{\beta\dot\alpha})$.}
\[
\int d^4\lambda_2\,
\delta^4(p_1+p_2+P)
=2\pi \delta\bigbrk{\sbra{1}P\cket{1}+P^2 } \,.
\]
The spinor $\lambda_2,\tilde\lambda_2$ is fixed up to a phase
\[
\sket{2}=xP\cket{1},
\qquad
\cbra{2}=\tilde x\sbra{1}P,
\qquad
x\tilde x=-\frac{1}{\sbra{1}P\cket{1}},
\qquad \tilde x=\pm x^{\ast}.
\]
We then substitute the appropriate momenta and note that
$\sbra{j+k}P\rng{k}{j}\cket{j+k}=\sinv{k+1}{j}-\sinv{k}{j}$.
The resulting delta function subsequently localises the integral over $\alpha$
\<
\int d\alpha\,d^4\lambda\indup{c}\,
\delta^4(p\indup{a}+p\indup{c}+P\rng{k}{j})
\eq
2\pi \int d\alpha\,
\delta\bigbrk{(\sinv{k+1}{j}-\sinv{k}{j})\cos^2\alpha+\sinv{k}{j}}
\nln\eq
\frac{\pi}{\bigabs{\sinv{k+1}{j}-\sinv{k}{j}}\sin\alpha\cos\alpha}\,.
\>
The variables are fixed at%
\footnote{In fact, for real $0\leq\alpha\leq\pi/2$ we must assume $0\leq \cos^2\alpha\leq 1$
implying an energy signature $(\pm\pm\mp)$ of the three particles.
Here we also allow for the ranges $\cos^2\alpha<0$ and $1<\cos^2\alpha$.
These additional ranges correspond to the energy signatures
$(\mp\pm\pm)$ and $(\pm\mp\pm)$ of the three particles
contributed by the 2 cyclic images in \eqref{eq:AnomalyMain}.}
\[
\lambda\indup{c}^\beta=
x(P\rng{k}{j})^{\beta\dot\alpha}\varepsilon_{\dot\alpha\dot\gamma}\tilde\lambda^{\dot\gamma}_{j+k},
\qquad
\cos^2\alpha=\frac{\sinv{k}{j}}{\sinv{k}{j}-\sinv{k+1}{j}}\,.
\]
The fermionic integral on the second line of \eqref{eq:SbarTriangle}
can be evaluated by expressing $Q$ in a basis of $\lambda\indup{a}$
and $\lambda\indup{c}$ and using \eqref{eq:deltasplit}
to split up the $\delta^8$ into the product of
two $\delta^4$
\<
\earel{}
\int d^4\eta\indup{c}\,d^4\eta'\, \eta'^B
\delta^8(q\indup{a}+q\indup{c}+Q\rng{k}{j})
\nln\eq
\bigbrk{\sin\alpha\cos\alpha\sprod{j+k}{\mathrm{c}}}^3
\bigbrk{\cos^2\alpha\sprod{j+k}{\mathrm{c}}\eta_{j+k}^B
-\lambda\indup{c}^{\delta}\varepsilon_{\delta\epsilon}(Q\rng{k}{j})^{\epsilon B}}.
\>
The rational spinor function on the third line of \eqref{eq:SbarTriangle} reads
\<
\earel{}
\frac{\sprod{j-1}{j}}{\sprod{j-1}{\mathrm{c}}\sprod{\mathrm{c}}{j}}\,
\frac{\sprod{j+k-1}{j+k}\sprod{j+k}{j+k+1}}
{\sprod{j+k-1}{\mathrm{a}}\sprod{\mathrm{a}}{\mathrm{c}}\sprod{\mathrm{c}}{\mathrm{b}}\sprod{\mathrm{b}}{j+k+1}}
\nln\eq
\frac{\sprod{j-1}{j}}{(\sin\alpha\cos\alpha\sprod{j+k}{c})^2\sprod{j}{\mathrm{c}}\sprod{j-1}{\mathrm{c}}}\,.
\>

We assemble and simplify these expressions and altogether we find
\[
(T\rng{k}{j})^B_{\dot\alpha}=
A^{(0)}_{n}
\varepsilon_{\dot\alpha\dot\gamma}\tilde\lambda_{j+k}^{\dot\gamma}
\frac{\sprod{j-1}{j}
\bigbrk{\tilde\lambda_{j+k}^{\dot\kappa} \varepsilon_{\dot\kappa\dot\lambda}
(P\rng{k}{j})^{\delta\dot\lambda} \varepsilon_{\delta\epsilon}(Q\rng{k}{j})^{\epsilon B}
-\sinv{k}{j}\eta_{j+k}^B}
}{
\sbra{j}P\rng{k}{j}\cket{j+k}\,
\sbra{j-1}P\rng{k}{j}\cket{j+k}
}\,.
\]
%

\section{Details of Six-Point NMHV Amplitude}
\label{sec:NMHVapp}
Here we give some more details relevant to the six-point NMHV amplitude
\[
A\indup{6;NMHV}^{(1)}=A\indup{6;MHV}^{(0)} \left(
R_{146}F_6^{[1]}+ R_{251}F_6^{[2]}+R_{362}F_6^{[3]}
+R_{413}F_6^{[1]}+R_{524}F_6^{[2]}+R_{635}F_6^{[3]}
\right).
\]
Explicit expressions of the $R$ invariants can be found in \cite{Drummond:2008vq}.
For the six-point amplitudes they can be written as, e.g.\
\[
R_{146}=\frac{\sprods{3}{4}\sprods{5}{6}\sprods{6}{1}\sprods{4}{5}\delta^{(4)}(\zeta_{456})}{x_{14}^2\langle 1|P_1^3|4]\langle 3|x_{36}|6][45][56]}
\]
where
\[
\zeta_{456}=\eta_4[56]+\eta_5[64]+\eta_6[45])~.
\]
A general relation amongst the $R$ structures which holds
for any amplitude is
\[
R_{r,r+2,s}=R_{r+2,s,r+1}
\]
and an important  relation which holds for the specific case of the six-point amplitude is
\[
\label{eq:threetermid}
R_{146}+R_{135}+R_{136}=R_{624}+R_{625}+R_{635}~.
\]
Thus, using $R_{251}=R_{625}$, $R_{362}=R_{136}$, $R_{413}=R_{624}$, $R_{524}=R_{135}$,
 we could pick $R_{146}$, $R_{625}$, $R_{136}$, $R_{624}$ and $R_{135}$ as the independent
 structures (i.e.\ we remove $R_{635}$ in terms of the others).
 A useful expression for the sum of box functions which occur in $A^{(1)}_{6;\mathrm{NMHV}}$, from which one can extract the variation,  is:
\<
\label{eq:boxfn}
F^{[1]}_6\eq
-\frac{c_{\epsilon}}{2\epsilon^2}\sum_{i=1}^6
\left( \frac{-t^2_{i}}{\mu^2}\right)^{-\epsilon}
\nl
-\Big[ \log \frac{t_1^3}{t_1^2}  \log \frac{t_1^3}{t_2^2}
+  \log \frac{t_1^3}{t_4^2}  \log\frac{t_1^3}{t_5^2}-  \log\frac{t_1^3}{t_3^2}  \log\frac{t_1^3}{t_6^2}\Big]
\nl
+\frac{1}{2}\Big[  \log \frac{t_1^2}{t_4^2}  \log\frac{t_2^2}{t_5^2}
+ \log \frac{t_6^2}{t_1^2}  \log\frac{t_2^2}{t_3^2}+ \log \frac{t_3^2}{t_4^2}  \log\frac{t_5^2}{t_6^2}\Big]+\frac{\pi^2}{3}~.
\>
The remaining $F^{[i]}_6$ can be found by cyclicly permuting this expression.
The variation can be written as
\<
& &\bar{\gen{S}}^{(0)}_{1 \to 1}A^{(1)}\indup{6;NMHV}=A^{(0)}\indup{6;MHV}R_{146}
 \Bigg[
\sum_{i=1}^6\bar{\gen{S}}^{(0)}_{1 \to 1}
\Big[-\frac{c_{\epsilon}}{2\epsilon^2}\left(\frac{t_i^2}{-\mu^2}\right)^{-\epsilon}\Big]+
 \log t_1^3\,\bar{\gen{S}}^{(0)}_{1 \to 1}\log \Big[\frac{t_1^2 t_2^2 }{t_3^2  t_6^2} \Big]\nn\\
& &\qquad +\log \frac{t_1^2}{-\mu^2}\, \bar{\gen{S}}^{(0)}_{1 \to 1}\left(
-\frac{1}{2}\log \Big[ \frac{(t_2^2)^2 }{t_3^2 t_1^3}\Big]\right)
+\log \frac{t_2^2}{-\mu^2}\, \bar{\gen{S}}^{(0)}_{1 \to 1}
\left(-\frac{1}{2}\log \Big[ \frac{(t_1^2)^2}{t_6^2 t_1^3}\Big]\right)\nn\\
& & \qquad +\log \frac{t_3^2}{-\mu^2}\, \bar{\gen{S}}^{(0)}_{1 \to 1}\left(
+\frac{1}{2}\log \Big[ \frac{t_1^2 }{t_1^3}\Big]\right)
+\log \frac{t_4^2}{-\mu^2}\, \bar{\gen{S}}^{(0)}_{1 \to 1}\left(
-\frac{1}{2}\log \Big[ \frac{t_2^2 }{ t_6^2}\Big]\right)\nn\\
& & \qquad +\log \frac{t_5^2}{-\mu^2}\, \bar{\gen{S}}^{(0)}_{1 \to 1}\left(
-\frac{1}{2}\log \Big[ \frac{t_1^2 }{ t_3^2}\Big]\right)
+\log \frac{t_6^2}{-\mu^2}\, \bar{\gen{S}}^{(0)}_{1 \to 1}\left(
-\frac{1}{2}\log \Big[ \frac{t_1^3 }{t_2^2}\Big]\right)\Bigg]+\dots
\>
where we have made use of the relations
\[
R_{146}\bar{\gen{S}}^{(0)}_{1 \to 1}\log \frac{t_5^2}{t_1^3}=R_{146}\bar{\gen{S}}^{(0)}_{1 \to 1}\log \frac{t_4^2}{t_1^3}=0~.
\]
The remaining terms can again be found by cyclic permutations.

\section{One-Loop Anomaly of MHV-4 Amplitude}
\label{sec:5NMHV}

In this appendix we consider the one-loop superconformal invariance
of amplitudes using the approach of \cite{Sever:2009aa}.
Although the deformation of the superconformal generators itself
does not make reference to the CSW rules \cite{Cachazo:2004kj},
their application to amplitudes is hard to define properly without them.
Here we perform an explicit calculation for the four-particle
MHV amplitude pointing out an ambiguity and how it may be resolved.

In the proposal \cite{Sever:2009aa}
the one-loop superconformal anomaly for $n$-particle MHV amplitudes
is compensated by the action of $\bar{\gen{S}}_{2\to 1}$ \eqref{eq:G21G30}
on $(n+1)$-particle NMHV amplitudes.
We now consider the simplest case of $4$-particle MHV amplitudes.
We apply $\bar{\gen{S}}_{2\to 1}$ directly
to the 5-particle NMHV amplitude without the use of CSW rules
in order to understand the subtleties concerning collinear configurations.
The 5-particle NMHV amplitude reads \cite{Drummond:2008bq}
\[
A\supup{NMHV}_5=
\frac{\delta^4(P)\,\delta^8(Q)\,\delta^4(\eta_3\cprods{4}{5}+\eta_4\cprods{5}{3}+\eta_5\cprods{3}{4})}
{\cprods{1}{2}\cprods{2}{3}\cprods{3}{4}\cprods{4}{5}\cprods{5}{1}\sprods{1}{2}^4}\,.
\]
We wish to act with $\bar{\gen{S}}_{2\to 1}$ on legs $4$ and $5$.
In order to gain access to the collinear divergence,
we express $\tilde\lambda_4$ in a basis of $\tilde\lambda_3$ and
$\tilde\lambda_5$ using \eqref{eq:spinorbasis}
(for simplicity we shall work in $(2,2)$ signature
where $\lambda$ and $\tilde\lambda$ are independent)
\[
A\supup{NMHV}_5=
\int \frac{d\tilde y}{\tilde y}\,\frac{d\tilde z}{\tilde z}\,
\delta^2(\tilde z\tilde\lambda_3+\tilde y\tilde\lambda_5-\tilde\lambda_4)\,
\delta^4(\tilde z\eta_3+\tilde y\eta_5-\eta_4)\,
\frac{\cprods{3}{5}^2\,\bigabs{\cprods{3}{5}}\,\delta^4(P)\,\delta^8(Q)}
{\cprods{1}{2}\cprods{2}{3}\cprods{5}{1}\sprods{1}{2}^4}\,.
\]
We then multiply by the vertex $\bar S_3$ in the form of \eqref{eq:firstSbA3}
\<
\bar S^B_{\dot a}\eq
-\varepsilon_{\dot\alpha\dot\gamma}
(\tilde\lambda_5^{\dot\gamma}\eta^B_4-\tilde\lambda_4^{\dot\gamma}\eta^B_5)
\int \frac{dx_4}{x_4}\,\frac{dx_5}{x_5}\,d\tilde x_4\,d\tilde x_5\,
\delta(1+x_4\tilde x_4+x_5\tilde x_5)
\nl\qquad\cdot
\delta^2(\lambda_4-x_4\lambda_{45})\,
\delta^2(\lambda_5-x_5\lambda_{45})
\nl\qquad\cdot
\delta^2(\tilde\lambda_4+\tilde x_4\tilde\lambda_{45})\,
\delta^2(\tilde\lambda_5+\tilde x_5\tilde\lambda_{45})\,
\delta^4(x_4\eta_4+x_5\eta_5-\eta_{45}).
\>
and integrate out $\Lambda_4$ and $\Lambda_5$.
Here we make sure that at first only the integrations over spinors are performed;
the integrals over the auxiliary variables are left untouched.
This yields
\<
((\bar{\gen{S}}_{2\to 1})\rng{1}{4})^B_{\dot\alpha}A\supup{NMHV}_5\eq
-\int dx_4\,dx_5\,d\tilde x_4\,d\tilde x_5\,
d\tilde y\,d\tilde z\,
\delta(1+x_4\tilde x_4+x_5\tilde x_5)\,
\delta(\tilde z)\,\delta(\tilde y-\tilde x_4/\tilde x_5)
\nl\qquad\cdot
\frac{(x_4\tilde x_5\tilde y+x_5\tilde x_5)^3}{x_4 x_5 \tilde x_5\tilde y (x_4\tilde x_4+x_5\tilde x_5)^2}\,
\frac{\delta^4(P-(1+x_4\tilde x_4+x_5\tilde x_5)p_4)\,\delta^8(Q)}
{\sprods{1}{2}\sprods{2}{3}\sprods{3}{4}\sprods{4}{1}}
\nl\qquad\cdot
\frac{\varepsilon_{\dot\alpha\dot\gamma}\tilde\lambda_{4}^{\dot\gamma}}{\cprods{3}{4}}
\lrbrk{
\frac{x_4\tilde x_4+x_5\tilde x_5}{\tilde x_5}\,\eta^B_3
+\frac{\tilde y-\tilde x_4/\tilde x_5}{\tilde z}\,\eta^B_4
}.
\>
The second term in the brackets is undetermined because it
equals $0/0$ on the support of the delta functions;
let us replace it by some undetermined expression $\ast$.
The remaining term is well-defined and we can now
perform the integrals over the auxiliary variables
\[
((\bar{\gen{S}}_{2\to 1})\rng{1}{4})^B_{\dot\alpha}A\supup{NMHV}_5=
-\int \frac{dx_4}{x_4}\,\frac{dx_5}{x_5}\,\frac{d\tilde x_4}{\tilde x_4}\,\frac{d\tilde x_5}{\tilde x_5}\,
\delta(1+x_4\tilde x_4+x_5\tilde x_5)
\frac{\varepsilon_{\dot\alpha\dot\gamma}\tilde\lambda_{4}^{\dot\gamma}\,(\eta^B_3+\ast\eta^B_4)}{\cprods{3}{4}}\,
A\supup{MHV}_4.
\]
After performing the integrals over the $x$'s in proper $(3,1)$ Minkowski signature
making sure that the energies of particles $4$ and $5$ have equals signs
in agreement with \eqref{eq:G21G30}
we end up with
\[
((\bar{\gen{S}}_{2\to 1})\rng{1}{4})^B_{\dot\alpha}A\supup{NMHV}_5=
-16\pi^2\int_{0}^{\pi/2}\frac{d\alpha}{2\sin\alpha\cos\alpha}\,
\frac{\varepsilon_{\dot\alpha\dot\gamma}\tilde\lambda_{4}^{\dot\gamma}\,(\eta^B_3+\ast\eta^B_4)}{\cprods{3}{4}}\,
A\supup{MHV}_4.
\]
The integral is clearly divergent. This divergence is of infrared type,
and it is expected from \cite{Sever:2009aa}.
In fact it looks similar
to the action of $\bar{\gen{S}}^{(1)}_{2\to2}$ defined in \eqref{eq:planardiagonalDSSb}.
Luckily, we can adjust the undetermined coefficient $\ast$ in order to
match the structure precisely.
Noting an identity for valid for physical four-particle configurations
\[
\frac{\tilde\lambda_4\eta_3}{\cprods{3}{4}}
=
\frac{\tilde\lambda_4\eta_1}{\cprods{1}{4}}
+
\frac{\tilde\lambda_4\eta_4}{\cprods{1}{4}}
\frac{\sprods{2}{4}}{\sprods{2}{1}}
=
\frac{\tilde\lambda_4\eta_1}{\cprods{1}{4}}
-
\frac{\tilde\lambda_4\eta_4}{\cprods{3}{4}}
\frac{\sprods{2}{4}}{\sprods{2}{3}}
\]
we set $\ast\to \half(\sprods{2}{4}/\sprods{2}{3})$ so that
the spinor structure becomes
\[
\frac{\varepsilon_{\dot\alpha\dot\gamma}\tilde\lambda_{4}^{\dot\gamma}(\eta^B_3+\half(\sprods{2}{4}/\sprods{2}{3})\eta^B_4)}{\cprods{3}{4}}\,
A\supup{MHV}_4
=
\lrbrk{
\frac{\varepsilon_{\dot\alpha\dot\gamma}\tilde\lambda_{4}^{\dot\gamma}\eta^B_3}{2\cprods{3}{4}}
+
\frac{\varepsilon_{\dot\alpha\dot\gamma}\tilde\lambda_{4}^{\dot\gamma}\eta^B_1}{2\cprods{1}{4}}
}
A\supup{MHV}_4
.
\]
The expression agrees with \eqref{eq:planardiagonalDSSb} up to cyclic
permutations of the four particles.
Even the prefactor appears to agree once one imposes some ad-hoc regulator.

In conclusion we see that the alternative proposal \cite{Sever:2009aa}
does appear to give analogous results
up to interpreting terms of the kind 0/0 in a suitable fashion.
This is presumably achieved by the CSW rules.
Regularising divergent terms is another (separate) issue.
In our proposal, cf.\ \secref{sec:MHVex},
all expressions are well-defined in dimensional regularisation
(or any other suitable scheme), however, at the cost of having
a substantially more involved deformation than just
$\bar{\gen{S}}_{2\to 1}$.

\section{Conventions and Identities}

In this appendix we list a few of the basic conventions
and identities used in this paper.

\paragraph{Coupling Constant.}
We define the coupling constant $g$ which we use for the loop expansion:
\[
g^2=
\frac{\lambda}{16\pi^2}\,,
\qquad
\lambda=g\indup{YM}\cas.
\]
The 't~Hooft coupling is written using the adjoint Casimir
which equals $\cas=N\indup{c}$ for a $\grp{SU}(N\indup{c})$ gauge group.
We employ a dimensional regularisation scheme with
straight minimal subtraction.
The undesirable contributions of Euler's gamma constant
and $\log 4\pi$'s are absorbed into a constant $c_\epsilon$
which typically dresses poles in $\epsilon$
\[\label{eq:ceps}
c_\epsilon=
(4\pi)^{\epsilon}\frac{\gammafn(1+\epsilon)\,\gammafn(1-\epsilon)^2}{\gammafn(1-2\epsilon)}
=\exp\bigbrk{(\log 4\pi-\gamma)\epsilon-\sfrac{1}{12}\pi^2\epsilon^2+\order{\epsilon^3}}
=1+\order{\epsilon}.
\]
The precise form of $c_\epsilon$ has no physical significance whatsoever.

\paragraph{Complex Integrals.}
For performing integrals over the complex plane we use the
convention that $z=(x+iy)/\sqrt{2}$.
We can then write two-dimensional integrals as
simple products of one-dimensional integrals as follows
\[
d^2z = dz\,d\bar z.
\]
This proves particularly useful when Wick rotating to
two independent real coordinates $z,\bar z$.
Similarly, the corresponding delta functions factorise
and the holomorphic anomaly reads
\[
\delta^2(z)=\delta(z)\delta(\bar z),
\qquad
\frac{\partial}{\partial \bar z}\,\frac{1}{z}=2\pi \delta(z)\delta(\bar z).
\]

\paragraph{Gauge Generators.}
The generators $T^a$ and structure constants $f^{abc}$
of the $\grp{U}(N\indup{c})$ gauge group are normalised such that
\[
\comm{T^a}{T^b}=if^{abc}T^c,
\qquad
\Tr(T^aT^b)=\delta^{ab}.
\]
This leads to the following identities in traces
\[T^aXT^a=\Tr X,
\qquad
T^a \Tr(T^aX)=X.\]

\paragraph{Vectors and Spinors.}

For vectors we choose $(-,+,+,+)$ as the signature of the Minkowski metric,
hence the mass shell condition for a massive particle is $p^2=-m^2$.

The conversion between vector and spinor indices
is normalised such that for two
light-like momenta $p_1,p_2$
\[
(p_1+p_2)^2=
2p_1\cdot p_2=
p_1^{\delta\dot\alpha}
\varepsilon_{\dot\alpha\dot\gamma}
p_2^{\beta\dot\gamma}
\varepsilon_{\beta\delta}
=
\Tr(\varepsilon p_1\varepsilon p_2^{\scriptscriptstyle\mathrm{T}})
=
\cprod{1}{2}\sprod{2}{1}.
\]
Moreover for a generic momentum $P$ one has
$P\varepsilon P^{\mathrm{\scriptscriptstyle T}}=-P^2\varepsilon$.

\bibliographystyle{nb}
\bibliography{scloop}

\end{document}